\documentclass{jfm}

\usepackage{graphicx,graphics,color,colortbl,xcolor}
\usepackage{natbib}
\usepackage{array} 
\usepackage{booktabs}
\usepackage{tabularx}
\usepackage{amsmath}
\usepackage{etoolbox} 
\usepackage{breqn}

\ifCUPmtlplainloaded \else
  \checkfont{eurm10}
  \iffontfound
    \IfFileExists{upmath.sty}
      {\typeout{^^JFound AMS Euler Roman fonts on the system,
                   using the 'upmath' package.^^J}%
       \usepackage{upmath}}
      {\typeout{^^JFound AMS Euler Roman fonts on the system, but you
                   dont seem to have the}%
       \typeout{'upmath' package installed. JFM.cls can take advantage
                 of these fonts,^^Jif you use 'upmath' package.^^J}%
      }
  \else
  \fi
\fi

\ifCUPmtlplainloaded \else
  \checkfont{msam10}
  \iffontfound
    \IfFileExists{amssymb.sty}
      {\typeout{^^JFound AMS Symbol fonts on the system, using the
                'amssymb' package.^^J}%
       \usepackage{amssymb}%
         \let\leq=\leqslant
         \let\geq=\geqslant
      }{}
  \fi
\fi

\ifCUPmtlplainloaded \else
  \IfFileExists{amsbsy.sty}
    {\typeout{^^JFound the 'amsbsy' package on the system, using it.^^J}%
     \usepackage{amsbsy}}
    {\providecommand\boldsymbol[1]{\mbox{\boldmath $##1$}}}
\fi

\providecommand\bnabla{\boldsymbol{\nabla}}

\newcommand{\N}[1]{\textcolor{black}{#1}}

\newsavebox{\astrutbox}
\sbox{\astrutbox}{\rule[-5pt]{0pt}{20pt}}

\newcommand\md{\mathrm{d}}

\title[Compressible Convection Playground]{A playground for compressible natural convection with a nearly uniform density}

\author[T. Alboussi\`ere, J. Curbelo, F. Dubuffet, S. Labrosse and Y. Ricard]%
{Thierry Alboussi\`ere$^a$\thanks{Email address for correspondence: thierry.alboussiere@ens-lyon.fr},
 Jezabel Curbelo$^b$, Fabien Dubuffet$^a$, St\'ephane Labrosse$^a$ and Yanick Ricard$^a$}

\affiliation{$^a$Univ. Lyon, Laboratoire de G\'eologie de Lyon, CNRS, ENS-Lyon, Universit\'e Lyon 1, France \\
$^b$Departament de Matem\`atiques, Universitat Polit\`ecnica de Catalunya, Barcelona, Spain}

\pubyear{}
\volume{}
\pagerange{}

\date{?; revised ?; accepted ?. - To be entered by editorial office}

\begin{document}

\maketitle

\begin{abstract}
	In the quest to understand the basic universal features of compressible convection, one would like to disentangle genuine consequences of compression from spatial variations of transport properties. Unfortunately this is sometimes difficult: for instance, one may choose to consider a fluid with uniform dynamic viscosity, but then, compressible effects will often generate a density gradient and consequently the kinematic viscosity will not be uniform. Similarly, a uniform thermal conductivity leads to a gradient of thermal diffusivity. In the present work, we consider a very peculiar equation of state, whereby entropy is solely dependent on density, so that a nearly isentropic fluid domain is nearly isochoric. Within this class of equations of state, there is a thermal adiabatic gradient and a key property of compressible convection is still present, namely its capacity to viscously dissipate a large fraction of the thermal energy involved, of the order of the well-named dissipation number. In a series of anelastic approximations, under the assumption of an infinite Prandtl number, the number of governing parameters can be brought down to two, the Rayleigh number and the dissipation number. This framework is proposed as a playground for compressible convection, an opportunity to extend the vast corpus of theoretical analyses on the Oberbeck-Boussinesq equations regarding stability, bifurcations or the determination of upper bounds for the turbulent heat transfer. Here, \N{in a two-dimensional geometry,} we concentrate on the structure of upward and downward plumes depending on the dissipation number, on the heat flux dependence on the dissipation number and on the ratio of dissipation to convective heat flux. For all Rayleigh numbers, a change in the vertical temperature profile is observed in the range of dissipation number between $0$ and less than $0.4$, associated with the weakening of ascending plumes. For larger dissipation numbers, the heat flux dependence on this number is found to be well predicted by Malkus's model of critical layers. For dissipation numbers of order unity, in the limit of large Rayleigh numbers, dissipation becomes related to the entropy heat flux at each depth, so that the vertical dissipation profile can be predicted, and consequently so does the total ratio of dissipation to convective heat flux.


\end{abstract}

\begin{keywords}
	Rayleigh-B\'enard, Equation\N{s} of state, Compressible convection, Anelastic Approximations
\end{keywords}

\section{Introduction}

Natural convection is the response of a fluid with a specific equation 
 of state subject to a thermal or compositional buoyancy forcing -- for instance an imposed temperature difference in a gravity field -- while conservation laws of mass, momentum and energy
apply. Compressibility effects are inevitable, but in a famous approximation due to \cite{oberbeck} and \cite{boussinesq}, pressure effects are relegated to a secondary role. The Oberbeck-Boussinesq model is so simple and has become so popular that most theoretical studies of natural convection are made in its framework. We will concentrate on the Rayleigh-B\'enard configuration (\cite{benard,rayleigh}), mostly relevant to stars and planets. In these large natural objects, where compressibility plays a large role, fewer theoretical results have been derived and we think this is essentially due to the absence of a simple set of equations which could be used as a playground for studies of compressible convection. 

In a simple geometry, the Oberbeck-Boussinesq model has just two dimensionless parameters, the Rayleigh $Ra$ and Prandtl $Pr$ numbers. The Prandtl number is only relevant to the inertial effects in the momentum equation. In the limit of infinite Prandtl numbers as it is the case for convection in the solid mantle of terrestrial planets, this parameter becomes irrelevant, so that there is a single governing parameter, the Rayleigh number $Ra$. Since the stability analysis of \cite{rayleigh}, a century of theoretical investigations were led and thousands of scientific papers have been published using the Oberbeck-Boussinesq model. As soon as compressibility effects are taken into account, the number of governing parameters jumps to six \citep[see][]{cdadlr2019}: $Pr$, $Ra$, $\alpha T$, $c_p / c_v$, $T_h / T_c$, $\alpha g H / c_p$, where the symbols $\alpha$, $c_p$, $c_v$, $T$, $T_h$, $T_c$, $g$, $H$ denote the coefficient of thermal expansion, heat capacity at constant pressure, heat capacity at constant volume, temperature, hot imposed temperature, cold imposed temperature, gravity and the height of the fluid layer, respectively. Depending on the equation of state considered, there can be fewer parameters ($\alpha T = 1$ for ideal gases) or more parameters needed to describe the fluid. This -- and numerical difficulties mentioned below -- explain why there are comparatively few studies devoted to compressible convection and stresses the need to propose simple approaches that might enable the community to identify basic features of compressible effects. Hopefully our work will contribute to this objective. 

\citet{carnot1824} was the first to suggest that the low temperature at high altitude were due to adiabatic decompression of air in ascending currents, while descending currents and adiabatic compression would bring air back to the higher temperature at sea level. This was later generalized by \citet{schwarzschild} for the temperature profile in convective regions of stars, while \citet{jeffreys} proved that the stability of compressible convection was governed by the superadiabatic Rayleigh number with the same threshold (for moderate compressibility) as obtained by \citet{rayleigh} in the Boussinesq approximation. Later, stability was studied by a number of authors \citep{spiegel,gs1970,bormann,busse1967,pc1987,flp1992}. More recently, we published a model of stability valid for any arbitrary equation of state and uniform \N{dynamic} viscosity and conductivity \citep{AR2017}. 

A difficulty with the fully compressible governing equations was soon spotted: they contain the fast sound wave and the convective timescales. In many instances those timescales are so different that the numerical task of computing convection is overwhelming. Anelastic approximations (AA) were developed for the atmosphere, Earth's core and stars (\cite{op1961,br1995,lf1999}), valid in convective regions, consisting in an expansion about an isentropic state. The simplified anelastic liquid approximation ALA was proposed in \citep{ajs05} in which the role of pressure fluctuations on other thermodynamic quantities is neglected. In the stably stratified cases, sound-proof models have also been developed \citep{durran1989,lipps1990,vlbwz2013} in the pseudo-incompressible approximation. \N{\citet{lecoanet2014} note that the pseudo-incompressible equation of state introduces some inaccuracies in the thermodynamic variables. }

When compressible effects are present, there is usually a significant range of temperatures in the system, since the adiabatic gradient is a key feature of compressible convection. The same is true for pressure, density, and so on. A consequence is that transport coefficients of heat or momentum -- thermal conductivity and \N{(dynamic)} viscosity -- are usually not uniform. It is then difficult to distinguish between consequences of compressibility and consequences of non-uniform transport properties. Even in the classical Boussinesq model can non-uniform transport coefficients be modelled, they are called the Non-Oberbeck-Boussinesq (NOB) effects, for instance in \cite{HSW13}. In the present paper, we try to minimize the NOB effects. For this reason, we choose uniform constant thermal conductivity and \N{(dynamic)} viscosity. However, when density varies so do kinematic viscosity and thermal diffusivity. Hence we make a peculiar choice of equation of state, such that density is constant when entropy is constant $s(\rho )$. This ensures that a nearly isentropic convective region is also a region of nearly uniform density and kinematic viscosity. We will see that the heat capacity $c_p$ and thermal diffusivity are also uniform where entropy is uniform.   

In the next section \ref{impossible}, we discuss the general validity of an equation of state and expand the case $s=s( \rho )$. Using that class of equations of state, section \ref{gov} is devoted to the description of the configuration and to writing the governing equations and anelastic approximations. In section \ref{start} we first show results of the initial phase of convection from rest, with a small superadiabatic Rayleigh number and a large dissipation number, in order to assess the validity of the different anelastic approximation models. We then show that a significant change in temperature profiles occurs at small dissipation number, in section \ref{smallD}, namely the disappearance of the top overshoot on the vertical averaged profile. Top and bottom asymmetry is further studied in section \ref{asymmetry} for larger values of the dissipation number. The basic model of critical boundary layer is applied to the compressible case in section \ref{HeatFlux} and provides a model for the change of heat flux when the dissipation number is increased. In section \ref{heatflux}, we introduce the expressions for the vertical heat flux in the different models (fully compressible and anelastic approximations), as well as that for the dissipation profile, under a form that will be suited to understand energy transfers in the final sections. The numerical results of global dissipation relative to the convective heat flux are shown for all models and a range of superadiabatic Rayleigh numbers and dissipation numbers in section \ref{DissHeatFlux}. A definite limit is observed at large superadiabatic Rayleigh numbers which is further studied in section \ref{profiles}. It is interpreted as a local mesoscale equilibrium state whereby the entropy flux contribution is found to correspond both to energy dissipation and to the main part of the heat flux. Finally, in section \ref{conf}, we consider the additional effects of inertia, cavity aspect ratio and boundary conditions, to show that another state of flow can be obtained which does not correspond to that local equilibrium and exhibits larger dissipation. However, those last boundary conditions with impermeable vertical walls are less relevant in the geophysical and astrophysical context. In conclusion (section \ref{conclusions}), our study gives support to the mesoscale equilibrium implying that the vertical profile of dissipation takes the form of the function $\alpha g / c_p$ in compressible convection in the limit of large \N{dissipation and} superadiabatic Rayleigh numbers. 

\N{Considering the very specific equation of state considered here, the condition of infinite Prandtl number, the two-dimensional geometry and the absence of rotation and magnetic effects, the results of this study should not be applied to stellar or planetary objects without further investigations. However, they provide a possible asymptotic behaviour for the large compressibility, large Rayleigh number, limit. It remains to determine under which conditions that behaviour will be observed.} 

\section{Impossible EoS $\rho (T)$ and possible EoS $\rho (s)$}
\label{impossible}

From gases to solids, a wide range of equations of state \N{(EoS in short)} is possible. From a theoretical point of view, one can wonder what should be a possible EoS and when a tentative EoS is impossible. One answer is that one should just start from a fundamental EoS under the form
\begin{equation}
	e=e(s, \rho). \label{fund_eos}
\end{equation}
where $s$, $e$ and $\rho$ are the specific entropy, specific energy and density, respectively. From Gibbs equation $\mathrm{d}e = T\mathrm{d}s + P/\rho ^2 \mathrm{d} \rho $ ($T$ temperature, $P$ pressure), one just needs $T$ to be positive, if we want to consider real existing conditions
\begin{equation}
	\left. \frac{\partial e}{\partial s} \right| _{\rho} > 0. \label{cond_poss}
\end{equation}
However, one rarely starts from a fundamental EoS \N{(\ref{fund_eos})}. Usually, one expresses density $\rho$ as a function of pressure $P$ and temperature $T$. The first obvious idea when one wishes to get rid of compressible effects -- and jump immediately into the Boussinesq approximation -- is to state that density is independent of pressure
\begin{equation}
	\rho=\rho(T). \label{eos_rhoT}
\end{equation}
We now investigate the consequences of this assumption \N{(\ref{eos_rhoT})} (see also \cite{gp2020}). We derive a general relationship, equation (A7) in \cite{AlRi13}, on the partial derivative of enthalpy \N{$h= e + P / \rho $} with respect to pressure at constant temperature
\begin{equation}
	\left. \frac{\partial h}{\partial P} \right| _T = \frac{1-\alpha T}{\rho} , \label{dhdP}
\end{equation}
\N{which is obtained from Gibbs equation under several forms (using the differential of $h$ and that of Gibbs free energy $g=h-Ts$) and deriving Maxwell relations}.
\N{Note that} the right-hand side -- assuming (\ref{eos_rhoT}) and hence $\alpha = -\rho ' / \rho $ where the prime denotes derivative with respect to the single variable $T$ -- is a function of $T$ only, that we denote $A$
\begin{equation}
	A = \frac{1}{\rho } + \frac{\rho ' T }{\rho ^2}, \label{AofT}
\end{equation}
Equation (\ref{dhdP}) is integrated to give an expression for the enthalpy
\begin{equation}
	h = A P + B, \label{formh}
\end{equation}
where $B$ is another function of temperature. This expression is used to write $\mathrm{d}h$ which is then substituted in Gibbs equation $\mathrm{d}h = T \mathrm{d} s + \mathrm{d}P / \rho $ leading to
\begin{equation} 
	\mathrm{d} s = \frac{A' P + B' }{T} \mathrm{d} T + \frac{\rho '}{\rho ^2} \mathrm{d} P .  \label{ds}
\end{equation}
\N{Considering that $A' = T (\rho '/ \rho ^2 )'$ from (\ref{AofT}), equation (\ref{ds})} implies the following form for $s$
\begin{equation}
	s = \frac{\rho '}{\rho ^2} P + C , \label{formofs}
\end{equation}
where $C$ is yet another function of $T$ which turns out to be an integral of $B'/T$. We can now write an expression for the Gibbs free energy $g = h -Ts$
\begin{equation}
	g = \frac{ P}{\rho } + B - T C . \label{formofg} 
\end{equation}
A condition of stability of a material substance \citep{bazarov} is that its Gibbs free energy $g$ should be a concave function of $P$ and $T$. If not, the substance would split into two different phases that have together a larger entropy, as for instance in the phase-change region of the Van der Waals model. Locally, a necessary condition is that the Hessian of $g$ (its matrix of second partial derivatives) is a negative-definite matrix, {\it i.e.} has alternatively negative and positive leading principal minors according to Sylvester's criterion \citep{sylvester}. The Hessian matrix is
\begin{equation}
	\left[ \begin{array}{l l}\left. \frac{\partial ^2 g }{ \partial T^2} \right | _P & \frac{\partial ^2 g }{ \partial T \partial P } \\ \frac{\partial ^2 g }{ \partial P \partial T } & \left. \frac{\partial ^2 g }{ \partial P^2} \right | _T  \end{array} \right]  \label{hessian}
\end{equation}
The first leading principal minor is $\partial ^2 g / \partial T^2$ at constant pressure. From Gibbs equation $\mathrm{d}g = -s \mathrm{d} T +  \mathrm{d}P / \rho$, we have 
\begin{equation}
	\left.  \frac{\partial g }{ \partial T} \right| _P = -s , \hspace{2 cm}	\left.	\frac{\partial g }{ \partial P} \right| _T = \frac{1}{\rho} , \label{dgdP}
\end{equation}
and hence 
\begin{equation}
	\left.	\frac{\partial ^2 g }{ \partial T^2} \right|_P = - \left.  \frac{\partial  s }{ \partial T} \right|_P = - \left( \frac{\rho '}{\rho ^2} \right) ' P - C' =  - \left( \frac{\rho '}{\rho ^2} \right) ' P - \frac{B'}{T} , \label{d2gdT2}
\end{equation}
from (\ref{formofs}), which can indeed be made negative for an appropriate choice of the function $\rho (T)$ and $B (T)$. Now, the second and last leading principal minor (in dimension two) is the determinant of the whole Hessian matrix. From (\ref{dgdP}), we can see that 
the second derivative of $g$ with respect to $P$ will be zero. The determinant of the Hessian matrix is then just equal to 
\begin{equation}
	\det \left[ \begin{array}{l l}\left. \frac{\partial ^2 g }{ \partial T^2} \right| _P & \frac{\partial ^2 g }{ \partial T \partial P } \\ \frac{\partial ^2 g }{ \partial P \partial T } & \left. \frac{\partial ^2 g }{ \partial P^2} \right| _T \end{array} \right] = - \left( \frac{\partial ^2 g }{ \partial P \partial T } \right) ^2 = - \left[ \left. \frac{\partial }{ \partial T} \left( \frac{1}{\rho} \right) \right| _P   \right] ^2 . \label{hessiandet}
\end{equation}
It is negative, meaning that $g$ is not a concave function of $T$ and $P$. 
The only way to save that equation of state would be to make this determinant zero: since it is the partial derivative of $1/\rho$ with respect to $T$ at constant $P$, it is zero only when $\rho $ is a constant. Such an equation of state is not interesting for thermal convection as no buoyancy would exist. Hence we consider equation (\ref{eos_rhoT}) as an impossible equation of state. 
Another related aspect can be noted from Mayer's relationship
\begin{equation}
	c_p - c_v = - \frac{T}{\rho ^2} \left. \frac{\partial P }{ \partial T} \right| _\rho \left. \frac{\partial \rho }{ \partial T} \right| _P .  \label{mayerT}
\end{equation}
That difference is infinite since ${\partial P }/{ \partial T}$ is infinite at constant $\rho$, \N{from Euler's chain rule $\left. \partial P / \partial T \right| _\rho \left. \partial T / \partial \rho \right| _P \left. \partial \rho / \partial P \right| _T = -1$}. Hence the choice of meaningful heat capacities is impossible. 

We now investigate another simple form of EoS, such that density is a function of entropy only and show that it satisfies marginally the criterion of stability, as noted in \cite{scott2001}. 
Let us identify all possible equations of state such that density is solely a function of entropy, or reciprocally such that entropy is solely a function of density
\begin{equation}
	s = s( \rho ) . \label{sofrho}
\end{equation}
The thermodynamics Gibbs equation can be written
\begin{equation}
	\md e = T \md s - P \md v = \left( T s' + \frac{P}{\rho ^2} \right) \md \rho ,   \label{gibbs}
\end{equation}
where $v$ is the specific volume ($v = 1 / \rho$) and the primes now denote the usual derivative with respect to the single variable $\rho$. It follows from the previous equation that $e$ is also solely a function of $\rho$, 
\begin{equation}
        e = e( \rho ) . \label{eofrho}
\end{equation}
The next consequence, by definition, is that the heat capacity at constant volume $c_v$ is zero
\begin{equation}
	c_v  \equiv \left. \frac{\partial e}{\partial T} \right| _v = \left. \frac{\partial e}{\partial T} \right| _\rho =  0 . \label{cv0}
\end{equation}
This shows that our choice is a limit case of valid equations of state, a negative $c_v$ would not be realistic. \N{Instead of considering that entropy is a function of density only, had we added a tiny dependence on temperature, we would probably have been able to obtain a strictly positive and small value for $c_v$ and that equation of state would have been perfectly valid.} \N{ Equation (\ref{gibbs})} \N{can also be written}
\begin{equation}
	T s' = e' - \frac{P}{\rho ^2}. \label{eqGibbs}
\end{equation}
Multiplying (\ref{eqGibbs}) by $\rho ^2$ and deriving with respect to temperature $T$ at constant pressure $P$ leads to the following expression for the coefficient of thermal expansion
\begin{equation}
	\alpha \equiv - \frac{1}{\rho } \left. \frac{\partial \rho }{\partial T} \right| _P = \frac{\rho s'}{-(\rho ^2 e' )' + T (\rho ^2 s')'} , \label{alpha}
\end{equation}
Using Gibbs equation (\ref{eqGibbs}) to extract $P / \rho $, we express the specific enthalpy $h$ as follows
\begin{equation}
	h \equiv e + \frac{P}{\rho} = e + \rho e' - \rho T s' . \label{enthalpy}
\end{equation}
From (\ref{enthalpy}) and (\ref{alpha}), after straightforward but slightly tedious steps, we derive an expression for the heat capacity at constant pressure 
\begin{equation}
	\N{c_p \equiv  \left. \frac{\partial h }{\partial T} \right| _P = - \rho T \alpha  s' . \label{cp}}
\end{equation}
At this point, from (\ref{alpha}) and (\ref{cp}), we note that $\alpha T / c_p$ -- which multiplied by gravity $g$ expresses the adiabatic gradient -- is solely a function of $\rho$
\begin{equation}
	\frac{\alpha T}{c_p} = - \frac{1}{\rho s'}, \label{alphaTsurcp}
\end{equation}
so that, in an isentropic region under a uniform gravity field, one can expect to observe a uniform adiabatic temperature gradient. The condition $s' < 0$ is needed to avoid a negative $\alpha$ or worse a negative $c_p$ according to equation (\ref{alphaTsurcp}). However, $c_p$ and $\alpha T$ are functions of $\rho$ and $T$, hence will not be uniform in an isentropic region. In order to avoid complexity, we assume -- in addition to $s$ being a function of $\rho$ -- that $(\rho ^2 e' )'$ is zero, hence 
\begin{equation}
        e = \frac{ K }{\rho } , \label{energy2}
\end{equation}
up to an irrelevant additive constant, and where the multiplicative constant $K$ is a parameter \N{whose value can be freely specified.} This eliminates the temperature dependence of $\alpha T$ and $c_p$. 
\N{With equation (\ref{energy2})} we have
\begin{align}
	\alpha T &= \frac{\rho s'}{\left( \rho ^2 s' \right) '}, \label{alphaT2} \\
	c_p &= - \frac{\rho ^2 {s'}^2}{\left( \rho ^2 s' \right) '} . \label{cp2}
\end{align}
The condition of stability on the leading principal minors of the Hessian matrix of $g$ is now examined. \N{With our choice for energy (\ref{energy2}) and the form of $h$ in (\ref{enthalpy}), Gibbs free energy $g \equiv h - Ts$ takes the form $g = -T (\rho s )'$. }
Using (\ref{dgdP}), we obtain
\begin{align}
	\left. \frac{\partial ^2 g }{\partial T^2} \right| _P &= - \left. \frac{\partial s}{\partial T} \right| _P = - s' \left. \frac{\partial \rho}{\partial T} \right| _P , \label{d2gdTT} \\
	\left. \frac{\partial ^2 g }{\partial P^2} \right| _T &= - \frac{1}{\rho ^2} \left. \frac{\partial \rho}{\partial P} \right| _T , \label{d2gdPP} \\
	\frac{\partial ^2 g }{\partial T \partial P}  &= - \left. \frac{\partial s}{\partial P} \right| _T = - s' \left. \frac{\partial \rho}{\partial P} \right| _T . \label{d2gdTP}
\end{align}
In order to evaluate these second derivatives, we need expressions for the partial derivatives of density with respect to temperature and density. From (\ref{eqGibbs}), we get 
\begin{equation}
	\left. \frac{\partial P }{\partial \rho } \right| _T = \left( \rho ^2 e' \right) ' - T \left( \rho ^2 s' \right) ' = - T \left( \rho ^2 s' \right) ', \label{drhodP} 
\end{equation}
owing to our choice $\left( \rho ^2 e' \right) ' = 0$. The inverse of (\ref{drhodP}) provides ${\partial \rho}/{\partial P}$ while (\ref{alphaT2}) is used to express ${\partial \rho}/{\partial T}$. When substituted in (\ref{d2gdTT}), (\ref{d2gdPP}) and (\ref{d2gdTP}), we obtain the Hessian matrix
\begin{equation}
	\left[ \begin{array}{l l} \left. \frac{\partial ^2 g }{\partial T^2} \right| _P & \frac{\partial ^2 g }{\partial T \partial P} \\ \frac{\partial ^2 g }{\partial T \partial P} & \left. \frac{\partial ^2 g }{\partial P^2} \right| _T \end{array} \right] = \left[ \begin{array}{l l} \frac{\rho ^2 {s'}^2}{T \left( \rho ^2 s' \right) '} & \frac{s'}{T \left( \rho ^2 s' \right) '} \\ \frac{s'}{T \left( \rho ^2 s' \right) '} & \frac{1}{\rho ^2 T \left( \rho ^2 s' \right) '} \end{array} \right], \label{hessiangS}
\end{equation}
The first leading principal minor is negative when $\left( \rho ^2 s' \right) ' < 0$, so that this condition must be fulfilled. The second minor is the whole determinant of the Hessian matrix and it is easy to check that it is zero. In that sense the equation of state $s (\rho )$ is just marginally stable.


There is still a large set of possibilities since we are free to consider any function $s(\rho )$, provided $\left( \rho ^2 s' \right) ' < 0$ and $s' < 0$ if one wishes to restrict the analysis to positive values of $\alpha$. Let us choose a set of such decreasing functions, defined as one of the following up to an irrelevant additive constant
\begin{align}
	s (\rho ) &= a \ln (\rho ), \label{slogr} \\
	\mathrm{or}\ \ 	s (\rho ) &= a \rho ^n, \hspace*{1 cm} \mathrm{for}\ n>0 \ \mathrm{a\ real\ number} , \label{spower} 
\end{align}
where $a<0$ is a negative constant real parameter. With the logarithm function $s \sim \ln (\rho )$, we have $\alpha T =1$ and a constant $c_p = -a$. With a power law $s \sim \rho ^n$, we have a constant $\alpha T$ between 1 and 0 whose value can be tuned by choosing the positive exponent $n$ and $c_p$ is a function of $\rho$ 
\begin{equation}
	\alpha T = \frac{1}{n+1}, \hspace*{1 cm} c_p = -a \frac{n}{n+1} \rho ^n,  \hspace*{1 cm} \mathrm{for}\ n > 0 . \label{alphaTetcppower}
\end{equation}
Other relations will be needed, namely the expressions of $P$ and $h$
\begin{align}
	P &= - K - a T \rho, \hspace*{1 cm} h = -a T, \label{Phlog} \\
	\mathrm{or}\ \ P &= - K - n a T \rho ^{n+1}, \hspace*{1 cm} h = - n a T \rho ^{n},  \hspace*{1 cm} \mathrm{for}\ n>0 . \label{Phpower}
\end{align}
We first remark that one of these equations of state is an ideal gas equation: this is the case of (\ref{slogr}) when $K=0$, $a=-c_p$ and corresponds to an ideal gas with $c_v = 0$. The marginal stability of this equation of state is reflected by the infinite speed of sound that results from a finite $c_p$ and a null $c_v$. 

\N{Although our equation of state was built from theoretical arguments, one may try to find real substances with a similar behaviour, at least in some range of temperature and pressure: a monoatomic gas with large molar mass has a small $c_v$ for instance. Radon gas is a good example. Next, the ratio $c_p/c_v$ can be made large (diverging to infinity) near the critical point, so that radon near the critical point would have the expected behaviour concerning heat capacities. However, the thermal expansion coefficient also diverges near the critical point and that does not match our equation of state. }

Among the large class of equations of state such that entropy is a function of density, driven by a principle of simplicity, we have identified a set of such equations, with $\alpha T $ constant ranging from $1$ (log function) to zero asymptotically (power law with $n \rightarrow \infty$). For all of them, $c_p$ and the expected adiabatic gradient are functions of density only. 

\section{Rayleigh-B\'enard configuration and governing equations}
\label{gov}

\N{We define the geometric configuration and boundary conditions that will be investigated in this paper. Different convection models are considered : complete continuum thermodynamic and dynamic equations (FC for 'fully compressible'), anelastic approximation (AA), anelastic liquid approximation (ALA) and a further simplified model (SCA for 'simple compressible approximation'). } 

\subsection{\N{Fully compressible model}}
\label{exact}

For simplicity, we take the infinite Prandtl number approximation which eliminates inertia. \N{This limit has been studied mathematically \citep{wang2004} and used for the study of mantle dynamics \citep{RicardTOG} for which Prandtl numbers are estimated around $10^{25}$: the effective kinematic viscosity of solids is much larger than their thermal diffusivity. Other objects, like the Earth's core, the interior of stars and of gaseous planets have low Prandtl numbers \citep{schaeffer2017,garaud2020,fuentes2020}.} For infinite Prandtl numbers, the governing equations of thermal convection are the following
\begin{align}
	\frac{\mathrm{D} \rho}{\mathrm{D} t} &= - \rho {\bnabla} \cdot {\bf u}, \label{continuity}\\
	{\bf 0} &= - {\mathbf \bnabla} P + \rho {\bf g} + \eta  \left[ {\mathbf \bnabla}^2 {\bf u} + \frac{1}{3} {\mathbf \bnabla} \left(\bnabla \cdot {\bf u} \right) \right], \label{momentum} \\
	\rho T \frac{\mathrm{D} s}{\mathrm{D} t} &= \dot{\epsilon} : \tau + \bnabla \cdot \left( k {\mathbf \bnabla } T \right) , \label{entropy}
\end{align}
where ${\bf u}$ is the velocity field, ${\bf g}$ the gravity field, $\eta $ the \N{dynamic} viscosity of the fluid, $k$ its thermal conductivity, $\mathrm{D} / \mathrm{D}t = \partial / \partial t + {\bf u}\cdot {\mathrm \bnabla} $ is the material derivative, $\dot{\epsilon}$ denote the tensor of rate of deformation, $\tau$ the Newtonian stress tensor defined as
\begin{equation}
	\tau _{ij} = 2 \eta \left[ \dot{\epsilon}_{ij} - \frac{1}{3} \left( {\mathbf \bnabla} \cdot {\bf u} \right) \delta_{ij} \right], \label{tau}
\end{equation}
using Stokes' assumption regarding the bulk viscosity.

We consider a \N{two-dimensional} rectangular domain, with horizontal periodic boundary conditions. In a Cartesian frame $(x,z)$, the horizontal axis is $x$ while the vertical axis is $z$. The height of the cavity is $H$, its length $L$. The aspect ratio is set to $L/H = 4 \sqrt{2}$, corresponding to twice the horizontal period of the stability analysis in the Boussinesq approximation for an infinite layer. Gravity is uniform ${\bf g} = - g {\bf e}_z $ along the direction of the unit vertical vector ${\bf e}_z$.  
The thermal boundary conditions are that of a hot temperature $T_h$ at the bottom and a cold temperature $T_c$ at the top. At the top and bottom boundaries, the normal velocity component is zero and so is the tangential viscous stress, {\it i.e. } $\partial u_x / \partial z = 0$. Because there is no natural constraint on the horizontal velocity, we impose that the horizontal average of $u_x$ is zero on the top boundary. Finally, instead of imposing some pressure value, we impose that the average density in the domain is $\rho _0$
\begin{equation}
	\frac{1}{H L}	\int \rho\  \md x \md z  =  \rho _0 . \label{meanrho}
\end{equation}
\N{This condition is an initial condition and that integral cannot change in time with impermeable or periodic boundaries.} 

\begin{figure}
\begin{center}
	\includegraphics[width=13 cm, keepaspectratio]{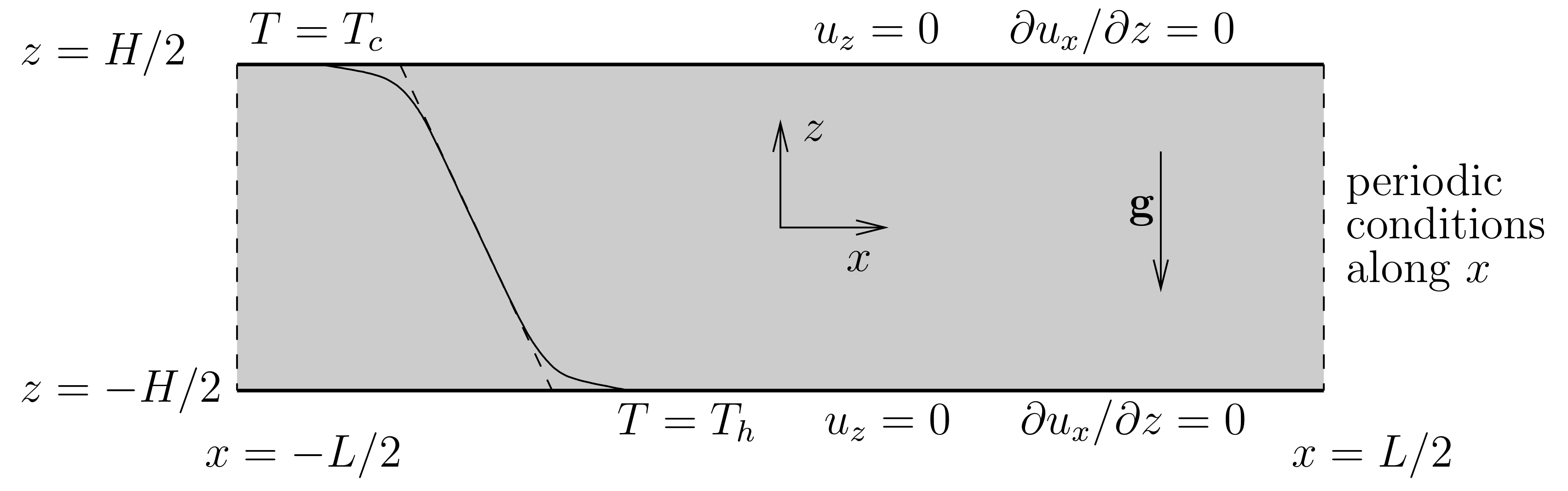}
	\caption{Geometry and boundary conditions. A typical vertical temperature profile is sketched (solid line) with an adiabatic, isentropic, temperature profile (dashed line). }
\label{geometry}
\end{center}
\end{figure}

The set of equations is complete when an equation of state is specified. In this paper, as we consider the class of EoS such that entropy is a function of density (\ref{sofrho}), we have $\mathrm{D} s / \mathrm{D} t = s' \mathrm{D} \rho / \mathrm{D} t$. Using the continuity equation (\ref{continuity}), equation (\ref{entropy}) can be written in the following form
\begin{equation}
	- \rho ^2 T s' {\mathbf \bnabla} \cdot {\mathbf u} = \dot{\epsilon} : \tau + \bnabla \cdot \left( k {\mathbf \bnabla } T \right) , \label{entropy2}
\end{equation}
which is now an elliptic equation for temperature. When $P $ is expressed in terms of $T$ and $\rho$ -- see equations (\ref{Phlog}) or (\ref{Phpower}) -- the Stokes's equation (\ref{momentum}) also becomes a Poisson equation for velocity (along with the continuity equation). By the way, it is already clear that the constant $K$ in the expression for the internal energy (\ref{energy2}) and in that for pressure (\ref{Phlog}) or (\ref{Phpower}) is completely irrelevant in the governing equations for convection: internal energy does not appear explicitly and taking the gradient of pressure eliminates $K$ from the momentum equation (\ref{momentum}).

The next step consists in defining dimensional scales and in writing the equations in dimensionless form. We have already mentioned a scale for density, $\rho _0$ which is the average density in the domain that remains constant with the imposed boundary conditions. Next, we define $T_0 = (T_h+T_c)/2$ the average temperature of the hot and cold boundaries. Then, we need to choose either a $\log$ or power law EoS along with an exponent $n$. We specify $c_{p0}$ the value of $c_p$ at the conditions $T=T_0$ and $\rho = \rho _0$, which is equivalent to specifying the constant $a$. From the logarithmic equation $s \sim \log \rho$, we have $c_{p0} = -a$ while for the power law $s \sim \rho ^n$ and (\ref{alphaTetcppower}), we have
\begin{equation}
	 c_{p0}  = -a \frac{n}{n+1} \rho _0^n,  \hspace*{1 cm} \mathrm{for}\ n>0 . \label{cp0Apower}
\end{equation}
From (\ref{slogr}) and (\ref{spower}), we derive an expression for $s'$ which is valid for both the logarithmic ($n=0$) and power-law ($n>0$) EoS
\begin{equation}
	s' (\rho ) = - (n+1) \frac{c_{p0}}{\rho _0} \left( \frac{\rho}{\rho _0} \right) ^{n-1},  \hspace*{1 cm} \mathrm{for}\ n \geq 0 . \label{sprime}
\end{equation}
Similarly, a generic expression is obtained for the pressure gradient, from (\ref{Phlog}) and (\ref{Phpower})
\begin{equation}
	{\mathbf \bnabla} P = (n+1) c_{p0} \left( \frac{\rho}{\rho _0} \right) ^n \left[ (n+1)T {\mathbf \bnabla} \rho + \rho {\mathbf \bnabla} T  \right], \hspace*{1 cm} \mathrm{for} \ \ n \geq 0 .  \label{gradP}
\end{equation}

We consider a uniform thermal conductivity $k$, so that a scale for thermal diffusivity is $\kappa = k/(\rho _0 c_{p0})$. We now make all variables dimensionless using $H$, $\kappa / H$, $H^2 / \kappa$, $T_0$, $\rho _0$, $c_{p0}$, $\rho _0 c_{p0} T_0$, $\kappa/ H^2$, $\eta \kappa/ H^2$, for length, velocity, time, temperature, density, entropy, pressure, deformation rate and stress. Using the same symbols for dimensionless variables, the equations of continuity, momentum and entropy become 
\begin{align}
	\frac{\mathrm{D} \rho}{\mathrm{D} t} &= - \rho {\bnabla} \cdot {\bf u}, \label{continuitya}\\
	{\mathbf 0} &= - \frac{Ra_{sa} (n+1)}{\varepsilon } \left( \frac{ \rho ^n }{ \mathcal{D}}  \left[ (n+1) T {\mathbf \bnabla} \rho + \rho {\mathbf \bnabla} T  \right] + \rho {\mathbf e}_z \right) + {\mathbf \bnabla}^2 {\bf u} + \frac{1}{3} {\mathbf \bnabla} \left(\bnabla \cdot {\bf u} \right) , \label{momentuma} \\
	0 &= - (n+1) \rho ^{n+1} T \bnabla \cdot {\bf u} + \frac{\varepsilon \mathcal{D}}{Ra_{sa}} \dot{\epsilon} : \tau + \nabla ^2 T , \label{entropya}
\end{align}
where the following dimensionless numbers appear, the superadiabatic Rayleigh number $Ra_{sa}$, the dissipation number $\mathcal{D}$, the ratio of superadiabatic temperature difference over the average temperature $\varepsilon$ and implicitly the product $\alpha _0 T_0$, as a function of $n$: 
\begin{align}
	Ra_{sa} &= \frac{\rho _0 g \alpha _0 \Delta T_{sa} H^3}{\eta \kappa}, \label{Rasa} \\
	\mathcal{D} &= \frac{\alpha _0 g H}{c_{p0}}, \label{D} \\
	\varepsilon &= \frac{\Delta T_{sa}}{T_0}, \label{epsilon} \\
	\alpha _0 T_0 &= \frac{1}{n+1}. \label{alphaT}
\end{align}
\N{The dissipation number $\mathcal{D}$ is one of the possible measures for compressibility, of the same nature as the number of scale heights in astrophysics \citep{sv1960}. It was introduced by \citet{gebhart1962}, motivated by the context of cooling turbine blades by natural convection. Interestingly, the dissipation number can be defined in the framework of the Boussinesq approximation, although compressibility is absent and despite the fact that its value has no effect on the solutions. Moreover, it can be shown rigorously from the Boussinesq equations that the integral of viscous dissipation is equal to the product of the dissipation number $\mathcal{D}$ and the convective heat flux in a Rayleigh-B\'enard cavity \citep{howard63}. }
The superadiabatic temperature difference $\Delta T_{sa}$ is equal to the difference between the imposed hot and cold temperatures minus the temperature difference along the adiabat
\begin{equation}
	\Delta T_{sa} = T_h - T_c - \frac{\alpha _0 g T_0 H}{c_{p0}}. \label{dTsa}
\end{equation}
When writing the dimensionless momentum equation (\ref{momentuma}), we \N{use} (\ref{gradP}) to express the pressure gradient in terms of density and temperature gradient. When writing the dimensionless entropy equation (\ref{entropya}), we \N{use} (\ref{entropy2}) and (\ref{sprime}). It can be checked that the final set of dimensionless equations (\ref{continuitya}), (\ref{momentuma}) and (\ref{entropya}) takes a generic form for any real value of $n \geq 0$: the case $n=0$ corresponds to the logarithmic relationship (\ref{slogr}) while the cases $n>0$ correspond to the power laws (\ref{spower}). The choice of $n$ amounts to choosing the product $\alpha _0 T_0$, see (\ref{alphaT}). 

As initial conditions, we set the velocity to zero, and density, pressure and, temperature fields satisfying the (potentially unstable) hydrostatic conduction regime, with an additional random temperature field of magnitude $10^{-6}$.  
The boundary conditions on the velocity and temperature fields are the following
\begin{align}
	\frac{\partial u_x}{\partial z} &= 0, \hspace{1 cm} \mathrm{when} \hspace{1 cm} z=\pm \frac{1}{2}, \label{uxbc}\\
	u_z &= 0, \hspace{1 cm} \mathrm{when} \hspace{1 cm} z=\pm \frac{1}{2}, \label{uzbc} \\
	& \N{\int_{-L/(2H)}^{L/(2H)} u_x \left( x, z=\frac{1}{2} \right) \md x = 0, \label{meanux}} \\ 
	{T} &= \frac{T_h}{T_0}, \hspace{1 cm} \mathrm{when} \hspace{1 cm} z=-\frac{1}{2}, \label{Thbc}\\
	{T} &= \frac{T_c}{T_0}, \hspace{1 cm} \mathrm{when} \hspace{1 cm} z=\frac{1}{2}. \label{Tcbc}
\end{align}
\N{The stress-free, non-penetrative conditions (\ref{uxbc}) and (\ref{uzbc}) do not constrain the mean horizontal velocity, hence an arbitrary condition of zero average horizontal velocity (\ref{meanux}) is imposed on the upper boundary.} 
The imposed temperature ratios are linked to the values of the dissipation number and the superadiabatic temperature coefficient
\begin{align}
	\frac{T_h}{T_0} &= 1 + \frac{\mathcal{D} + \varepsilon }{2} , \label{temph} \\ 
	\frac{T_c}{T_0} &= 1 - \frac{\mathcal{D} + \varepsilon }{2} . \label{tempc}
\end{align}

As shown in \cite{cdadlr2019}, the equations of convection with infinite Prandtl number are subjected to viscous relaxation, and the associated relaxation time  \N{limits} the time-steps to $\mathcal{D}/{Ra_{sa}}$ for numerical calculations. Let us determine here the expression for this relaxation time scale, for our particular class of EoS. We consider a small planar disturbance with respect to the steady solution $(\rho =1,\ T=1-\mathcal{D} z ,\ {\bf u} = {\bf 0})$ with $\epsilon = 0$ 
\begin{equation}
	\rho ' = \tilde{\rho} e^{i k x + \omega t}, \hspace{1 cm} T ' = \tilde{T} e^{i k x + \omega t}, \hspace{1 cm} u_x' = \tilde{u}_x e^{i k x + \omega t}, \label{disturb} 
\end{equation}
where $\tilde{\rho}$, $\tilde{T}$ and $\tilde{u}_x$ are scalars. 
The governing equations (\ref{continuitya}), (\ref{momentuma}) and (\ref{entropya}), are linearized near $z=0$ (the steady solution is nearly constant $T=1$) and lead to
\begin{align}
	\omega \tilde{\rho} &= - i k \tilde{u}_x , \label{lincont} \\
	0 &= - \frac{Ra_{sa} (n+1)}{\varepsilon \mathcal{D}} \left[(n+1) i k \tilde{\rho} + i k \tilde{T} \right] - \frac{2}{3} k^2 \tilde{u}_x, \label{linmom} \\
	0 &= - (n+1) i k \tilde{u}_x - k^2 \tilde{T} . \label{linent}
\end{align}
Eliminating $\tilde{u}_x$ and $\tilde{T}$, leads to a single equation for $\tilde{\rho}$
\begin{equation}
	0 = - \frac{Ra_{sa} (n+1)^2}{\varepsilon \mathcal{D}} \left[k \tilde{\rho} + \frac{\omega}{k} \tilde{\rho} \right] - \frac{2}{3} k \omega \tilde{\rho} , \label{linfinal}
\end{equation}
admitting non-trivial solutions when 
\begin{equation}
	\omega = - \frac{3}{2} \frac{Ra_{sa} (n+1)^2}{\varepsilon \mathcal{D}} \left[1+ \frac{\omega }{k^2} \right] , \label{timescale}
\end{equation}
implying that the magnitude of the rate of decay $| \omega |$ is \N{bounded from above as}  
\begin{equation}
	\N{	| \omega | < } \frac{3}{2} \frac{Ra_{sa} (n+1)^2}{\varepsilon \mathcal{D}}, \label{maxomega}
\end{equation}
irrespective of the wavenumber $k$. 
In practice, we make it slightly safer by changing the prefactor from $3/2$ to $1$, and our numerical scheme is always found to be stable with time-steps $\delta t$ smaller than 
\begin{equation}
	\delta t \leq \frac{\varepsilon \mathcal{D}}{Ra_{sa} (n+1)^2} = \frac{\varepsilon \mathcal{D} \left( \alpha _0 T_0 \right) ^2}{Ra_{sa}}. \label{dt}
\end{equation}
This makes it difficult to calculate flows with large superadiabatic Rayleigh numbers, small dissipation numbers, small superadiabatic parameters $\varepsilon$ or small products $\alpha T_0$ (large $n$). \\



We are now going to write a series of anelastic models from the most to the least faithful approximation of the fully compressible equations. 

\subsection{Anelastic Approximation AA}
\label{AA}

The first model is called simply the anelastic model and corresponds to the early models by \citet{op1961} for the atmosphere, \citet{lf1999} for stellar convection and \citet{br1995} for the Earth's core. It corresponds to a first order expansion modelling of thermodynamic variables with respect to a hydrostatic isentropic state. In our case, the structure of the well-mixed isentropic region is simple, with a uniform density and uniform temperature gradient: in dimensionless form 
\begin{align}
	\rho _a &= 1, \label{rhoa} \\
	T_a &= 1 - \mathcal{D} z , \label{Ta}
\end{align}
where $T_a$ is the isentropic profile. We have set arbitrarily $T_a = 1 $ at $z=0$ (mid-height) but we will see later that this does not constrain the anelastic solution. Let us denote with tildes the two-dimensional and time-dependent departures of each variable from its isentropic counterpart. From the standard procedure of linearization of the functions of state about the adiabatic profile \citep{ajs05}, and \N{with a change in the dimensional scale for temperature ($\Delta T_{sa}$ instead of $T_0$), pressure ($\rho _0 c_{p0} \Delta T_{sa}$ instead of $\rho _0 c_{p0} T_0$) and entropy ($c_{p0} \Delta T_{sa}/T_0$ instead of $c_{p0}$)}, we get the following dimensionless anelastic equations
\begin{align}
	{\mathbf \bnabla} \cdot {\bf u} &= 0 , \label{AAcont} \\
	{\bf 0} &= -\frac{Ra_{sa}}{\mathcal{D}} {\mathbf \bnabla} \tilde{P} + Ra_{sa} \tilde{s} {\hat{\mathbf{e}}}_z + {\mathbf \bnabla}^2 {\bf u} , \label{AAmomentum} \\
	\frac{ \mathrm{D} }{\mathrm{D} t} (T_a \tilde{s} ) &= - \mathcal{D} u_z  \tilde{s} + \frac{\mathcal{D}}{Ra_{sa}} \dot{\epsilon} : \tau + \nabla ^2  \tilde{T}. \label{AAentropy} 
\end{align}
For our equation of state, $\tilde{s}$ is proportional to $\tilde{\rho}$ -- see (\ref{sprime}) --  and linearizing equations (\ref{Phlog}) or (\ref{Phpower}) leads to
\begin{align}
	\tilde{s} &= - (n+1) \tilde{\rho} , \label{AAsrho} \\
	\tilde{\rho } &= - \frac{\tilde{T}}{(n+1) T_a} + \frac{\tilde{P}}{(n+1)^2 T_a}, \label{AArho}
\end{align}
\N{and therefore 
\begin{equation}
	\tilde{s} = \frac{\tilde{T}}{T_a} - \frac{\tilde{P}}{(n+1)T_a}. \label{AAs}
\end{equation}
}


Because of the new temperature scale $\Delta T_{sa}$, the temperature boundary conditions become
\begin{equation}
	\tilde{T} \left( z = \pm \frac{1}{2} \right) = \mp \frac{1}{2} . \label{AAbc}
\end{equation}

The boundary conditions for pressure are obtained from the condition of mass conservation. With our choice in (\ref{rhoa}), the (uniform) adiabatic density profile corresponds already to the total mass in the fluid layer, the integral of the departure $\tilde{\rho}$ must be zero at all times. This might be achieved by imposing an appropriate value of pressure at the top or at the bottom of the cavity. However, an easier way is to impose that the mean value of $\tilde{P}$ on the top boundary is equal to the mean value at the bottom. This can be seen on equation (\ref{AAmomentum}) by integration along $z$. The integral of density in the cavity is zero, \N{the viscous term ${\mathbf \bnabla}^2 {\bf u}$ in (\ref{AAmomentum}) integrates into the difference of averaged viscous traction $\tau _{zz} = 2 \partial u_z / \partial z$ between top and bottom boundaries. The continuity equation (\ref{AAcont}) leads to $\tau _{zz} = - 2 \partial u_x / \partial x$ whose integral of each boundary is zero with periodic conditions on $x$}. The condition on pressure is \N{thus}
\begin{equation}
	\int _0^{L/H} \tilde{P} \left( x, z=\frac{1}{2} \right) - \tilde{P} \left( x, z=- \frac{1}{2} \right) \md x = 0 . \label{AApressure}
\end{equation}

An invariance property of these AA equations can be put in evidence. Since the anelastic equations have been obtained by linearization around the adiabatic profile, one expects that a shift in the superadiabatic temperature conditions should leave the solution unchanged, \N{with the same total mass}. From a (possibly time-dependent) solution $({\bf u}, \tilde{P}, \tilde{T})$ to the equations above, we just add a constant $c$ to the temperature boundary conditions, now becoming $\tilde{T} (z=\pm 1/2 ) = \mp 1/2 + c$. We can check that $({\bf u}, \tilde{P}+c/(n+1), \tilde{T}+c)$ is a solution to the AA equations with the shifted temperature boundary conditions.

\subsection{Anelastic Liquid Approximation ALA}
\label{ALA}

In that approximation, departures of entropy from the adiabatic profile are considered to be due only to temperature departures, while departures in pressure are neglected \N{in equation \ref{AAs}} \citep{ajs05}. The governing equations are still (\ref{AAcont}), (\ref{AAmomentum}) and (\ref{AAentropy}) \N{where $\tilde{s}$ is changed in each instance into 
\begin{equation}
	\tilde{s} = \frac{\tilde{T}}{T_a} . \label{ALAs}
\end{equation}
}

When applying the boundary conditions, we now find that it is not possible to impose the obvious temperature boundary condition (\ref{AAbc}): if we did so, it is most likely that the integral of $\tilde{T}/ T_a$ over the fluid domain would not be zero. However, because of the anelastic liquid approximation, entropy departures are linked to $\tilde{T}/ T_a$ which are themselves proportional to density departures, because of the equation of state. Hence a non-zero integral of $\tilde{T}/ T_a$ implies that the total mass of the fluid is not conserved at first order. Thus, we keep the condition (\ref{AApressure}) on pressure, which ensures total mass conservation and we consider that only the imposed temperature difference is meaningful between two isothermal boundaries
\begin{align}
	&\frac{\partial  \tilde{T}}{\partial x} = 0 \hspace*{1 cm} z = \pm \frac{1}{2}, \label{ALAtemp} \\
	&\tilde{T} \left( x,z= - \frac{1}{2} \right) - \tilde{T} \left( x,z= \frac{1}{2} \right) = 1 . \label{ALAtempdelta}
\end{align}
Coming back to the pressure constraint (\ref{AApressure}), any additive constant to $\tilde{P}$ is irrelevant since only the gradient of pressure plays a role in the ALA equations. In conclusion, we may decide to set the mean value of $\tilde{P}$ to zero (or any other constant) on both hot and cold boundaries. Equation (\ref{AApressure}) is changed for 
\begin{equation}
	\int _0^{L/H} \tilde{P} \left( x, z= \pm \frac{1}{2} \right)    \md x = 0 . \label{ALAP}
\end{equation}

The invariance mentioned for the solutions to the AA equations is no longer relevant in the ALA equations. Now, the mass balance imposes that the integral of $T/T_a$ must be zero over the whole domain because density fluctuations are solely functions of entropy fluctuations, which are themselves solely functions of temperature fluctuations in the ALA approximation. That cannot be changed by another choice of pressure offset.   

\subsection{Simple Compressible Approximation SCA}
\label{SCA}

We now introduce a new approximation aiming at getting a very simple system of equations where compressible effects are still present. In the ALA approximation, the adiabatic temperature profile appears explicitly in the equations and we consider replacing $T_a$ by a constant value equal to $1$, the value of $T_a$ in the mid-plane of the cavity. We certainly \N{lose connection} to thermodynamics with that move, but compressible work is still present and it will be interesting to investigate which compressible effects are still well accounted for in this approximation. Note that this SCA model is equivalent to a version of the "Extended Boussinesq Approximation", or EBA \citep{king}, where the background density is assumed to be uniform (with our EoS, this is the case of all our anelastic models) and where the background temperature is also assumed to be uniform.  The SCA model \N{is still defined by equations (\ref{AAcont}), (\ref{AAmomentum}) and (\ref{AAentropy}), where the expression for entropy (\ref{ALAs}) is changed for
\begin{equation}
	\tilde{s} = \tilde{T} , \label{SCAs}
\end{equation}
and $T_a$ is also changed for the constant value $1$ is the left-hand side term of (\ref{AAentropy}). Under this approximation, the equations are very similar to the classical Boussinesq equations, except for viscous heating and adiabatic cooling that play a significant role when the dissipation number is of order one.
}


The boundary conditions are similar to those for the ALA equations. Now, the average of the temperature departure $\tilde{T}$ \N{on the whole domain} is zero thanks to the condition on pressure (\ref{ALAP}). An important invariance is valid only in the case $\mathcal{D} = 0$. This corresponds to the Boussinesq equations with another change in pressure scale, from $\rho _0 c_{p0} \Delta T_{sa}$ to $\alpha _0 \Delta T_{sa} \rho _0 g H$. In that case only ($\mathcal{D} = 0$), the equations are invariant by symmetry about the mid-plane. More precisely, if $(u_x,u_z,\tilde{P},\tilde{T})$ is a solution (possibly time-dependent), then the fields $(u_x(x,-z,t),-u_z(x,-z,t),\tilde{P}(x,-z,t),-\tilde{T}(x,-z,t))$ constitute another solution. This implies that the solutions to the incompressible Rayleigh-B\'enard system are bottom-up invariant: ascending and descending plumes are statistically symmetrical. However, when $\mathcal{D} \neq 0$, that invariance does no longer hold, for none of the compressible models presented here, fully compressible (FC), anelastic approximation (AA), anelastic liquid approximation (ALA) nor the last simple compressible approximation (SCA). We will have the opportunity to investigate that non-invariance in the following sections.

%

All numerical results, FC, AA, ALA and SCA have been obtained with the software Dedalus \citep{dedalus}. For the fully compressible model a Runge-Kutta model of order 1 was used (RK111 in dedalus) and of order 4 for the anelastic models (RK443 in dedalus). The number of Fourier modes in the horizontal direction $n_x$ is four times that of Chebyshev modes $n_z$ in the vertical direction. \N{A dealiasing factor of $3/2$ has been used in all cases.} The value of $n_z$ we used goes from $32$ at low superadiabatic Rayleigh numbers to $512$ at $Ra_{sa}=10^9$. Time steps have been set by \N{half} the short viscous relaxation time in the fully compressible model and by a {C}ourant condition for anelastic models \N{with a safety factor set to $0.9$. Noise on the initial conduction temperature field, of magnitude $10^{-6}$ has been added in all models to trigger convection.}

\N{The complete set of equations FC, AA, ALA and SCA, with boundary conditions, are written in their explicit forms in appendix \ref{A2}. The parameters of all simulations are listed in appendix \ref{A3}. The files used for each convection model FC, AA, ALA and SCA are provided as supplemental materials or available at github.}

\section{From rest to steady rolls at $Ra_{sa} = 10^4$, $\mathcal{D}=1.5$}
\label{start}

\begin{figure}
\begin{center}
	\includegraphics[width=14 cm, keepaspectratio]{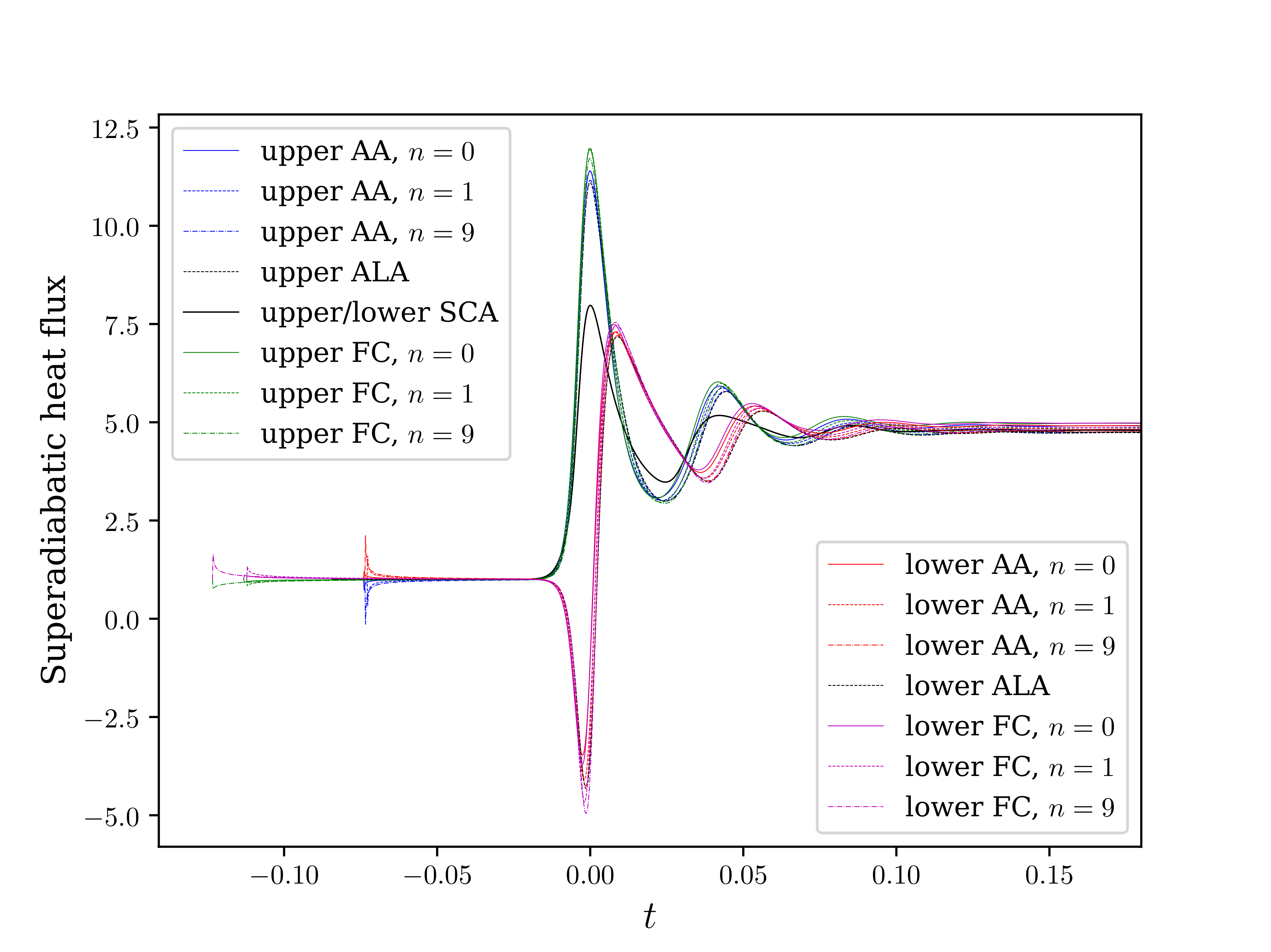}
	\caption{Upper and lower heat flux \N{(Nusselt number)} during the initial transient from rest to steady rolls, at $Ra_{sa} = 10^4$ and $\mathcal{D}= 1.5$. }
\label{bothflux}
\end{center}
\end{figure}

In this section, we just analyze the transition from an unstable superadiabatic motionless initial state to steady rolls of convection, for a moderate superadiabatic Rayleigh number of $Ra_{sa}=10^4$ and a large dissipation number of $\mathcal{D}=1.5$. We do that for different values of the dimensionless parameter $\alpha _0 T_0 =  1$, $0.5$ and $0.1$ (with $n = 0$, 1 and 9) and the different models of convection: FC, AA, ALA and SCA. In figure \ref{bothflux}, we plot the upper and lower heat fluxes (on the top and bottom boundaries) for all values of $\alpha _0 T_0$ and all models. Only the heat fluxes above the conduction flux along the adiabatic gradient are represented: this is straightforward in the anelastic model as the conduction flux along the adiabat is not computed, while for the fully compressible model we just remove the contribution of conduction along the adiabatic gradient. Then, the remaining part of the flux is scaled by the conduction heat flux driven by $\Delta T_{sa}$, the superadiabatic temperature difference: again, this is natural in the anelastic models where temperature intervals are already scaled by $\Delta T_{sa}$, while in the fully compressible model the temperature scale is $T_0$ and the flux has to be rescaled by $\Delta T_{sa}$ corresponding to a division by the superadiabatic fraction $\varepsilon = \Delta T_{sa}/ T_0$. In the present FC calculations, the superadiabatic fraction $\varepsilon$ is set to $0.1$. The initial (unstable) conduction state corresponds to a heat flux unity, while when a convective steady state of convection is reached, the heat flux is around 5 or slightly less. \N{This dimensionless flux is the classical Nusselt number, which will also be used in the next sections.} The blue/green colors correspond to the heat flux at the upper boundary, while red/purple colors correspond to the heat flux on the lower boundary (for all values of $\alpha _0 T_0$ and FC, AA, ALA). The simple compressible model (SCA) is plotted with a black color: it can be shown easily that upper and lower heat fluxes coincide at all times for this model. All plots have been shifted in time so that the maximum upper flux is at $t=0$. The real starting time of the simulation depends on the model considered and is made visible by a small transient period, as we have started the simulations from our approximations of the conductive hydrostatic state. 
The time needed to develop the convective instability depends on the model of convection, and weakly on $\alpha _0 T_0$ for FC. For all models FC, AA and ALA, the curves of upper and lower heat fluxes are rather similar, we will come back to the small differences below. In all cases, in the beginning of the convective instability, the heat flux on the upper boundary grows rapidly to a large value (around 11), while the heat flux on the lower boundary decreases rapidly to negative values (around -4). Then follows a series of oscillations of decreasing amplitude, with a phase shift of approximately $\pi / 2$ between upper and lower fluxes, until a steady state is reached with equal upper and lower fluxes (around 5 or slightly less). 

\begin{figure}
\begin{center}
	\hspace*{-0.0 cm}\includegraphics[height=3.5 cm, keepaspectratio]{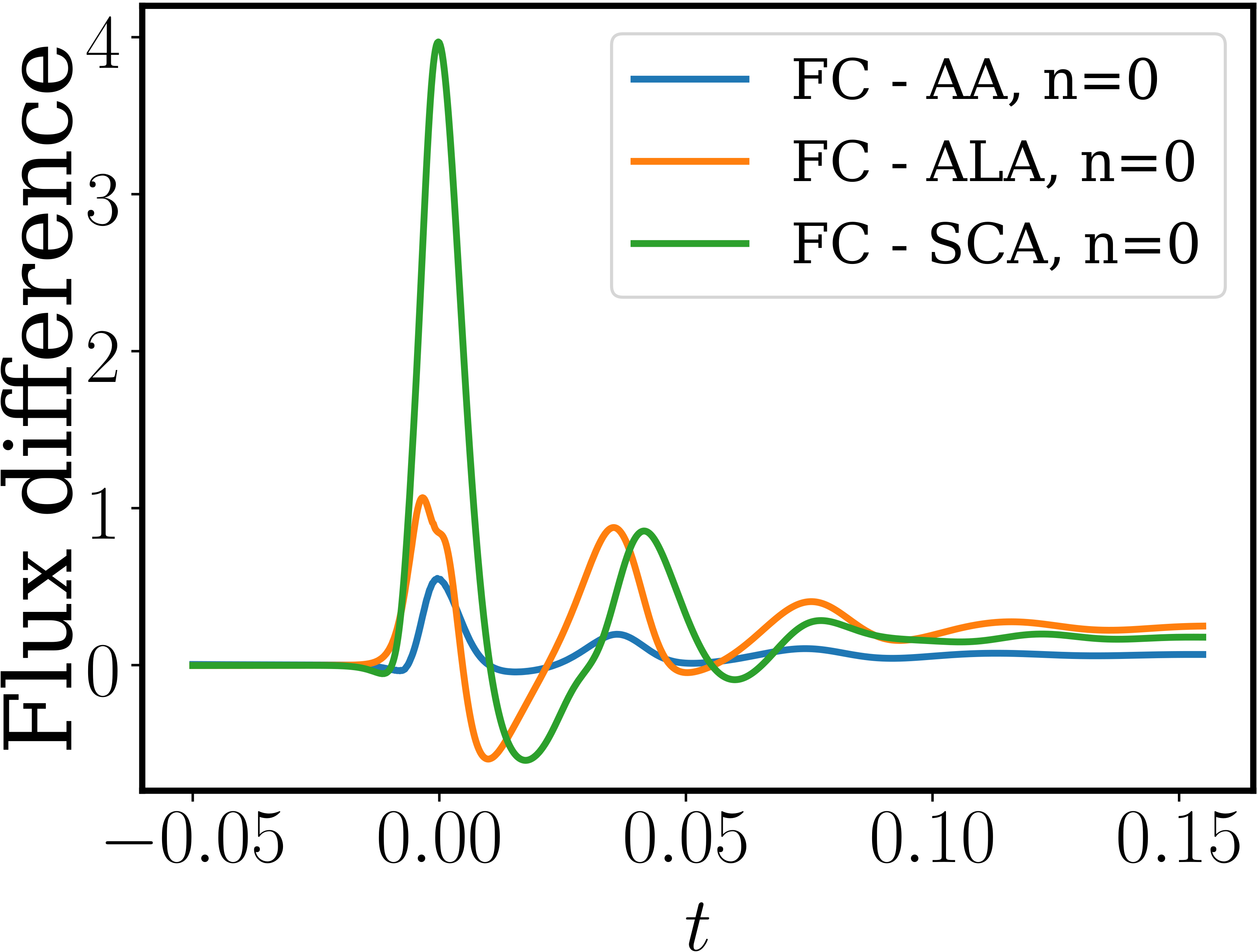}\hspace*{0.1 cm}\includegraphics[height=3.5 cm, keepaspectratio]{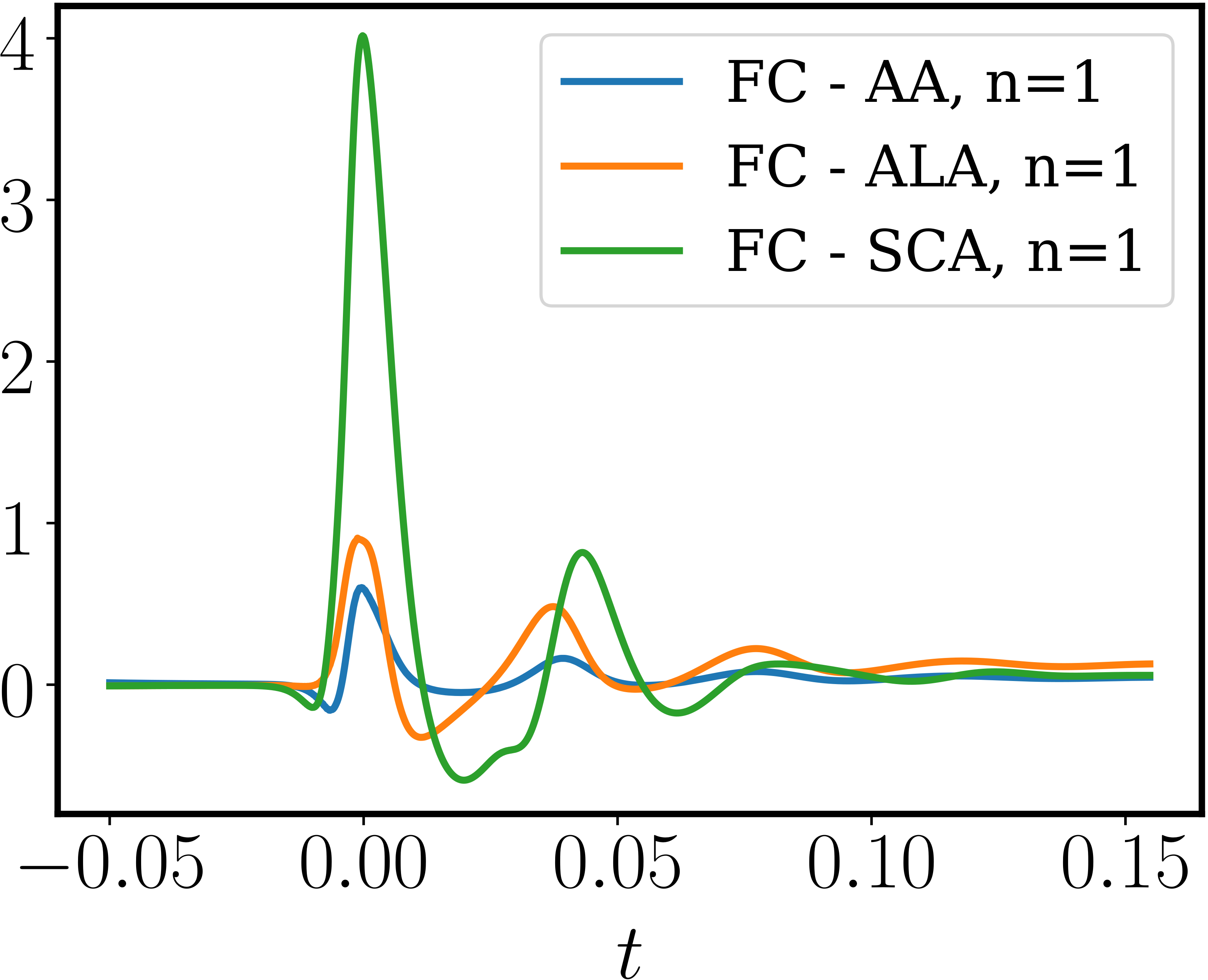}\hspace*{0.1 cm}\includegraphics[height=3.5 cm, keepaspectratio]{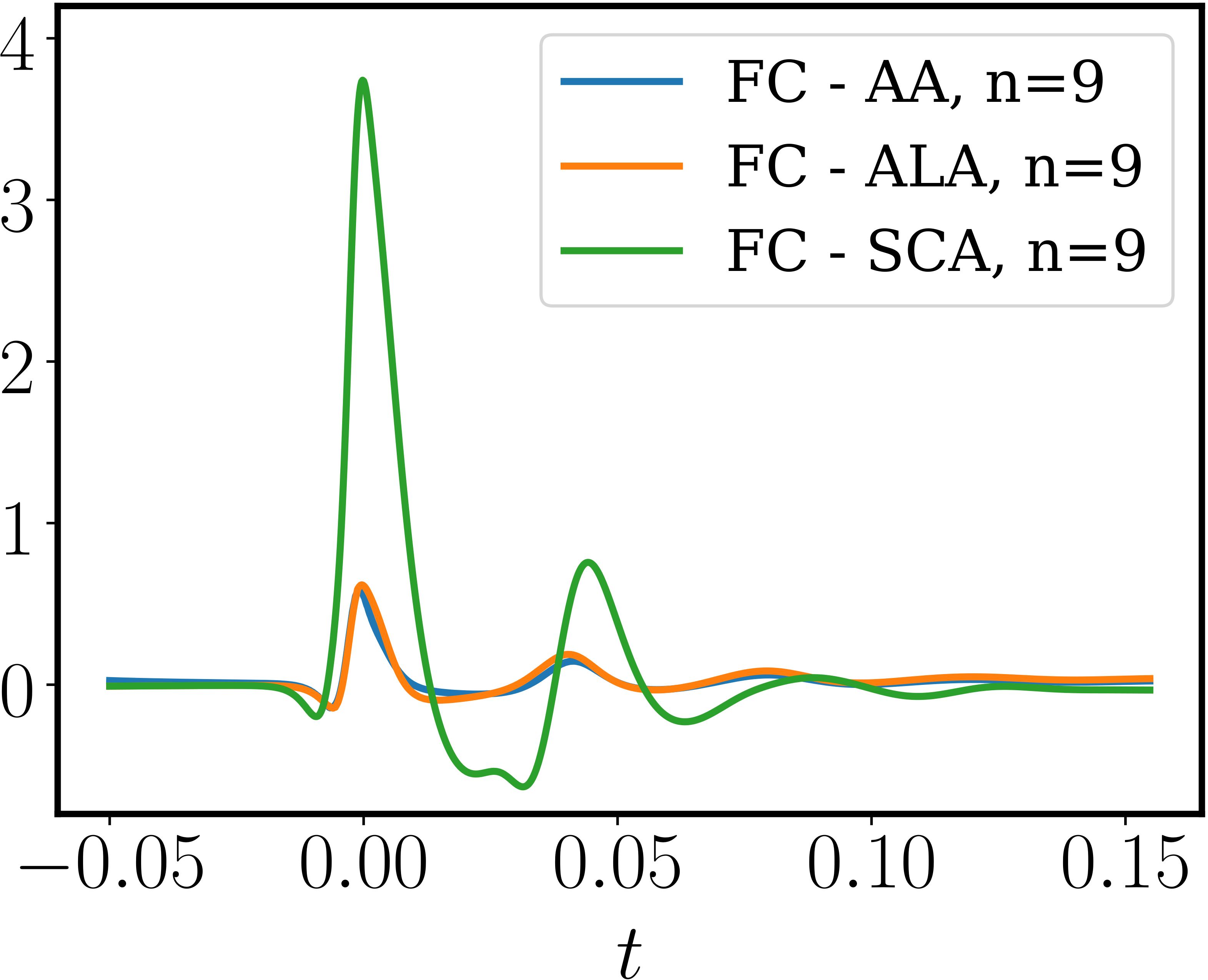}
	\caption{Difference between the heat flux \N{(Nusselt number)} on the upper boundary obtained with FC and AA, ALA and SCA respectively, at $Ra_{sa} = 10^4$ and $\mathcal{D} = 1.5$, for $n=0$ ($\alpha _0 T_0=1$), $n=1$ ($\alpha _0 T_0=0.5$) and $n=9$ ($\alpha _0 T_0=0.1$) from left to right. }
\label{difffluxModel}
\end{center}
\end{figure}

On Fig.~\ref{difffluxModel}, we take the difference on the upper flux between the FC model and the approximations AA, ALA and SCA. Unsurprisingly, the smallest difference is obtained with the AA approximation, followed by the ALA and finally the SCA approximation. We also observe that the difference between AA and ALA approximations becomes smaller as the product $\alpha _0 T_0$ is decreased. This was expected from \N{equation (\ref{AAs})} as the effect of pressure is divided by $n+1$, {\it i.e.} decreases with $\alpha _0 T_0$. 

\section{\N{Top/bottom asymmetry}}
\label{asymmetryG}

\N{The top/bottom symmetry is observed to hold for all models in the limit of vanishing compressibility effect $\mathcal{D} \longrightarrow 0$. In the case of the fully compressible model (FC) one must also have a top/bottom temperature ratio close to one, but that condition has a small effect on the asymmetry compared to that of the dissipation parameter. However, when $\mathcal{D}$ is non-zero we will see that a distinct difference appears between the top and bottom parts of the average temperature profile, or between raising and descending plumes. Perhaps surprisingly, the asymmetry becomes very clear from relatively small values of the dissipation number, $\mathcal{D} \sim 0.1$ (section \ref{smallD} below), and continues to exist when $\mathcal{D}$ is further increased (section \ref{asymmetry} below). Also, increasing the superadiabatic Rayleigh number does not seem to change that asymmetry. }
\N{From this point until the end of the paper, the value of $\alpha _0 T_0$ is set to $1$ ($n=0$).}

\subsection{Change of temperature profile with moderate compressibility}
\label{smallD}

\begin{figure}
\begin{center}
	\hspace*{-5mm}\includegraphics[width=6.9 cm, keepaspectratio]{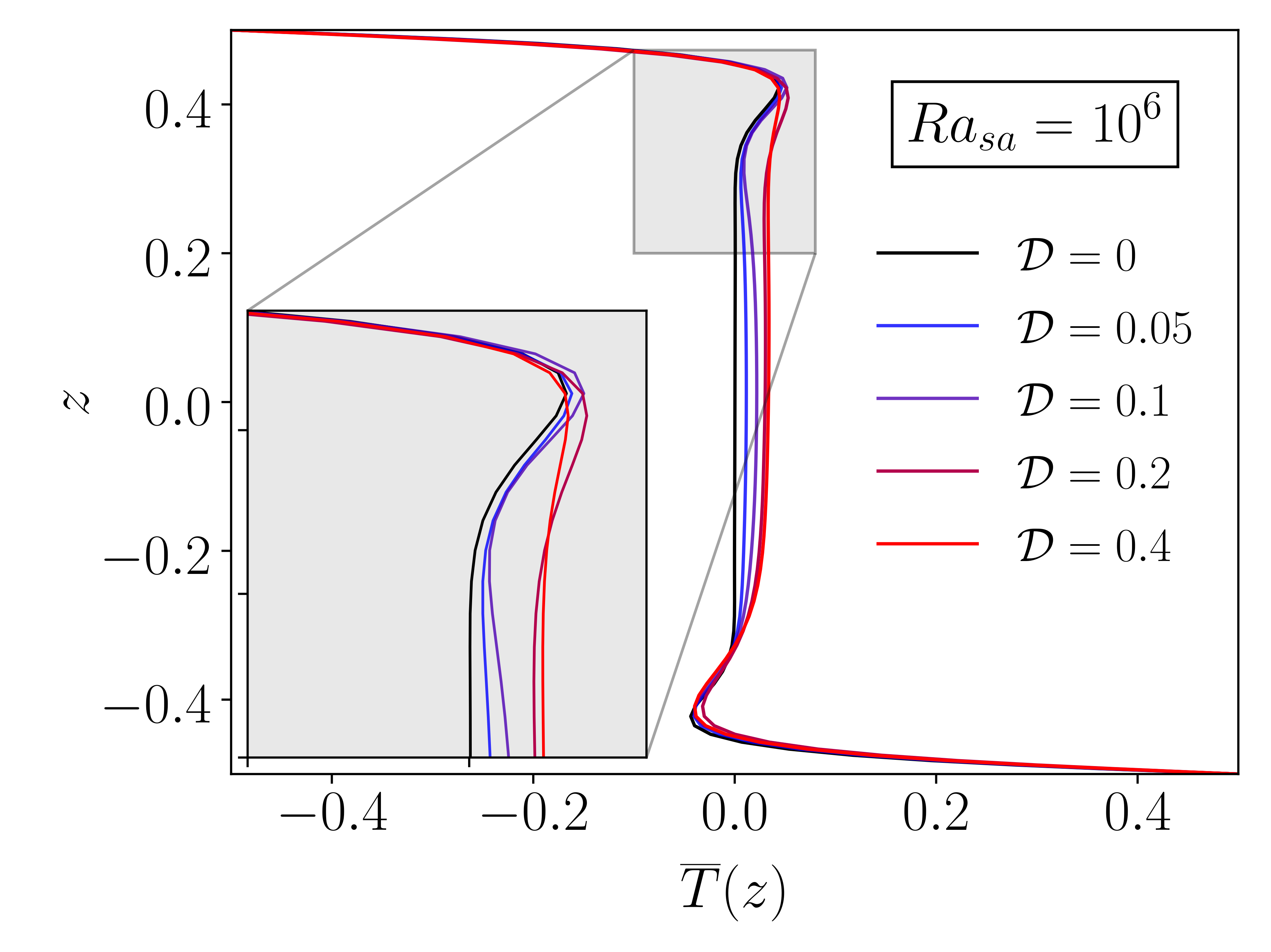} \includegraphics[width=6.9 cm, keepaspectratio]{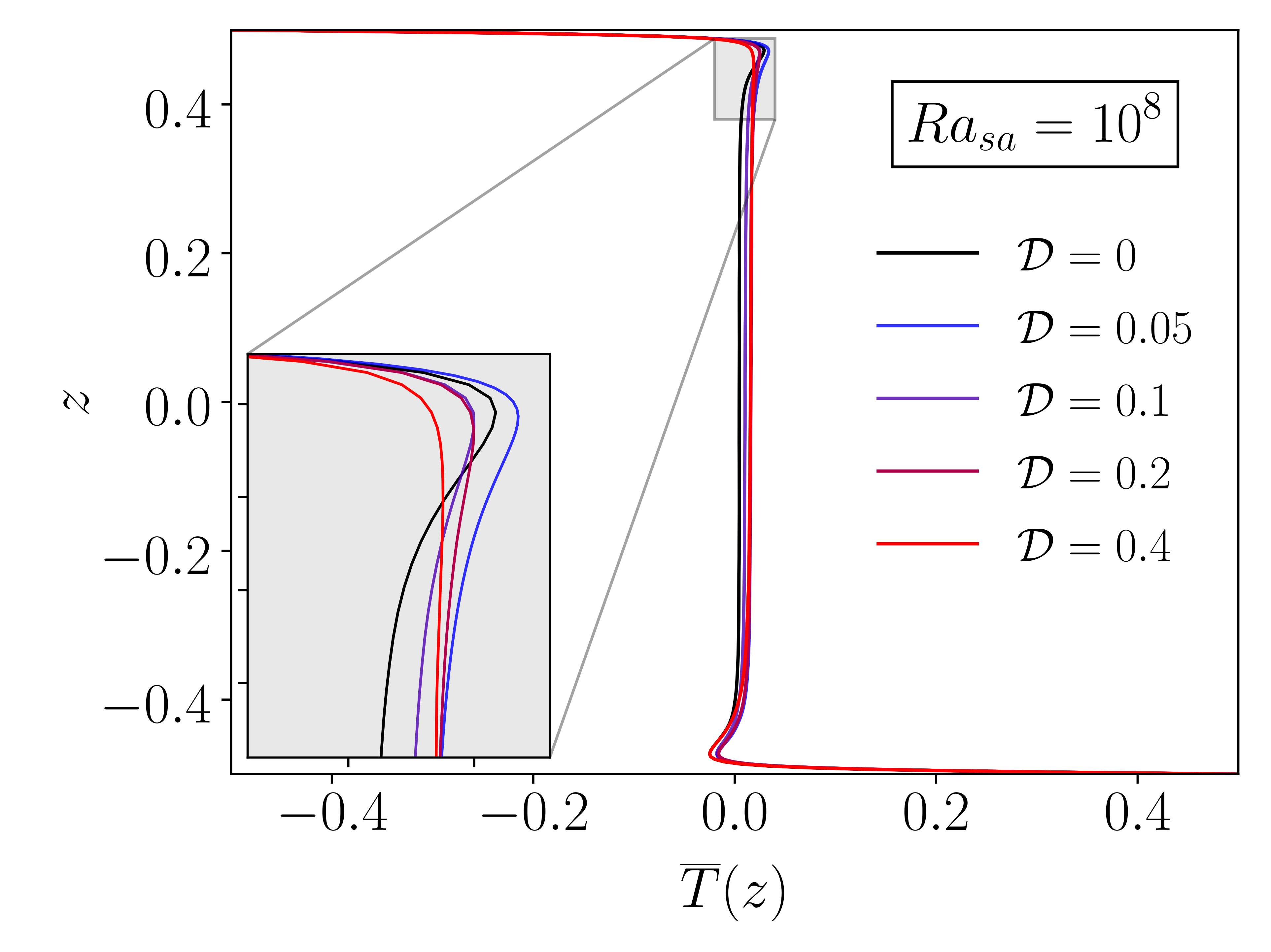} 
	\caption{Time and horizontal averaged superadiabatic temperature profiles along the vertical direction $z$, for $Ra_{sa}=10^6$ (left) and $Ra_{sa}=10^8$ (right), for $\mathcal{D}$ between $0$ and $0.4$, $\alpha _0 T_0 = 1$, obtained in the anelastic approximation AA.}
	\label{topovershoot}
\end{center}
\end{figure}

We examine now the effect of a small compressibility on the structure of convection. \N{From this point until the end of the paper, the value of $\alpha _0 T_0$ is set to $1$ ($n=0$).} When the dissipation number $\mathcal{D}$ is increased from $0$ to a moderate value of $0.1$ to $0.4$, a change in the averaged vertical temperature profile is observed. In the absence of compressible effects ($\mathcal{D} \simeq 0$), the temperature profile is symmetrical with respect to the horizontal mid-plane as a result of the invariance of the Boussinesq equations under the transformation $T(x,z) \rightarrow -T(x,-z)$, $u_x(x,z) \rightarrow - u_x(x,-z)$ and $u_z(x,z) \rightarrow u_z(x,-z)$. It is also well-known that overshoots in the temperature profile occur near the top and bottom thermal boundary layers \citep{sl99}. These overshoots have an amplitude (and extent) decreasing with increasing Rayleigh numbers, but are always present. We observe here that the overshoot near the top is nearly eliminated when the dissipation number $\mathcal{D}$ exceeds 0.2. In Fig.~\ref{topovershoot}, $\mathcal{D}$ is increased from $0$ to $0.4$ in AA calculations and the time and horizontally averaged superadiabatic temperature profiles are plotted along the vertical direction for two values of the superadiabatic Rayleigh number $Ra_{sa} = 10^6$ and $Ra_{sa} = 10^8$. The inset shows a close-up view of the top overshoot in the temperature profile. A tiny value of $\mathcal{D} = 0.05$ already has a noticeable effect, and when $\mathcal{D} = 0.2$ most of the change has been made. Conversely, the bottom overshoot is nearly unchanged. As a result, the nearly constant mean value of temperature is increased, an observation that will be related to the behaviour of the Nusselt number in section \ref{HeatFlux}.

\begin{figure}
\begin{center}
        \includegraphics[width=12 cm, keepaspectratio]{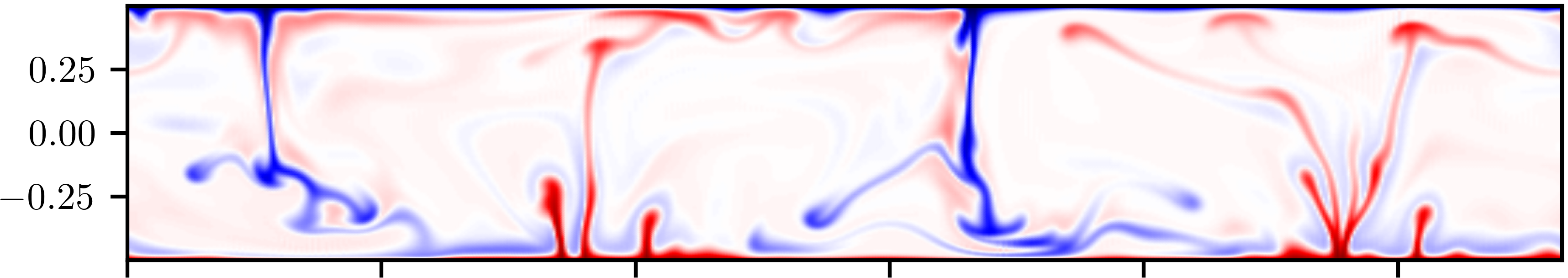} \includegraphics[width=12 cm, keepaspectratio]{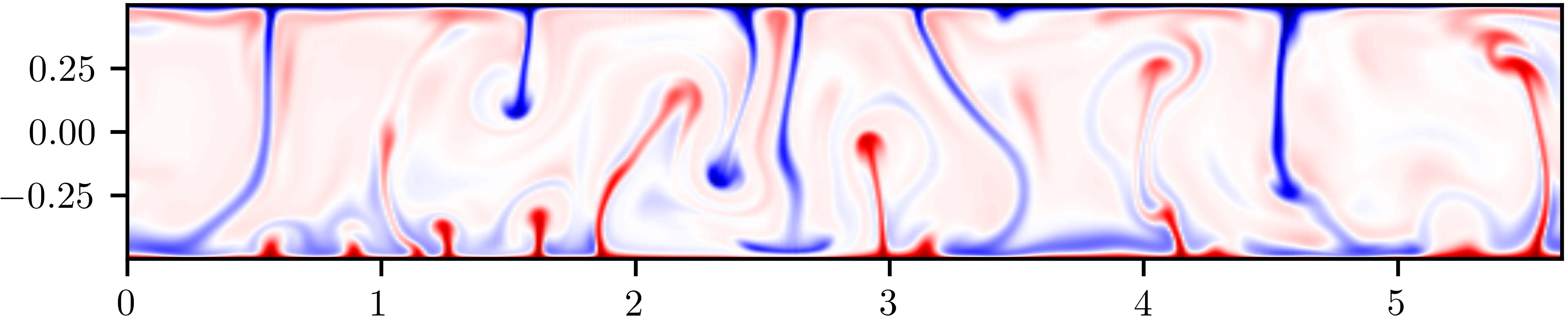}
	\caption{Snapshots of superadiabatic temperature fields for $Ra_{sa}=10^7$, \N{$\alpha _0 T_0 = 1$,} with $\mathcal{D}=0.05$ (top) and $\mathcal{D}=0.2$ (bottom) obtained in the anelastic approximation AA.}
	\label{plumestop}
\end{center}
\end{figure}

The disappearance of the top superadiabatic temperature overshoot under moderate compressibility can be connected to the fact that ascending plumes fail to reach the top of the cavity. In Fig.~\ref{plumestop}, we show snapshots of the superadiabatic temperature fields, obtained in the anelastic approximation AA, for $Ra_{sa} = 10^7$ and two values of the dissipation number $\mathcal{D}=0.05$ and $\mathcal{D}=0.2$. In the first case, ascending plumes initiated at the bottom reach the top boundary and spread horizontally, creating the top temperature overshoot. Similarly descending plumes initiated at the top reach the bottom and spread. When $\mathcal{D}=0.2$ however, most ascending plumes get mixed with the surrounding fluid before they reach the top, hence the absence of top overshoot in the temperature profile. Descending plumes can still reach the bottom.

\N{One should not think that compressible effects are always associated with stronger descending plumes. This property seems to be related to the equation of state. In another paper \citep{ralcd22}, we consider an equation of state suitable for planetary mantles and in that case, the opposite is true: with compressible effects, ascending plumes are stronger.}

\subsection{Asymmetry at larger compressibility}
\label{asymmetry}

\begin{figure}
\begin{center}
	\includegraphics[width=7 cm, keepaspectratio]{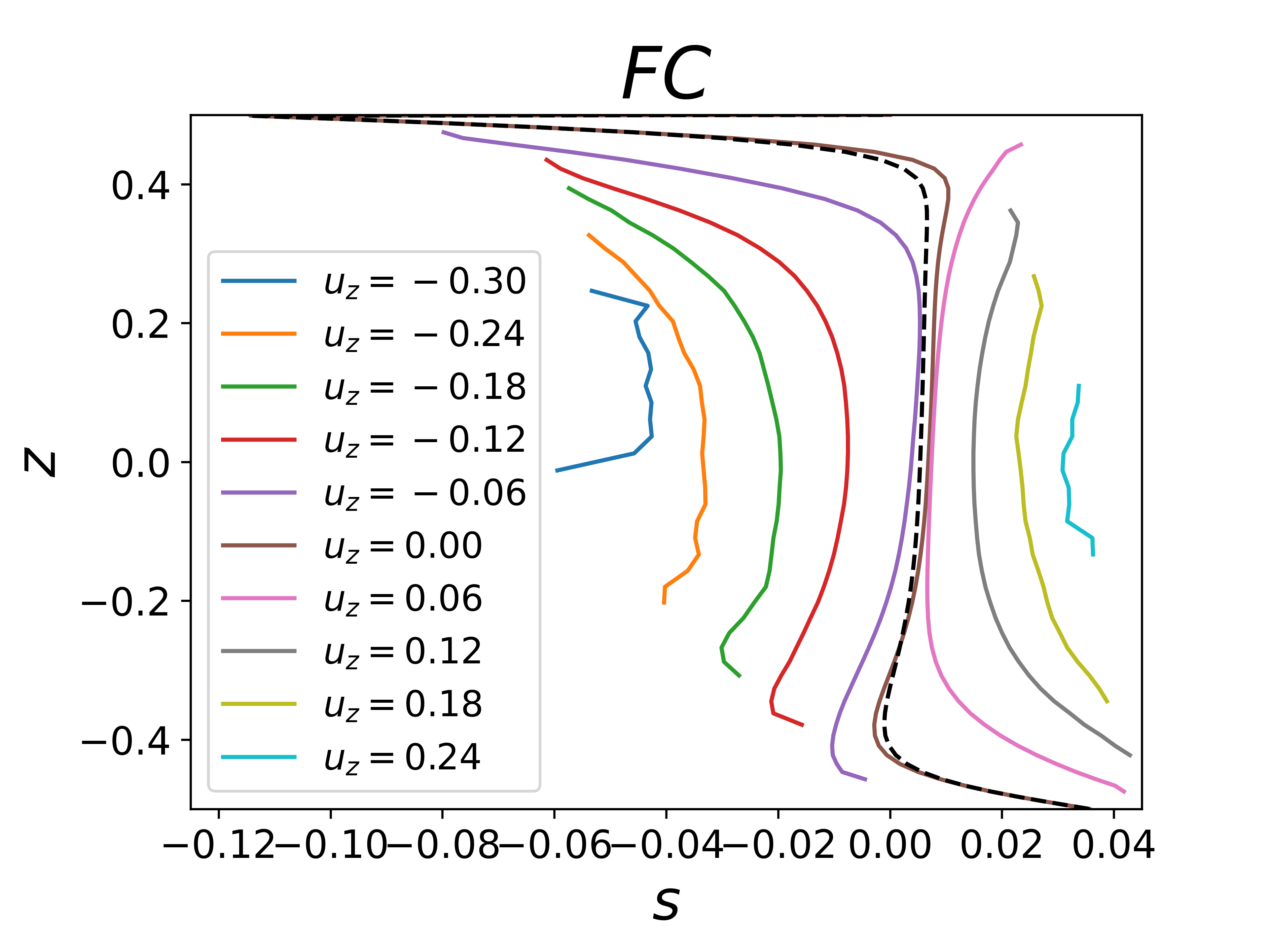} \hspace*{-8 mm} \includegraphics[width=7 cm, keepaspectratio]{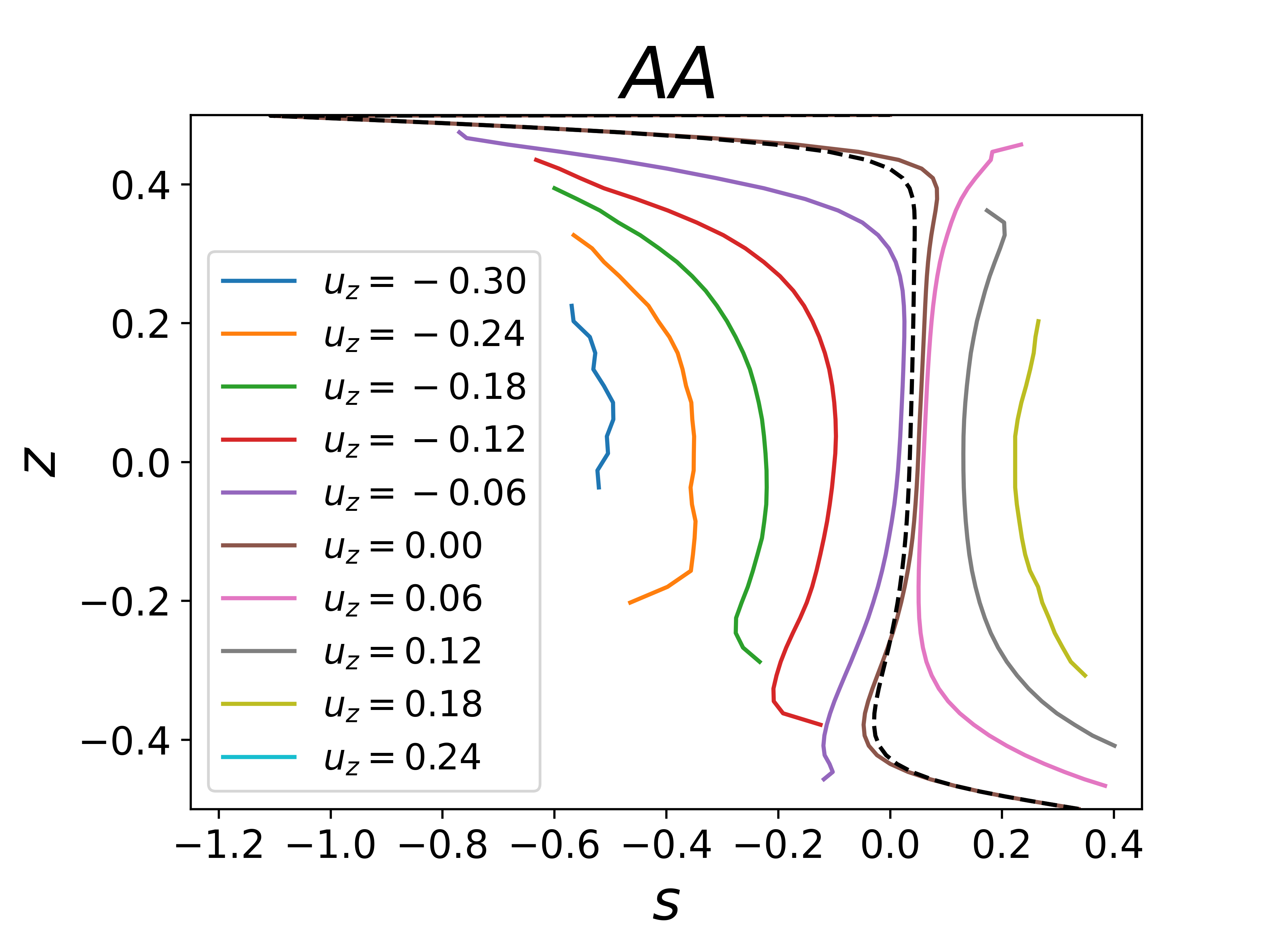}
	\includegraphics[width=7 cm, keepaspectratio]{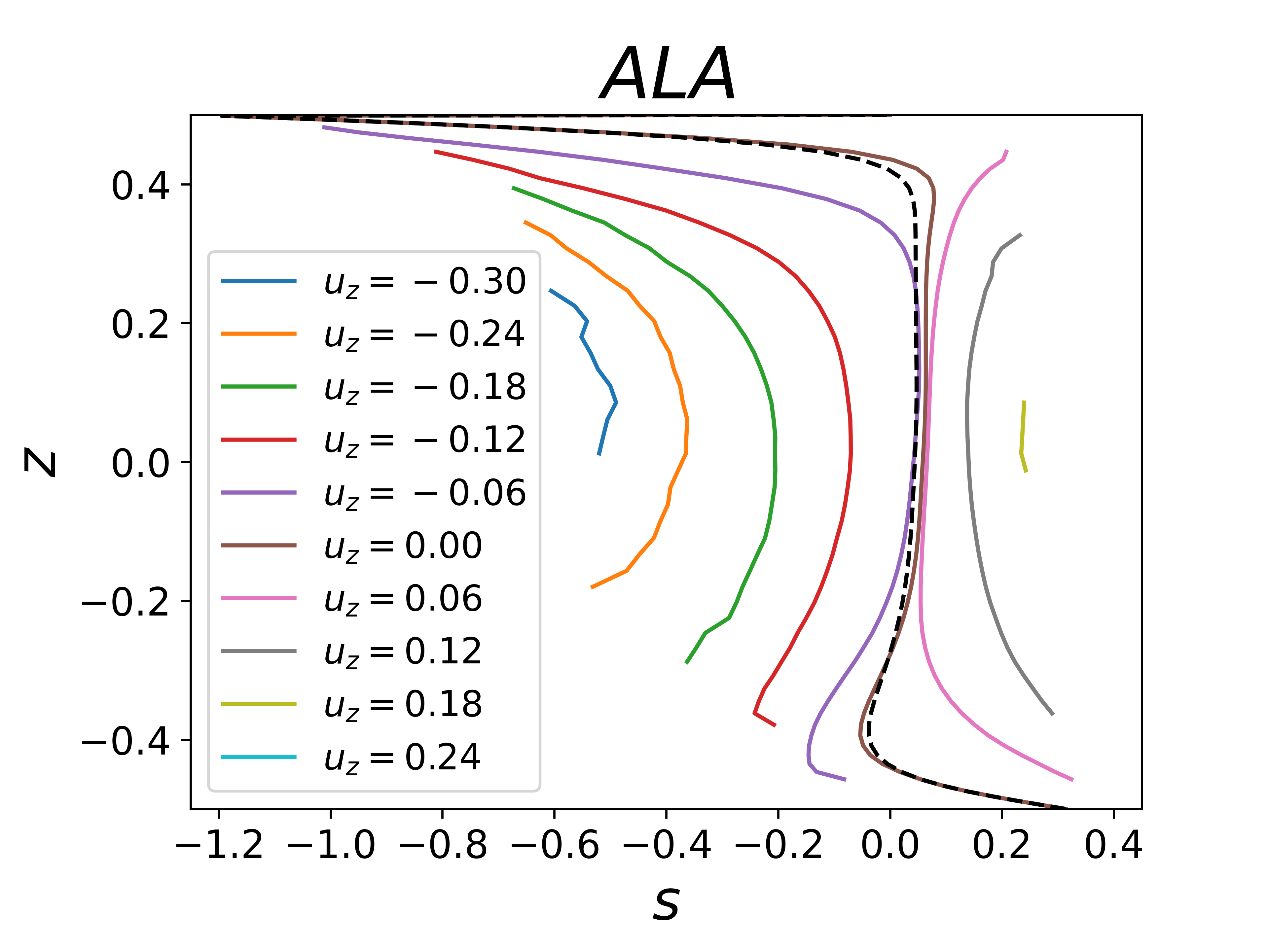} \hspace*{-8 mm} \includegraphics[width=7 cm, keepaspectratio]{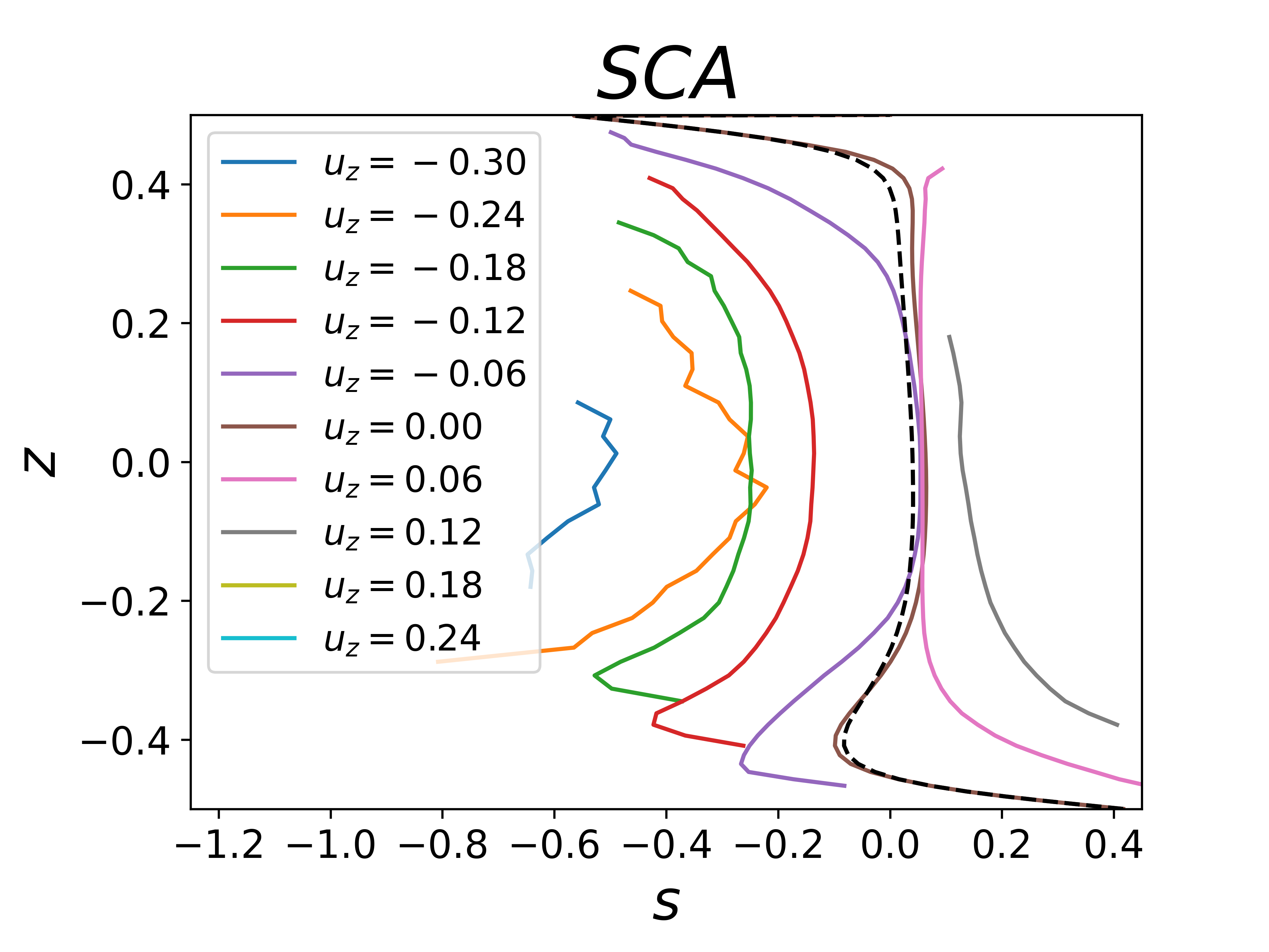}

	\caption{Vertical profiles of conditional entropy, depending on the vertical velocity scaled by $Ra_{sa}^{2/3}$, for the fully compressible equations FC and anelastic models AA, ALA and SCA, $Ra_{sa}=3\, \times \, 10^5$, $\mathcal{D}=1.2$ and \N{$\alpha _0 T_0 = 1$}. For FC, a value $\epsilon = 0.1$ has been taken for the superadiabatic temperature difference. The dashed profile is the overall mean entropy profile.  }
\label{s_profiles}
\end{center}
\end{figure}

We have seen in the previous section that a relatively moderate dissipation number of $\mathcal{D} = 0.2$ introduces a top-bottom asymmetry in thermal convection. In this section, we consider a large dissipation number $\mathcal{D} = 1.2$ and investigate the asymmetry of convection depending on the four models presented earlier: fully compressible FC, anelastic approximation AA, anelastic liquid approximation ALA and simple compressible approximation SCA. In Fig.~\ref{s_profiles}, averaged entropy profiles are shown. They are conditional profiles obtained for selected values of the vertical velocity in bins centred around the indicated values, {\it i.e.} entropy profiles for parcels of a given vertical velocity. Negative velocity values put the emphasis on descending plumes, positive values on ascending plumes. We can see that there are more profiles with negative velocities. This is because there are strong descending plumes and weak ascending plumes, a tendency already seen in the previous section for moderate dissipation numbers. The FC results are  well recovered in the anelastic models AA and ALA which show perhaps a slight increase in the asymmetry with fewer curves corresponding to positive velocity. The entropy scale of the FC plot is exactly ten times smaller than in the other models as a result of the choice of the superadiabatic parameter $\epsilon =0.1$ in the FC calculations. This difference comes from the choice of temperature scale: it is $T_0$ for the FC model and $\Delta T_{sa}$ for the AA, ALA and SCA models.

\begin{figure}
\begin{center}
	\includegraphics[width=7 cm, keepaspectratio]{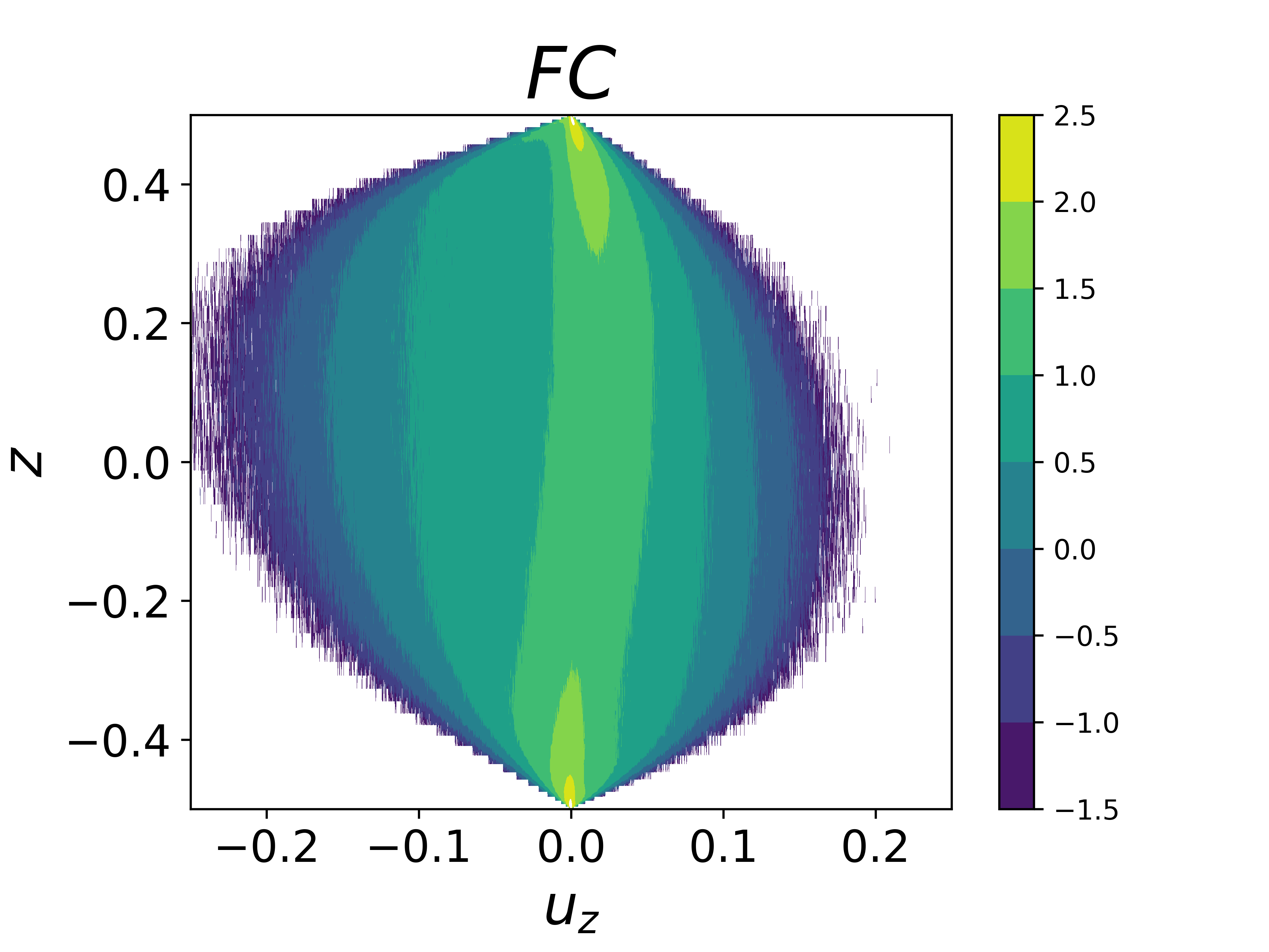} \hspace*{-8 mm} \includegraphics[width=7 cm, keepaspectratio]{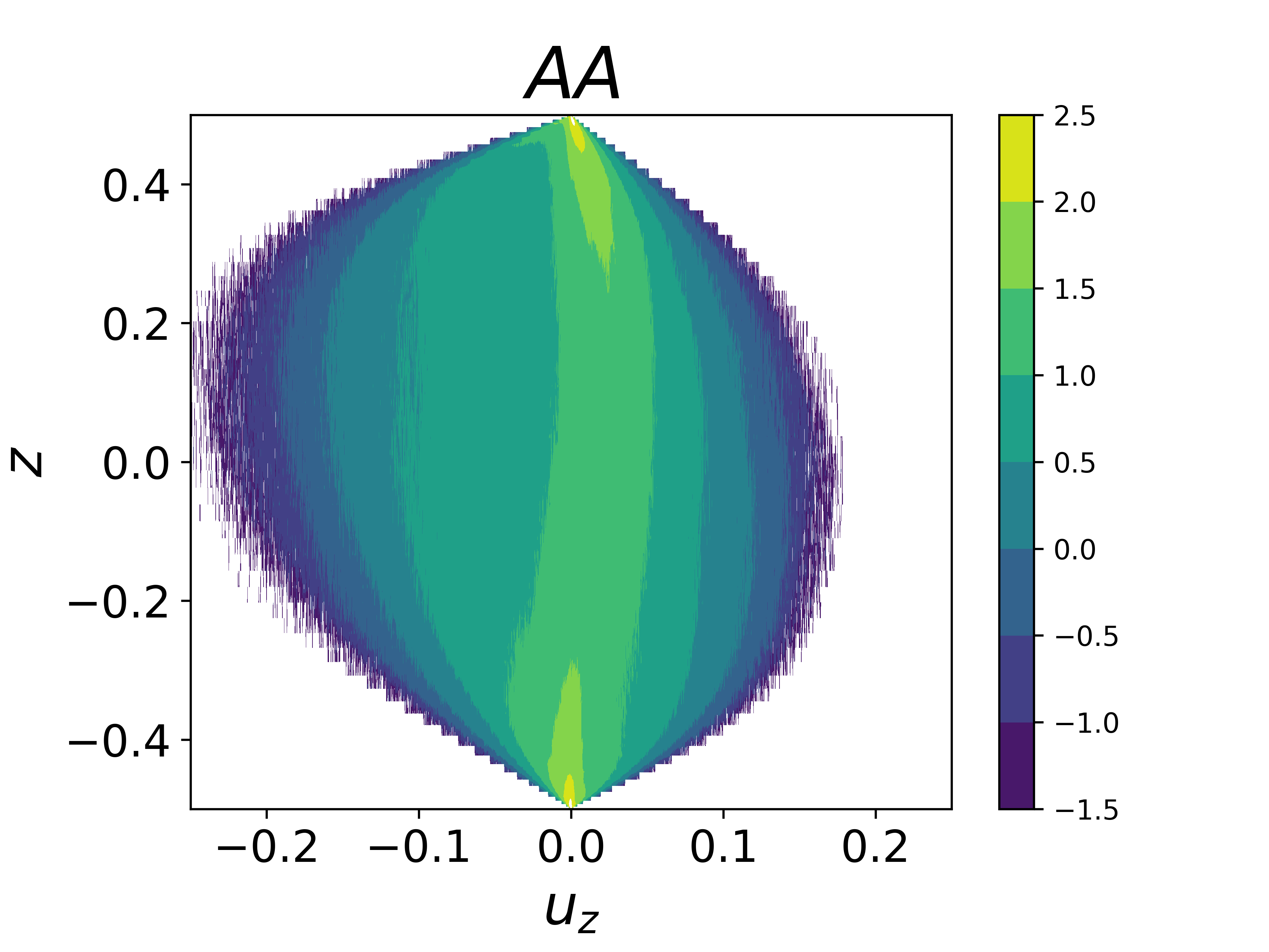}
	\includegraphics[width=7 cm, keepaspectratio]{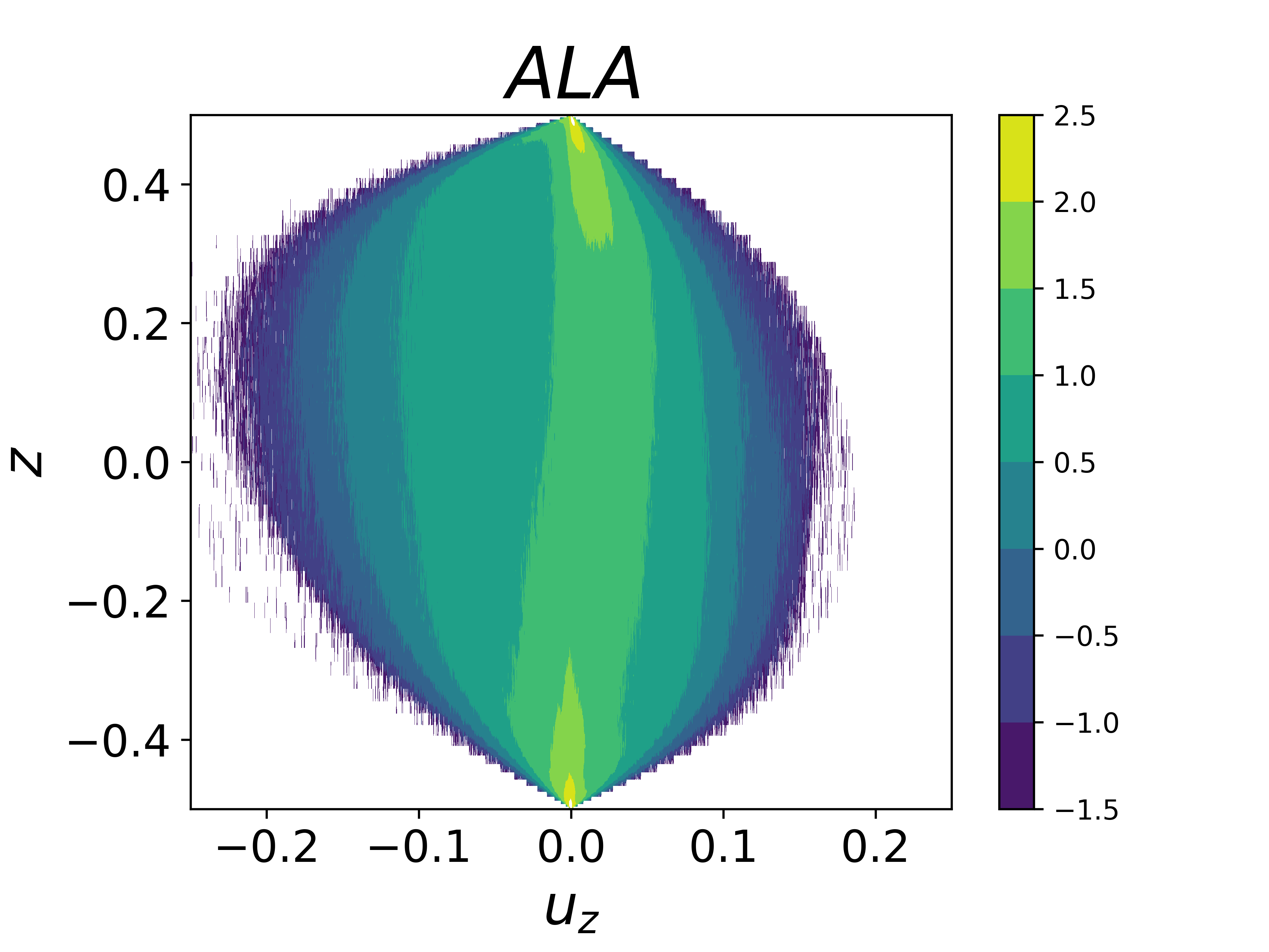} \hspace*{-8 mm} \includegraphics[width=7 cm, keepaspectratio]{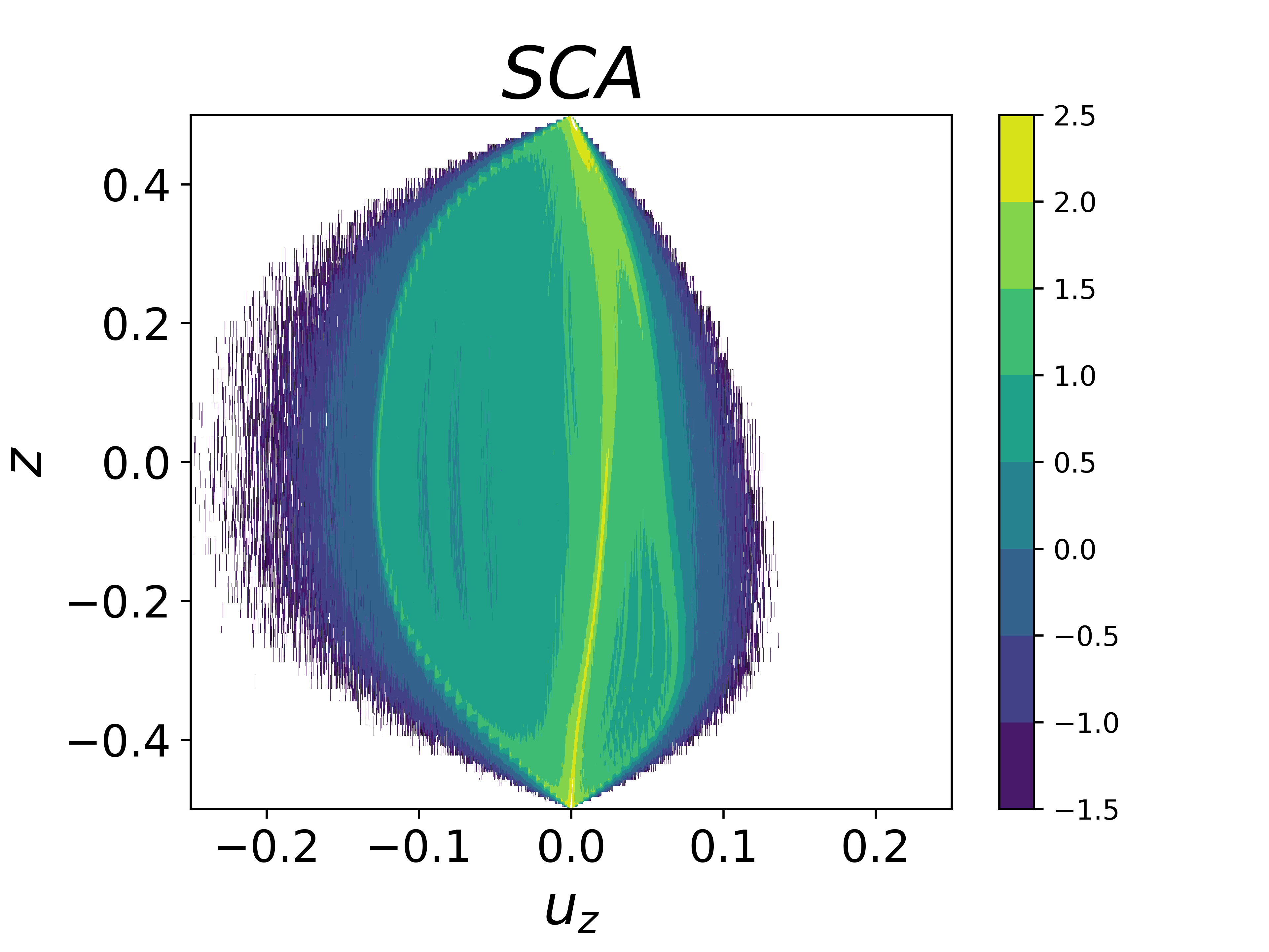}

	\caption{Probability density function (pdf) of the vertical velocity $u_z$ \N{in logarithmic scale}, depending on the altitude $z$, for FC, AA, ALA and SCA models, $Ra_{sa}=3\, \times \, 10^5$, $\mathcal{D}=1.2$ and \N{$\alpha _0 T_0 = 1$}. Velocity is scaled by $Ra_{sa}^{2/3}$. }
\label{uz_pdf}
\end{center}
\end{figure}

A striking feature of these profiles, and in particular the overall mean profiles (dashed lines) is that the top boundary layer has a much larger entropy drop compared to the bottom layer in the FC, AA and ALA cases. We will see in the next section that the superadiabatic temperature drop varies slowly with the dissipation number, and the entropy drop is essentially equal to the temperature drop divided by temperature. At $\mathcal{D} = 1.2$, the adiabatic temperature ratio is equal to $4$, hence an entropy drop nearly four times larger at the top compared to that at the bottom. Meanwhile, as can be seen in the anelastic equation (\ref{AAmomentum}), entropy is the driving force for convection. This explains why descending plumes are stronger than ascending plumes. In that respect, the situation gets somewhat similar to the case of convection cooled from the top and thermally insulated bottom, or equivalently to the case of volumic heating with fixed boundary temperatures \citep{sl99}. 

\N{Note however that other hand-waving arguments can be put forward indicating that ascending plumes (not descending) should be stronger in compressible convection. For instance, the increasing value of thermal expansion as altitude increases is thought be be such an argument leading to the strengthening of ascending plumes and weakening of descending plumes. This seems to apply in mantle convection \citep{ralcd22}. However, we have the opposite effect in the present paper, although thermal expansion also increases with altitude.}

In terms of entropy jump, the asymmetry of the thermal boundary layers is very much reduced in the simple anelastic model SCA (see Fig.~\ref{s_profiles}) because the adiabatic gradient is ignored in this model and the entropy drop is identical to the temperature drop. In the SCA case, the top-bottom asymmetry still exists, but its origin has been shifted in the thermal equation (\ref{AAentropy}). In that equation, the term $-\mathcal{D}u_z \tilde{T}$ is negative in ascending and descending plumes, and consequently favours descending plumes while impeding rising plumes. This feature of the thermal equation is really specific to the SCA model. For the other anelastic models AA and ALA, this is not the case, because the left-hand side term and the first term on the right-hand side of equation (\ref{AAentropy}) combine to form a conservative term $\mathrm{D} ( T_a \tilde{s} ) / \mathrm{D} t$. 

In Fig.~\ref{uz_pdf}, we have plotted the distribution of vertical velocities for the same simulations \N{as} those in Fig.~\ref{s_profiles}. A symmetrical top-bottom configuration would lead to contour plots symmetrical with respect to the central point $(z=0,\ u_z=0)$. The asymmetry of convection under $\mathcal{D}=1.2$ is clear, with distributions extending further toward the negative velocities (descending plumes), while the returning ascending flow is more broadly distributed on smaller values of positive velocity. Here again, AA and ALA models capture very well the distribution of velocities obtained in the FC model, although maybe with a lower probability of large values than for the FC case. However, the SCA model displays an exaggerated asymmetry and stands clearly away from the other models.

\section{\N{Malkus-type model of heat flux}}
\label{HeatFlux}

The total heat flux across the fluid layer is split into the heat conducted along the adiabat and the superadiabatic heat flux. The superadiabatic heat flux is  split itself into the conduction heat flux due to the superadiabatic temperature difference and the convective heat flux. Finally, as we will see in section \ref{heatflux}, the convective heat flux is split into the convective transport of enthalpy and the power of viscous forces. 

Here, we consider the superadiabatic flux and scale it with the heat conducted along the superadiabatic temperature difference, which is known as the Nusselt number $Nu$. In the fully compressible model (FC), we have to subtract the adiabatic conduction heat flux ($\mathcal{D}$ in this paper) to the total flux, so that the Nusselt number is defined as
\begin{equation}
	Nu = \frac{Q_{FC} - \mathcal{D}}{\varepsilon}. \label{NusseltFC} 
\end{equation}



One of the simplest models for the heat flux in the Boussinesq approximation is that proposed by \citet{malkus1954} (see also \citet{howard64}), AKA the ``critical boundary layer'' model. In this model, heat flux is independent of the depth of the fluid layer. In dimensionless terms, this leads to $Nu \sim Ra^{1/3}$. Let us adapt this model to the case of compressible convection. Let $\delta _c$ and $\delta _h$ be the dimensionless thicknesses of the top and bottom boundary layers, while $\Delta T _c$ and $\Delta T _h$ are the dimensionless superadiabatic temperature jumps across these layers. The heat flux conducted through both layers must be equal to the superadiabatic heat flux, so that
\begin{equation}
	Nu \sim \frac{\Delta T _c}{\delta _c} = \frac{\Delta T _h}{\delta _h}, \label{heatBT}
\end{equation}
while obviously, the sum of both temperature jumps must be equal to the superadiabatic temperature difference, in dimensionless terms:
\begin{equation}
	\Delta T _c + \Delta T _h = 1. \label{sumDeltaT}
\end{equation}
Concerning local Rayleigh numbers at the scale of each boundary layer, we can build them from (\ref{Rasa}) keeping in mind that $\alpha _0$ should be changed accordingly (other parameters are uniform with our equation of state). From (\ref{alphaT}), we can relate the local value of $\alpha$ to the local adiabatic temperature. Finally the local Rayleigh numbers can be written
\begin{align}
	Ra _c &= Ra_{sa} \frac{\delta _c^3 \Delta T_c}{Ta_c} = Ra_{sa} \frac{\delta _c^3 \Delta T _c}{1 - \mathcal{D}/2}, \label{localRat} \\
	Ra _h &= Ra_{sa} \frac{\delta _h^3 \Delta T_h}{Ta_h} = Ra_{sa} \frac{\delta _h^3 \Delta T _h}{1 + \mathcal{D}/2}. \label{localRab}
\end{align}
Assuming that the local Rayleigh numbers remain equal to a number $Ra_{BL}$ independent of $\mathcal{D}$, combining (\ref{heatBT}), (\ref{sumDeltaT}), (\ref{localRat}) and (\ref{localRab}) leads to the following relationship
\begin{equation}
	Nu \sim  \left( \frac{Ra_{sa}}{Ra_{BL}} \right) ^{1/3} \left[ \left( 1 -\frac{\mathcal{D}}{2} \right) ^{1/4}  + \left( 1 +\frac{\mathcal{D}}{2} \right) ^{1/4}  \right] ^{-4/3}. \label{Nucrit}
\end{equation}
Assuming that the boundary layer Rayleigh $Ra_{BL}$ \N{does not depend on the dissipation number,} this expression (\ref{Nucrit}) shall provide a prediction on the effect of the dissipation number on the Nusselt number. \N{The Rayleigh $Ra_{BL}$ defined above should not be confused with the critical Rayleigh number for the onset of convection (based on the height of the cavity). The latter is close to Rayleigh's value $27 \pi ^4 / 4$ for a nearly uniform density \citep{AR2017}.}

\begin{figure}
\begin{center}
        \includegraphics[width=12 cm, keepaspectratio]{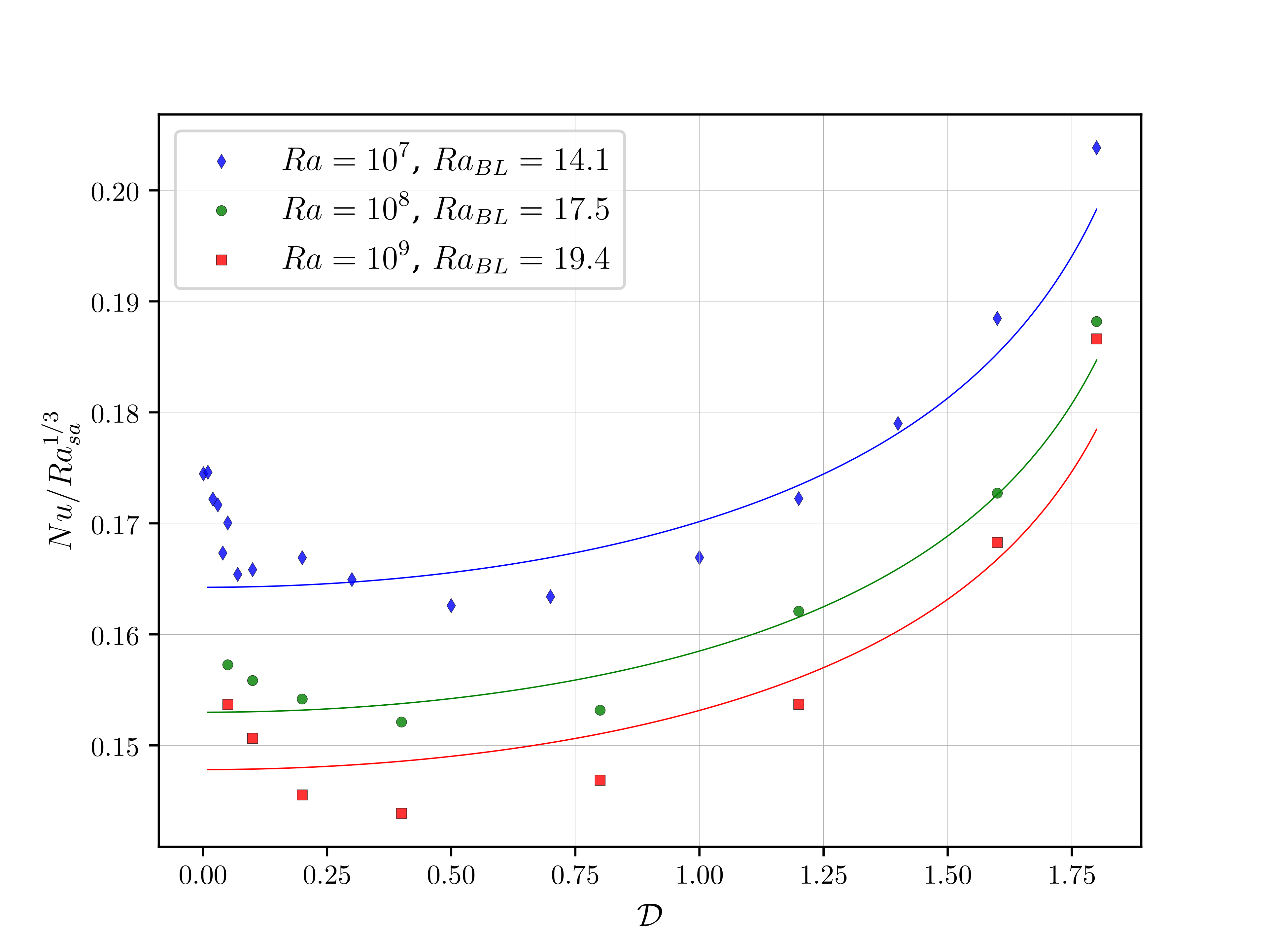}
	\caption{Nusselt number normalized by $Ra_{sa}^{1/3}$ in the anelastic model. The Nusselt number is defined as the ratio of the superadiabatic heat flux to the superadiabatic pure conduction flux. The solid lines correspond to the expression (\ref{Nucrit}) with boundary layer Rayleigh numbers equal to $Ra_{BL} = 14.1$, $Ra_{BL} = 17.5$ and $Ra_{BL} = 19.4$, such that they are quadratic best fits of the numerical Nusselt between $\mathcal{D}=0.4$ and $\mathcal{D}=1.8$.}
\label{nusselt}
\end{center}
\end{figure}

Figure \ref{nusselt} shows the Nusselt number as a function of the dissipation number, for three values of the superadiabatic Rayleigh number in the higher range $Ra = 10^7$, $Ra = 10^8$ and $Ra = 10^9$. From $\mathcal{D} = 0$ to approximately $\mathcal{D} = 0.4$, a reduction of the Nusselt number is observed: this should be understood as a consequence of the structure change discussed in section \ref{smallD} where the ascending plumes are shown to get weaker with the disappearance of the top overshoot of the superadiabatic temperature profile. Then for larger values of $\mathcal{D}$, the Nusselt number increases, as predicted by our model (\ref{Nucrit}) and even a little faster. The three lines in Fig.~\ref{nusselt} correspond to equation (\ref{Nucrit}) each with a boundary Rayleigh number $Ra_{BL}$ adjusted to fit the numerical Nusselt numbers at different $Ra$. A general good agreement is obtained from the model of boundary layer Rayleigh number (\ref{Nucrit}).

\N{It should be noted that a more sophisticated version of heat transfer model exists for compressible convection. The model of \citet{jmk2022} is designed as an extension of a model of heat transfer by \citet{GrLo00} valid in the Boussinesq approximation. These models provide expressions for the Nusselt number as a function of the Rayleigh number, Prandtl number and dissipation number in the compressible case. They apply in a domain of finite aspect ratio, {\it i.e.} bounded by vertical walls, as assumptions are made on the large-scale velocity field. The Prandtl number can be small or large, but not infinite. In the compressible case \citep{jmk2022}, the model takes a different form depending on whether dissipation occurs mostly in the boundary layers or in the bulk. Unfortunately, these models have been developed in the no-slip boundary condition and the case of dissipation mostly in the bulk is associated with small Prandtl numbers, and turbulence cascade. This makes it rather difficult to apply in our case of infinite Prandtl number and free-slip boundary conditions, even though dissipation is mainly in the bulk in the upper range of our superadiabatic Rayleigh numbers. Moreover, the model by \citet{jmk2022} has been developed for an EoS of a perfect gas. This model predicts a decrease of the Nusselt number with the dissipation number for a fixed value of the Rayleigh (and Prandtl) number, which is not compatible with our results. That discrepancy is however not relevant, given the mismatch between the conditions of validity of the model and our configuration. }

\section{Heat flux and Dissipation in the different models}
\label{heatflux}

The expression for the vertical heat flux across horizontal planes takes a specific form for each model of convection. In the fully compressible model (FC), we take equation (\ref{entropya}), integrate by parts the viscous dissipation term and use the dot-product of equation (\ref{momentuma}) with velocity ${\bf u}$ to obtain an expression for the heat flux as a function of height $z$
\begin{equation}
	Q_{FC} (z) = (n+1) \overline{T \rho ^{n+1} u_z} - \frac{\varepsilon \mathcal{D}}{Ra_{sa}} \overline{ u_j \tau _{zj}} - \frac{ \mathrm{d} \overline{T}}{\mathrm{d} z} , \label{fluxFC}
\end{equation}
where the overline $\overline{ \cdot  }$ denotes the average over horizontal planes, or constant-z surfaces, and over time. 
This is fully in accordance with the general expression for the heat flux (see for instance (4.5) in \cite{cdadlr2019})  
\begin{equation}
	Q_{FC} (z) = \overline{\rho h u_z} - \frac{\varepsilon \mathcal{D}}{Ra_{sa}} \overline{ u_j \tau _{zj}} - \frac{ \mathrm{d} \overline{T}}{\mathrm{d} z} , \label{fluxFC2} 
\end{equation}
 as from (\ref{Phlog}) and (\ref{Phpower}), the dimensionless expression of $h$ is 
\begin{equation}
	h = (n+1) T \rho ^n, \hspace*{1 cm} \mathrm{for}\ n \geq 0. \label{hFC}
\end{equation}
In the statistically stationary case considered here, the function $Q_{FC} (z)$ must be independent of $z$, and we may equally denote it $Q_{FC}$. 

In the anelastic approximation (AA), \N{equations (\ref{AAentropy}), (\ref{AAmomentum}) and (\ref{AAs})} lead to the following expression for the superadiabatic heat flux
\begin{equation}
	Q_{AA} (z) = \overline{ \left( \tilde{T} + \frac{n}{n+1} \tilde{P} \right) u_z} - \frac{\mathcal{D}}{Ra_{sa}} \overline{ u_j \tau _{zj}} - \frac{\mathrm{d}  \overline{\tilde{T}}}{\mathrm{d} z}. \label{fluxAA}
\end{equation}
This is again compatible with the general anelastic expression (see for instance (4.6) in \cite{cdadlr2019}) 
\begin{equation}
	Q_{AA} (z) = \rho _a \overline{\tilde{h} u_z} - \frac{\mathcal{D}}{Ra_{sa}} \overline{ u_j \tau _{zj}} - \frac{\mathrm{d} \overline{\tilde{T}}}{\mathrm{d} z}, \label{fluxAA2}
\end{equation}
with $\rho _a = 1$, uniform within the class of equations of state considered in this paper, and the following linearized expression for enthalpy, from (\ref{hFC}) and using (\ref{AArho})
\begin{equation}
	\tilde{h} = \tilde{T} + \frac{n}{n+1} \tilde{P}, \hspace*{1 cm} \mathrm{for}\ n>0. \label{hAA}
\end{equation}
In statistically stationary cases, the time-averaged heat flux is independent of $z$ and its value can be obtained by integrating (\ref{fluxAA}) over the height of the fluid domain, leading to
\begin{equation}
	Q_{AA} = \left< \tilde{T} u_z \right> +  \frac{n}{n+1} \left< \tilde{P}  u_z \right> + 1, \label{globalFluxAA}
\end{equation}
since the power of viscous forces can be shown to integrate to zero, because of the continuity equation ${\bf \bnabla} \cdot (\rho _a {\bf u}) = {\bf \bnabla} \cdot {\bf u} = 0$ in the anelastic approximation, as $\rho _a = 1$.  

In the anelastic liquid approximation (ALA), the expression for the superadiabatic flux, obtained in the similar way as for AA, is 
\begin{equation}
	Q_{ALA} (z) = \overline{ \left( \tilde{T} + \tilde{P} \right) u_z}- \frac{\mathcal{D}}{Ra_{sa}} \overline{ u_j \tau _{zj}} - \frac{\mathrm{d} \overline{\tilde{T}}}{\mathrm{d} z}. \label{fluxALA}
\end{equation}
In a statistically stationary situation, the heat flux can be computed as
\begin{equation}
	Q_{ALA} = \left< \tilde{T} u_z \right> + \left<  \tilde{P} u_z \right> + 1. \label{globalFluxALA}
\end{equation}
These expressions correspond to the limit $n \rightarrow \infty$ of those obtained in the general anelastic approximation (AA), which also corresponds to the limit $\alpha T \rightarrow 0$.  

Finally, in the simple compressible approximation (SCA), the expressions for the heat flux are identical to those in the anelastic liquid approximation (ALA).

However, the heat flux can be written differently than in the expressions above. For instance, in the anelastic approximation (AA), instead of (\ref{fluxAA}), (\ref{fluxAA2}) or (\ref{globalFluxAA}), one can base it on the flux of entropy. From Gibbs equation $\mathrm{d} h = T \mathrm{d} s + \mathrm{d} P / \rho $, expression (\ref{fluxAA2}) can be written 
\begin{equation}
	Q_{AA} (z) = T_a \overline{\tilde{s} u_z} + \overline{\tilde{P} u_z} - \frac{\mathcal{D}}{Ra_{sa}} \overline{ u_j \tau _{zj}} - \frac{\mathrm{d} \overline{\tilde{T}}}{\mathrm{d} z}. \label{fluxEntAA}
\end{equation}
In the same time, from the horizontal and time average of the dot-product of Navier-Stokes with velocity, one obtains \N{the expression for the viscous dissipation at each vertical position}
\begin{equation}
	\frac{\mathcal{D}}{Ra_{sa}} \overline{\dot{\epsilon} : \tau } (z) = \mathcal{D} \overline{\tilde{s} u_z} - \frac{\mathrm{d}}{\mathrm{d} z} \left[ \overline{\tilde{P} u_z} - \frac{\mathcal{D}}{Ra_{sa}} \overline{ u_j \tau _{zj}}   \right]. \label{dissEntAA}
\end{equation}
Introducing 
\begin{equation}
	G(z) = \overline{\tilde{P} u_z} - \frac{\mathcal{D}}{Ra_{sa}} \overline{ u_j \tau _{zj}} , \label{F}
\end{equation}
equations (\ref{fluxEntAA}) and (\ref{dissEntAA}) can be rewritten
\begin{align}
	Q_{AA} (z) & = T_a \overline{\tilde{s} u_z} + G(z) - \frac{\mathrm{d} \overline{\tilde{T}}}{\mathrm{d} z}, \label{fluxEntAA2} \\
	\frac{\mathcal{D}}{Ra_{sa}} \overline{\dot{\epsilon} : \tau } (z) & = \mathcal{D}  \overline{\tilde{s} u_z} - \frac{\mathrm{d} G(z)}{\mathrm{d} z} . \label{dissEntAA2}
\end{align}
It can be noted that these expressions are valid for any general equation of state and can also be generalized when inertia and possible Lorentz forces are included: it suffices to add the flux of inertia to the expression of $G(z)$, as detailed in appendix \ref{A1}. In the anelastic liquid approximation, the expression for $G(z)$ is unchanged and the flux and dissipation profiles become
\begin{align}
	Q_{ALA} (z) & = \overline{\tilde{T} u_z} + G(z) - \frac{\mathrm{d} \overline{\tilde{T}}}{\mathrm{d} z}, \label{fluxEntALA2} \\
	\frac{\mathcal{D}}{Ra_{sa}} \overline{\dot{\epsilon} : \tau } (z) & = \mathcal{D} \frac{ \overline{\tilde{T} u_z} }{T_a} - \frac{\mathrm{d} G(z)}{\mathrm{d} z} . \label{dissEntALA2}
\end{align}
In the simplest SCA approximation, $G(z)$ is still the same and the flux and dissipation profiles are
\begin{align}
	Q_{SCA} (z) & = \overline{\tilde{T} u_z} + G(z) - \frac{\mathrm{d} \overline{\tilde{T}}}{\mathrm{d} z}, \label{fluxEntSCA2} \\
	\frac{\mathcal{D}}{Ra_{sa}} \overline{\dot{\epsilon} : \tau } (z) & = \mathcal{D} \overline{\tilde{T} u_z} - \frac{\mathrm{d} G(z)}{\mathrm{d} z} . \label{dissEntSCA2}
\end{align}

\subsection{\N{Global dissipation}}
\label{DissHeatFlux}

The total Dissipation is expressed as a fraction of the heat flux, more specifically the superadiabatic convective heat flux. The reference result is obtained in the Boussinesq approximation, where it has been shown that the integrated viscous dissipation is equal to the product of the dissipation number and the convective heat flux \N{\citep{howard63,HMKW75}}.
We express the total viscous dissipation as a fraction of the conduction heat flux along the superadiabatic gradient and denote it $D_\nu$. In the fully compressible model (FC), the integrated viscous dissipation divided by $\Delta T_{sa}$ has the following expression
\begin{equation}
	D_\nu = \frac{\mathcal{D}}{Ra_{sa} \, \varepsilon} \left< \dot{\epsilon} : \tau \right> . \label{dissipFC}
\end{equation}
In the approximated models (AA), (ALA) and (SCA), the expression is 
\begin{equation}
	D_\nu = \frac{\mathcal{D}}{Ra_{sa}} \left< \dot{\epsilon} : \tau \right> , \label{dissipA}
\end{equation}
because the dimensionless temperature is scaled using $\Delta T_{sa}$ for these models. Considering the expressions for the Nusselt number in section \ref{HeatFlux}, the ratio \N{$E$} of viscous dissipation to the convective heat flux takes the same expression for all models
\begin{equation}
	E = D_\nu / (Nu -1 ) . \label{DissVersusConv}
\end{equation}
The classical Boussinesq result on dissipation comes from integrating the dot product of Navier-Stokes with the velocity field. It follows that viscous dissipation is exactly equal to the convective heat flux multiplied by the dissipation number \citep{howard63}, hence $\N{E = }\ D_\nu / (Nu -1 ) = \mathcal{D}$ in the Boussinesq limit. For this reason, we call the quantity $E - \mathcal{D}$ the ratio of dissipation to convective flux in excess of $\mathcal{D}$. It is zero in the Boussinesq case and we compute it for compressible convection FC, AA, ALA and SCA.

\begin{figure}
\begin{center}
        \includegraphics[width=12 cm, keepaspectratio]{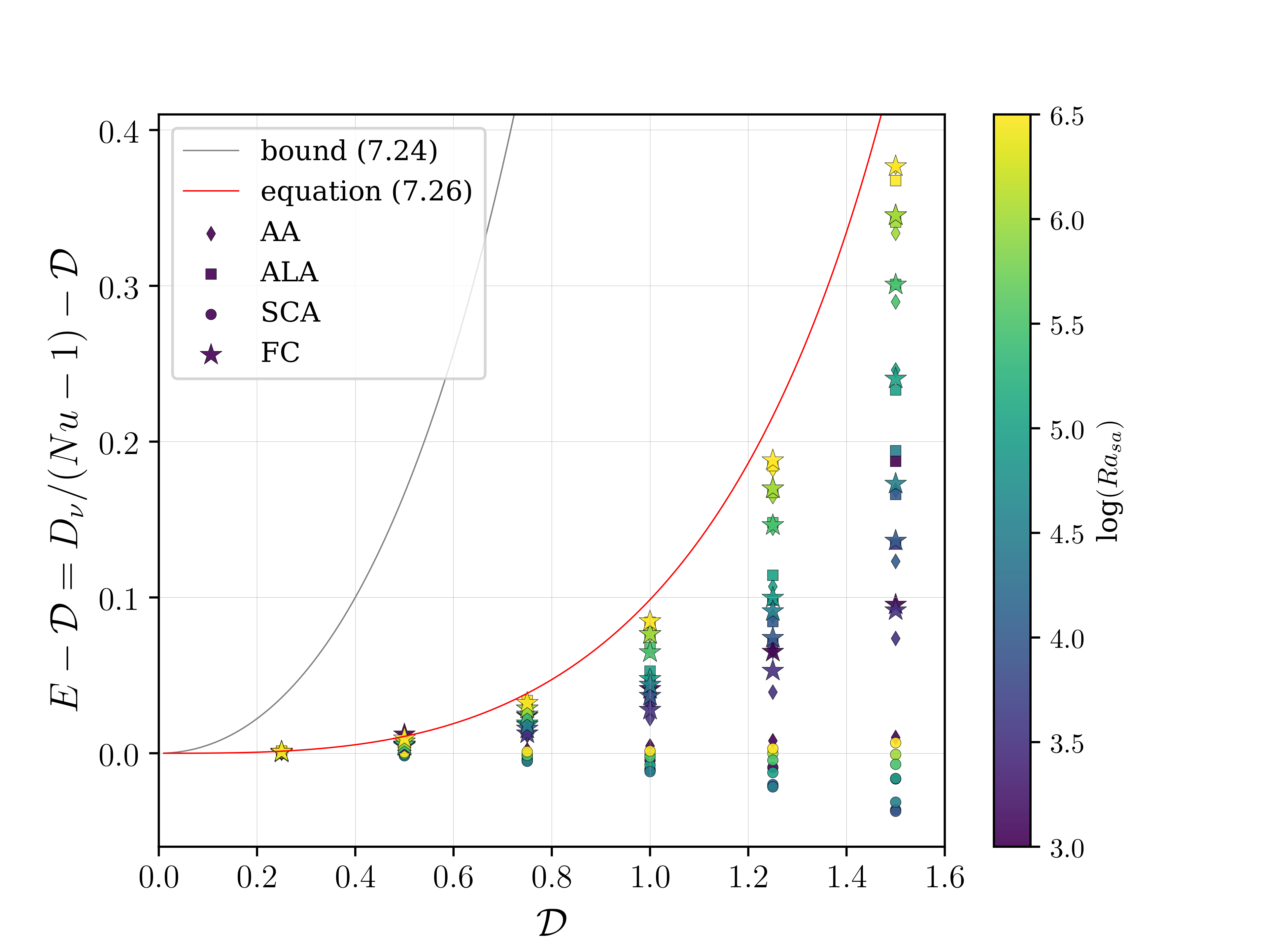}
	\caption{Excess of the ratio of dissipation to convective heat flux relative to $\mathcal{D}$, as a function of $\mathcal{D}$, for FC, AA, ALA, SCA models, \N{$\alpha _0 T_0 = 1$,} $\mathcal{D}$ in $[0.25,\, 0.5,\, 0.75,\, 1.0,\, 1.25,\, 1.5]$, $Ra_{sa}$ in $[10^3,\ 10^{3.5},\, 10^4,\ 10^{4.5},\, 10^5,\ 10^{5.5},\, 10^6,\ 10^{6.5}]$.}
\label{excessDiss}
\end{center}
\end{figure}

\begin{figure}
\begin{center}
        \includegraphics[width=12 cm, keepaspectratio]{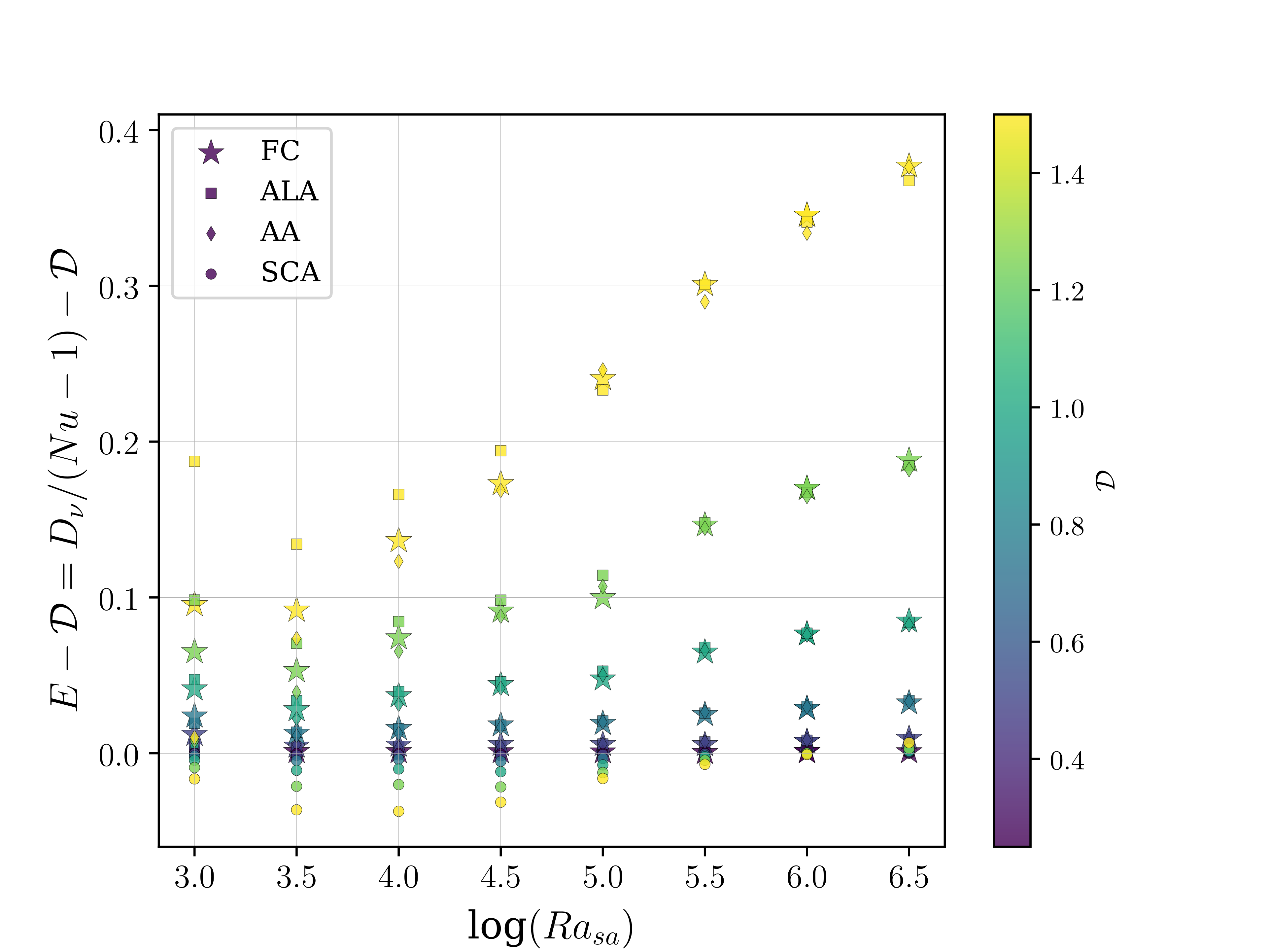}
	\caption{Excess of the ratio of dissipation to convective heat flux relative to $\mathcal{D}$, as a function of $Ra_{sa}$ (same values as in Fig.~\ref{excessDiss}).  }
\label{RaexcessDiss}
\end{center}
\end{figure}

The numerical results concerning viscous dissipation are shown in Fig.~\ref{excessDiss}. For small values of $\mathcal{D}$, all results are close to $0$, indicating that the Boussinesq results apply: dissipation is equal to the product of the convective heat flux and the dissipation number. When the dissipation number is increased, the simple compressible approximation model (SCA) goes to slightly negative values, whereas all other models go to positive values. So, for all models except (SCA), viscous dissipation becomes larger than predicted by the Boussinesq approximation. That departure increases with the dissipation number $\mathcal{D}$ and also with the superadiabatic Rayleigh number $Ra_{sa}$ (better seen in Fig.~\ref{RaexcessDiss}).

Let us consider some limits to the dissipation results. First, we have rigorous upper (and lower) bounds, since the papers of \citet{ba1975} \N{and \citet{HMKW75}}. Taking equation (\ref{AAentropy}), dividing by $T_a$ and integrating by parts leads to
\begin{equation}
	\frac{ \mathrm{D} \tilde{s} }{\mathrm{D} t} - {\bf \bnabla} \cdot \left( \frac{{\bf \bnabla} \tilde{T}}{T_a}  \right) = \frac{ \mathcal{D}}{Ra_{sa}} \frac{ \dot{\epsilon} : \tau}{T_a}  + \frac{{\bf \bnabla} \tilde{T} \cdot {\bf \bnabla} T_a }{T_a^2} , \label{AAtrueentropy}
\end{equation}
where we recognize the positive sources of entropy due to irreversible viscous dissipation and conduction on the right-hand side. The upper bound for total dissipation occurs for negligible conduction (as an entropy source) and in the case when dissipation takes place at the largest possible temperature $T_a = 1 + \mathcal{D}/2$, at the bottom of the domain. Integrating (\ref{AAtrueentropy}) over the whole domain leads to 
\begin{equation}
	\frac{ D_\nu }{1 + \frac{\mathcal{D}}{2}} <  Nu \left[\frac{1}{ 1 - \frac{\mathcal{D}}{2}} - \frac{1}{ 1 + \frac{\mathcal{D}}{2}} \right] , 
\end{equation}
implying the upper bound
\begin{equation}
	\frac{D_\nu}{Nu -1}  <  \mathcal{D} + \frac{\mathcal{D}^2}{ 2 - \mathcal{D} }, \label{AAubound}
\end{equation}
in the limit of a large Nusselt number $Nu \gg 1$. 

Another possible limit case will be shown to correspond to our numerical results at large Rayleigh number in the next section \ref{profiles}. That limit is that of a vanishing contribution of the $G(z)$ component of the heat flux defined in equation (\ref{F}) in the anelastic approximation. It corresponds to the case when the heat flux is carried entirely by the flux of entropy $T_a \overline{\tilde{s} u_z}$ (outside top and bottom conduction layers)  while the energy dissipation at each height is $\mathcal{D} \overline{\tilde{s} u_z}$, as can be seen from equations (\ref{fluxEntAA2}) and (\ref{dissEntAA2}). Under that assumption, the integral of energy dissipation is equal to
\begin{equation}
	D_\nu = Nu \int _{1/2}^{1/2} \frac{\mathcal{D}}{T_a} \mathrm{d} z , \label{entropyfluxdiss0}
\end{equation}
leading to the expression
\begin{equation}
	\frac{D_\nu}{Nu -1} = \log \left(\frac{1+\mathcal{D}/2}{1-\mathcal{D}/2} \right) , \label{entropyfluxdiss}
\end{equation}

The expressions for the upper bound (\ref{AAubound}) and for the ``entropy flux'' model (\ref{entropyfluxdiss}) are plotted in Fig.~\ref{excessDiss}. It can be seen that the upper bound curve (\ref{AAubound}) lies far above the numerical results and that the 'entropy flux' expression (\ref{entropyfluxdiss}) seems to correspond to the limit of the numerical results when the Rayleigh number is increased, for FC, AA and ALA models. For small (supercritical) Rayleigh numbers, dissipation is close to the 'Boussinesq' limit. For the SCA model, the behaviour is different:~starting from negative values at small $Ra_{sa}$ numbers, the 'Boussinesq' limit is reached for large $Ra_{sa}$ numbers. 
However, from a fundamental perspective, the SCA model does not behave differently from the other models if one considers the consequences of the 'entropy flux' assumption, because of the different expressions for the flux and dissipation. From (\ref{fluxEntSCA2}) and (\ref{dissEntSCA2}), it is clear that the assumption of a negligible $G$ contribution leads to the usual 'Boussinesq' limit for the SCA model, just because it is identical to the 'entropy flux' limit in that simple compressible approximation. Thus, in the limit of large $Ra_{sa}$, when the contribution of $G$ is small, it is expected that dissipation becomes close to the product of the dissipation number and the convective heat flux. This is what we observe on Figs.~\ref{excessDiss} and \ref{RaexcessDiss}.

It is striking that in Fig.~\ref{RaexcessDiss} at low Rayleigh number ALA, FC and AA results differ significantly. Dissipation is smallest with AA, then FC and highest with ALA. Typically AA excess dissipation goes to zero near the critical Rayleigh number, while ALA excess dissipation shows an increase to some finite value. FC results are intermediate. When the superadiabatic Rayleigh number becomes larger, the difference between AA, FC and ALA tends to shrink until the predicted dissipation is the same above $10^5$ or $10^6$. 





\subsection{Convergence of heat flux and dissipation profiles}
\label{profiles}

\begin{figure}
\begin{center}
	\includegraphics[width=7.3 cm, keepaspectratio]{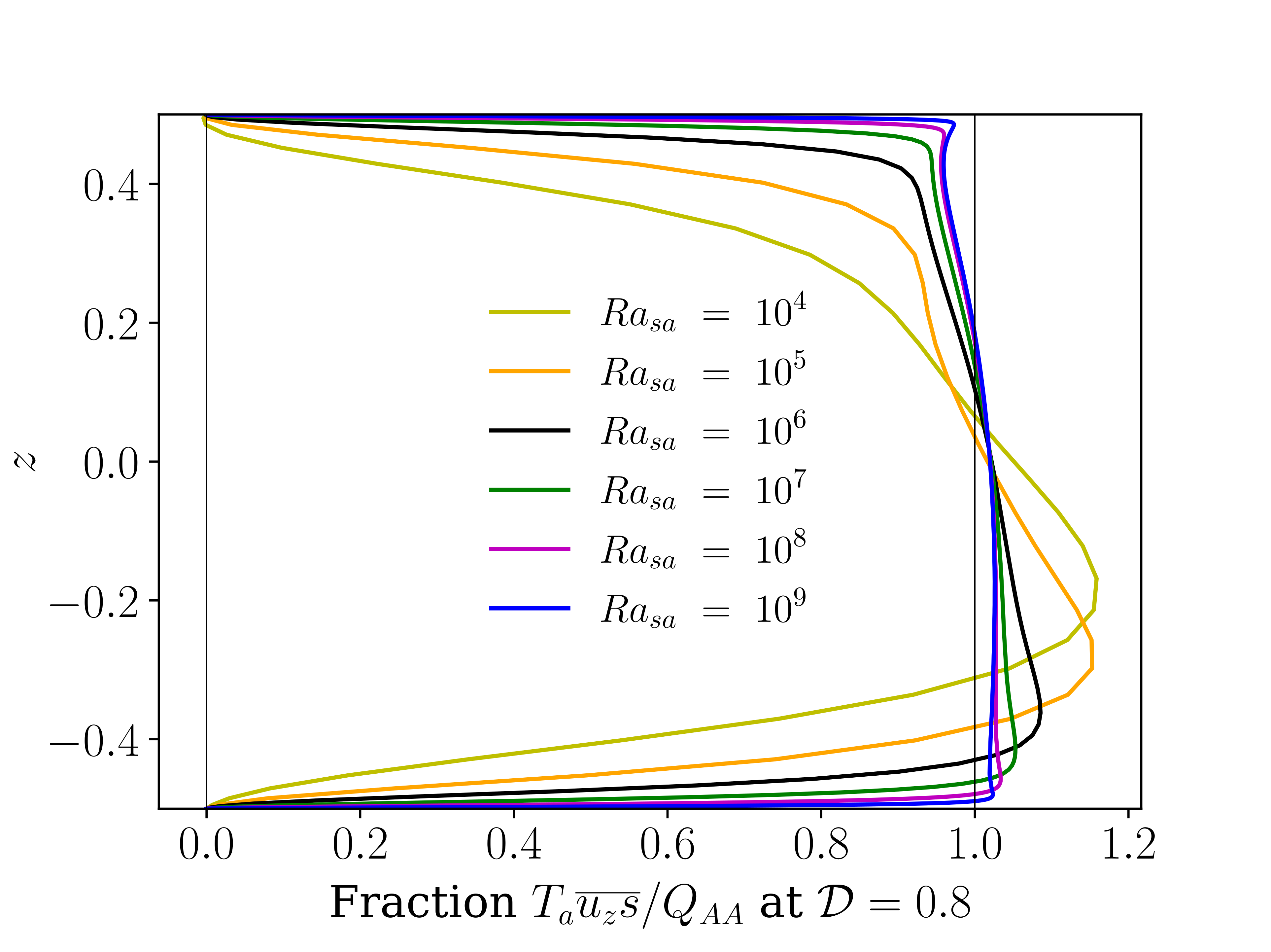} \hspace*{-0.7 cm} \includegraphics[width=7.3 cm, keepaspectratio]{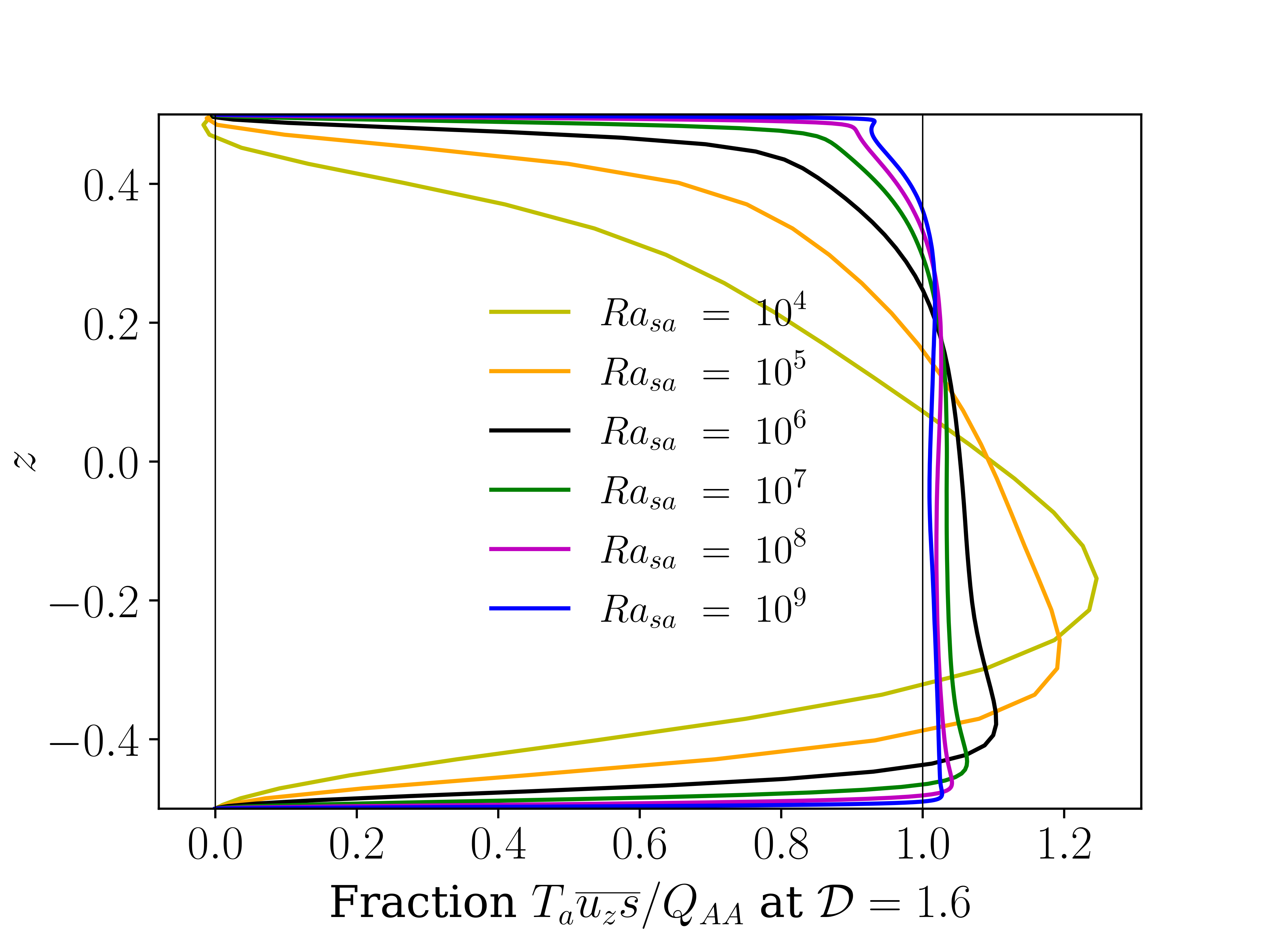}\hspace*{-0.6 cm}
	\caption{Entropy flux contribution for $\mathcal{D}$ equal to $0.8$ (left) and $1.6$ (right), for a superadiabatic Rayleigh number up to $10^9$ in the anelastic approximation and \N{$\alpha _0 T_0 = 1$}.  }
\label{FluxProf}
\end{center}
\end{figure}

\begin{figure}
\begin{center}
        \includegraphics[width=7.3 cm, keepaspectratio]{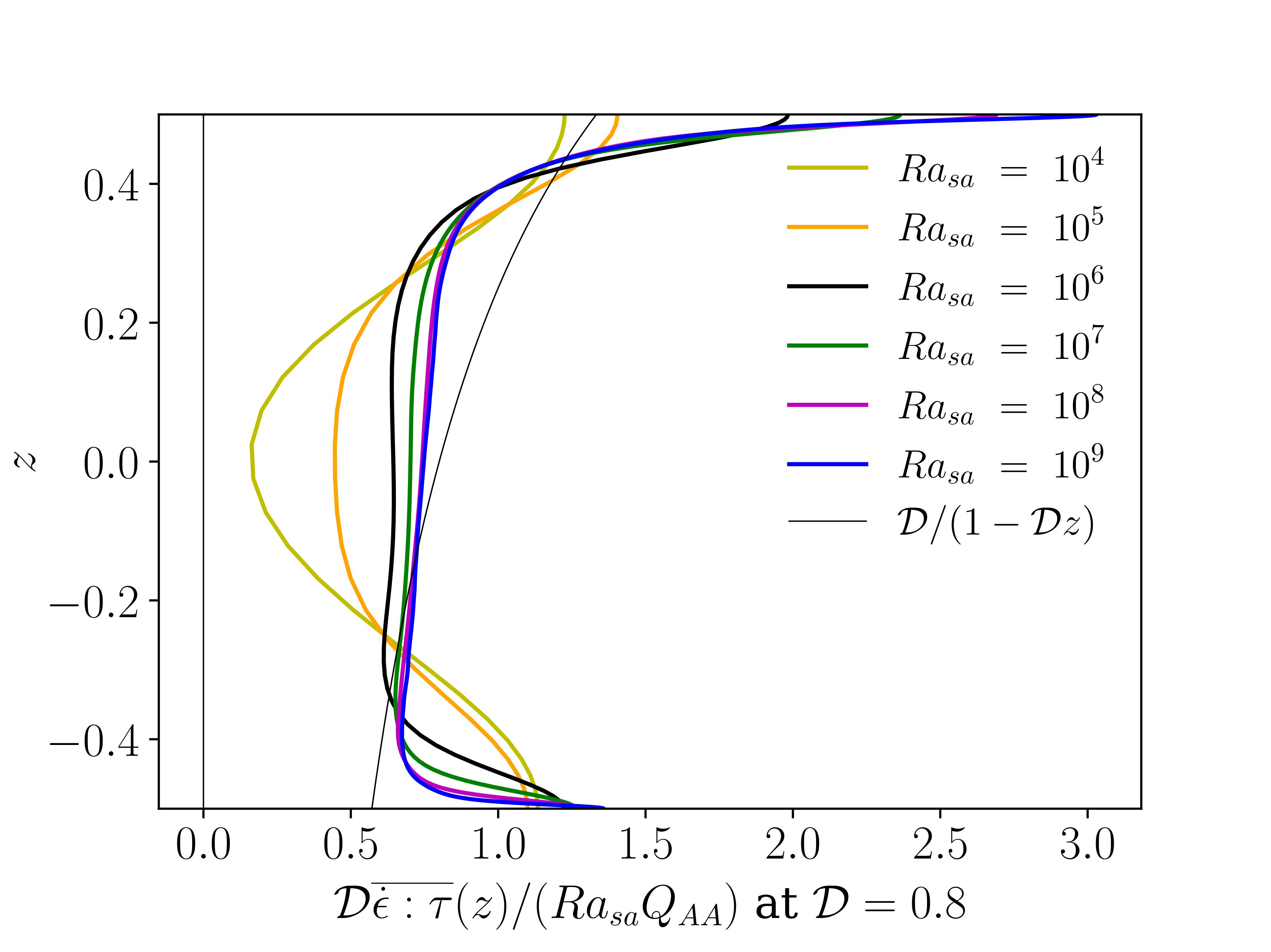} \hspace*{-0.7 cm} \includegraphics[width=7.3 cm, keepaspectratio]{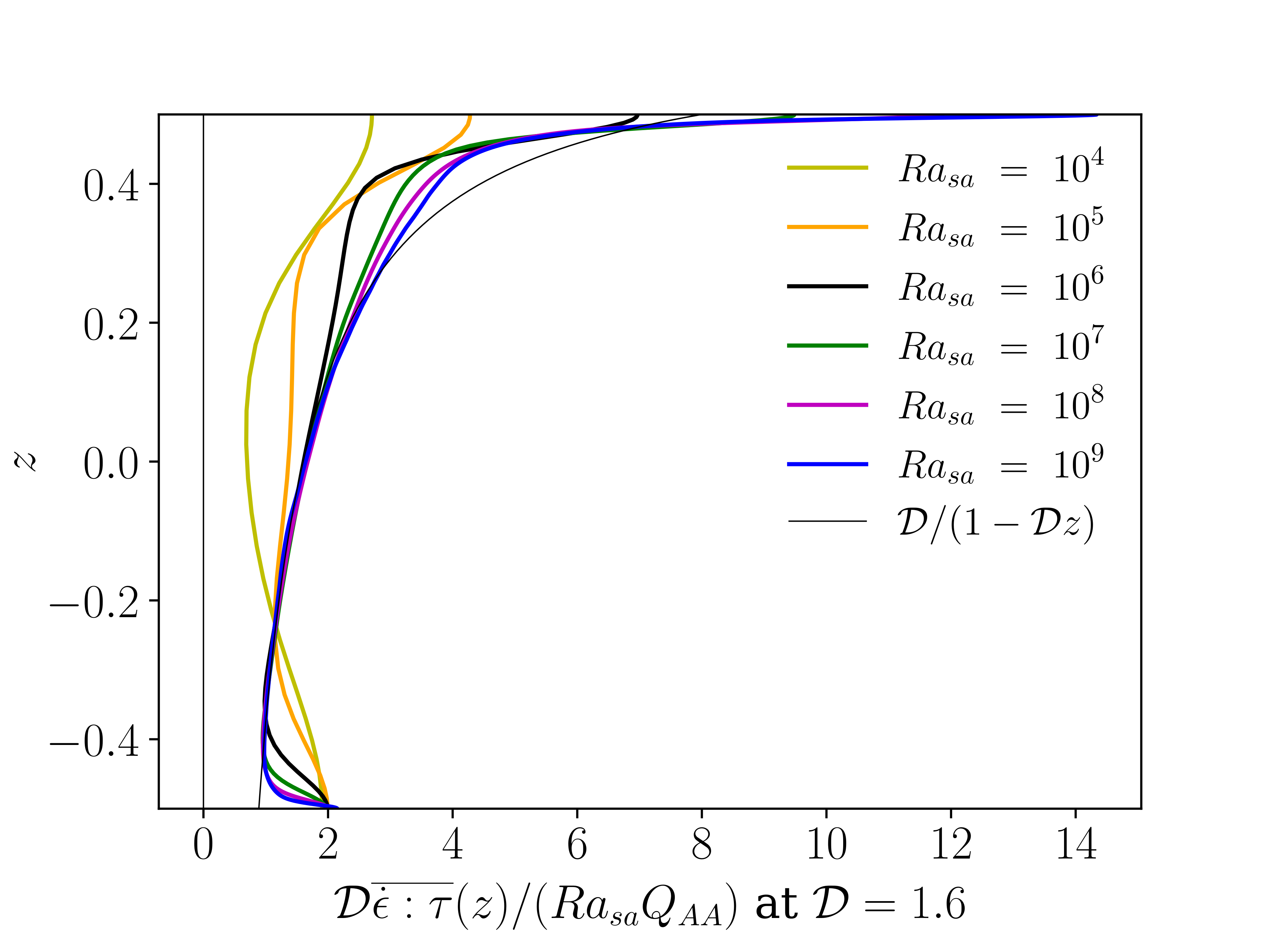}\hspace*{-0.6 cm}
	\caption{Dissipation profile for $\mathcal{D}$ equal to $0.8$ (left) and $1.6$ (right), \N{$\alpha _0 T_0 = 1$,} for a superadiabatic Rayleigh number up to $10^9$ in the anelastic approximation.  }
\label{DissProf}
\end{center}
\end{figure}

\begin{figure}
\begin{center}
	\hspace*{-1 cm} \includegraphics[width=8.1 cm, keepaspectratio]{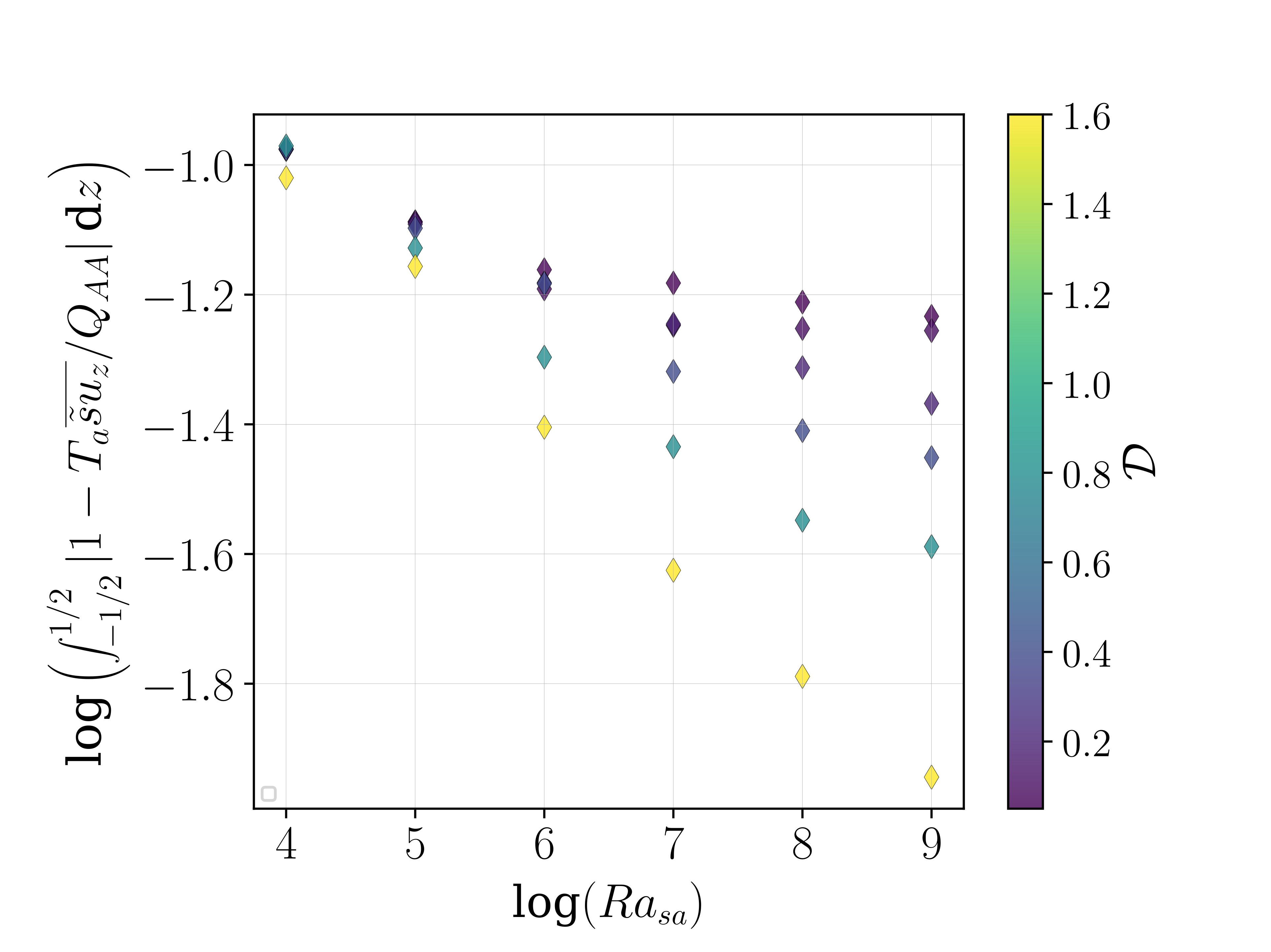} \hspace*{-2 cm} \includegraphics[width=8.1 cm, keepaspectratio]{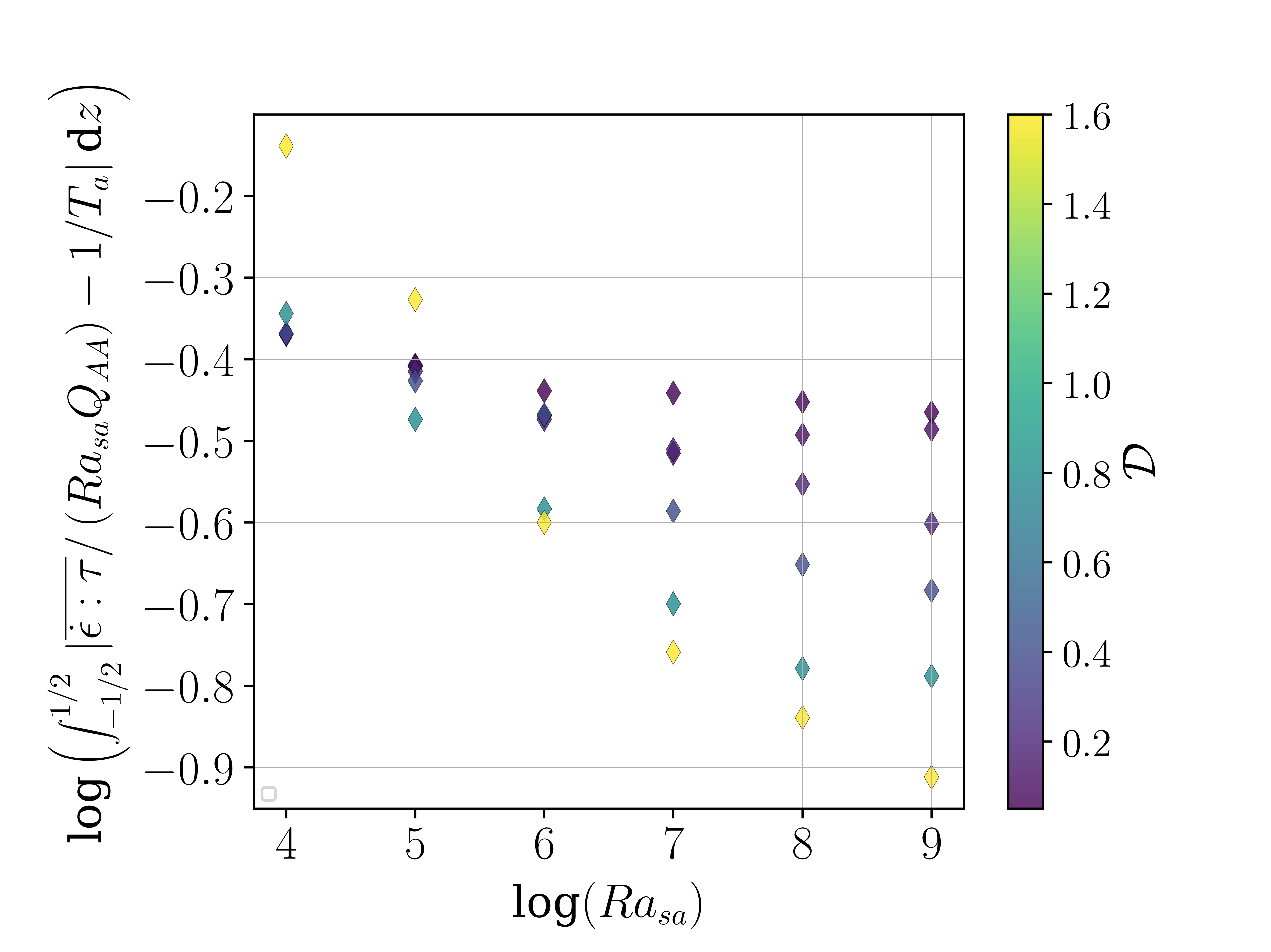} \hspace*{-1 cm}
	\caption{Relative distance ($L^1$-norm) of the entropy to total heat flux profile (left) and of the dissipation profile to the limit (\ref{dissprofiledim}). }
\label{convergence}
\end{center}
\end{figure}

In this section, we focus on the convergence of the entropy heat flux and of the dissipation profiles towards universal limit profiles, when the superadiabatic Rayleigh number becomes large. We have restricted our analysis to a maximum value of $Ra_{sa} = 10^9$, so that a spatial resolution of $512$ vertical and $2048$ horizontal modes was able to capture thin plumes and boundary layers. For simplification and ease of calculation, we consider the anelastic approximation AA only: from the previous section \ref{heatflux}, we have seen that the global amount of dissipation does not seem to depend on the model FC, AA or ALA, at large $Ra_{sa}$ numbers, so that we expect the same convergence for FC and ALA than that from AA (SCA is different, as we have seen). 

Figure \ref{FluxProf} shows the ratio of the entropy heat flux $T_a \overline{\tilde{s} u_z}$ to the uniform total heat flux $Q_{AA}$, see equation (\ref{fluxEntAA2}). For both values of the dissipation number $\mathcal{D}=0.8$ and $\mathcal{D}=1.6$, the entropy heat flux profile converges (slowly) towards the uniform value $1$, except in thin boundary layers: their thickness is of order $Ra_{sa}^{-1/3}$ and conduction is the only way to transfer heat in the vicinity of the top and bottom boundaries. This implies that the other flux contributions, called together $G(z)$, see (\ref{F}) and (\ref{fluxEntAA2}), become more and more negligible as the superadiabatic Rayleigh number is increased.

\begin{figure}
\begin{center}
        \includegraphics[width=13 cm, keepaspectratio]{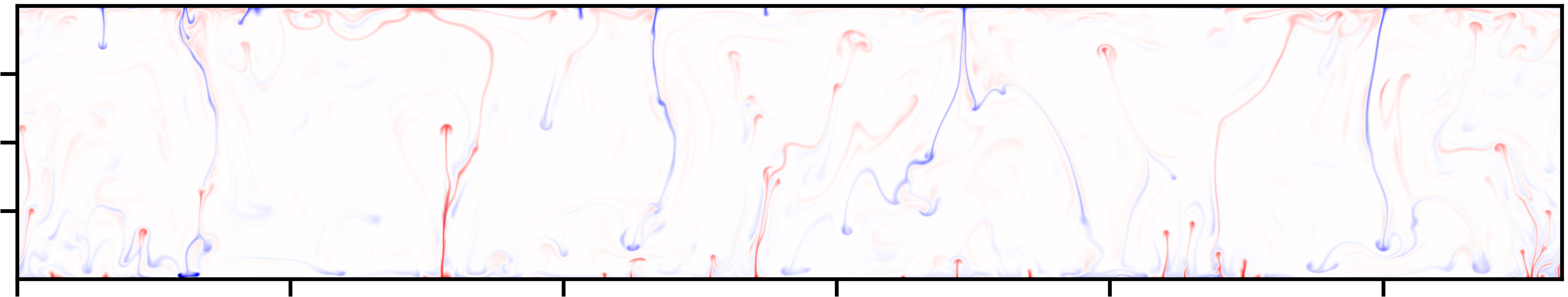} 
        \includegraphics[width=13 cm, keepaspectratio]{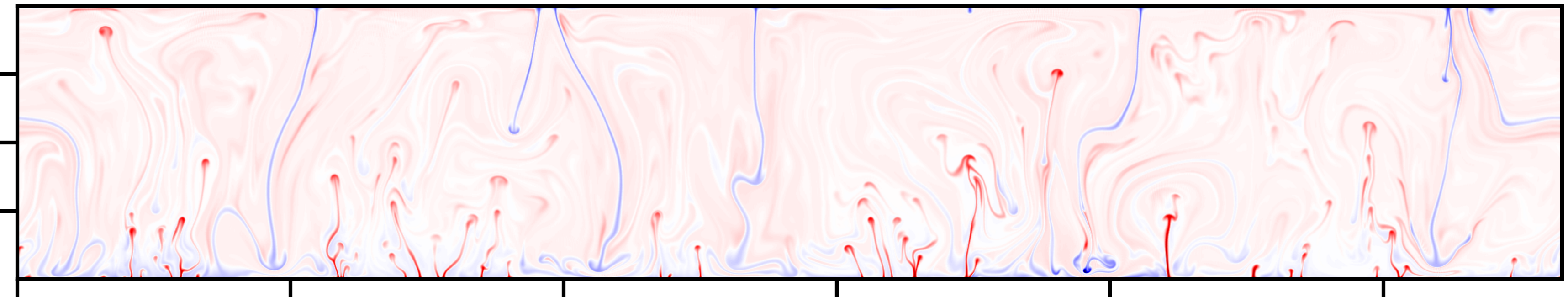} 
        \includegraphics[width=13 cm, keepaspectratio]{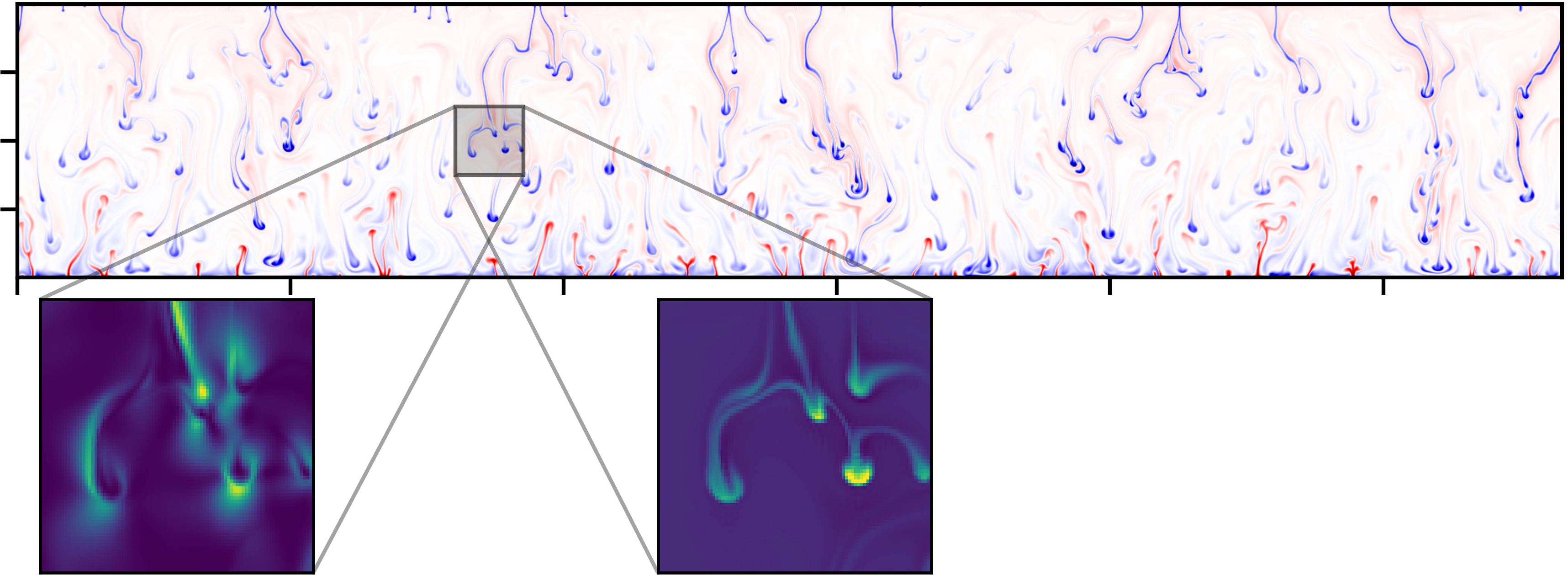} 
	\caption{Snapshot of the superadiabatic temperature field for a very small $\mathcal{D}=0.05$ (top), moderate  $\mathcal{D}=0.4$ (middle) to a large dissipation number $\mathcal{D}=1.6$ (bottom), for $Ra_{sa} = 10^9$, in the anelastic approximation \N{with $\alpha _0 T_0 = 1$}. In the close-up views, the distribution of viscous dissipation $\dot{\epsilon}:\tau$ (left) and entropy flux $\tilde{s} u_z$ (right) are shown. }
\label{example}
\end{center}
\end{figure}

As expected, as $G(z)$ becomes small (and without variations at some small scale), so does $\mathrm{d} G(z) / \mathrm{d}z$, implying that the profile of viscous dissipation becomes close to $\mathcal{D} \overline{\tilde{s} u_z}$, see (\ref{dissEntAA2}). As $T_a \overline{\tilde{s} u_z}$ is close to $1$ at large $Ra_{sa}$, we have a dissipation profile close to $1/T_a(z)$ (see appendix \ref{A1} for a general dimensional expression). Figure \ref{DissProf} shows the change in dissipation profiles as the superadiabatic Rayleigh number increases from $10^4$ to $10^9$, for two different values of the dissipation number, $\mathcal{D}=0.8$ and $\mathcal{D}=1.6$. As expected from (\ref{dissEntAA2}), the profile converges towards $1/T_a(z)$ in both cases, however the "entropy" component of dissipation being multiplied by $\mathcal{D}$, the convergence looks more obvious for the larger value of $\mathcal{D}$. 

We give a quantitative measure of the convergence of the entropy flux contribution (towards $1$) and dissipation profiles (towards $1/T_a(z)$) in Fig.~\ref{convergence}. The measure is defined in each case as the logarithm of the $\mathrm{L}^1$ distance. The larger the dissipation number, the faster the convergence is. 

In an attempt to understand how the heat flux contributions $G(z)$ become negligible as $Ra_{sa}$ is increased, we plot a snapshot of the superadiabatic temperature field for three different values of the dissipation number $\mathcal{D} = 0.05,\ 0.4\ \mathrm{and} \ 1.6$ at $Ra_{sa} = 10^9$ in Fig.~\ref{example}. At small $\mathcal{D}$ sparse plumes go from bottom to top or top to bottom and a velocity field is generated with a length-scale comparable to the height of the cavity, in the case considered here of infinite Prandtl number. At moderate $\mathcal{D}$ the top-bottom asymmetry is strong, more plumes are present and smaller scales are visible. At the largest $\mathcal{D}$, numerous plumes exist and they cannot cross the whole height of the cell (not even the descending ones) without their heads detaching from their tails and continue their course as isolated blobs. The length-scale of typical distance between adjacent plumes $l$ is reduced significantly compared to the height $H$ of the cavity. If $U$ is a typical scale for velocity, then we expect the local viscous dissipation $\overline{\dot{\epsilon} : \tau}$ to scale as $U^2/l^2$, while $u_j \tau _{zj}$ scales as $U^2/l$. Once averaged in time and horizontally, that quantity depends smoothly on $z$ on the global scale $H$, so that $\mathrm{d}/\mathrm{d} z (\overline{u_j \tau _{zj}})$ scales as $U^2/(lH)$, {\it i.e.} $l/H$ smaller than viscous dissipation. Extending this result on deviatoric stress work to pressure work, this explains that $\mathrm{d} G(z) / \mathrm{d} z$ becomes negligible compared to dissipation, see (\ref{dissEntAA2}).

\begin{figure}
\begin{center}
        \includegraphics[width=10 cm, keepaspectratio]{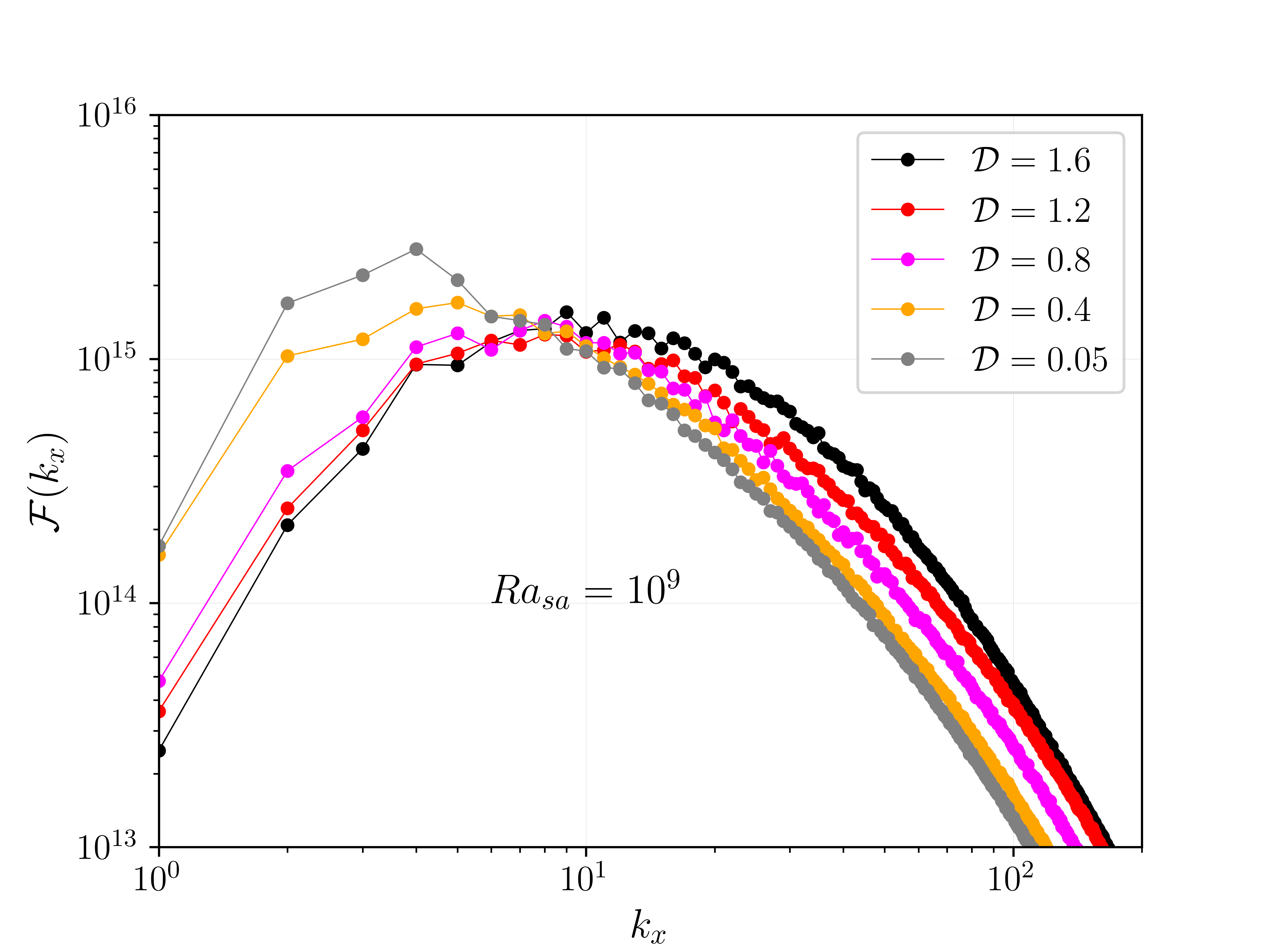} 
	\caption{Energy spectrum of $\partial u_z / \partial z$: the Fourier coefficients of this component of the velocity gradient are computed along the $x$ direction and the square of their complex magnitude $\mathcal{F} (k_x)$ is plotted as a function of the wavenumber ($k_x = 1$ corresponds to a signal of period $L$ the length of the domain along $x$, and for any value of $k_x$ the corresponding mode is $e^{i k_x 2 \pi x / L}$). Those Fourier coefficients are averaged over a central vertical extent from $z=-0.11$ to $z=0.11$. }
	\label{spectrum}
\end{center}
\end{figure}

In Fig.~\ref{spectrum}, we plot the energy spectrum of one component of the deformation rate tensor, $\partial u_z / \partial z$, for a large Rayleigh number $Ra=10^9$ and different dissipation numbers, in the anelastic approximation (AA). As the dissipation number is increased, a shift of the whole spectrum towards larger wavenumbers is observed. At small values of $\mathcal{D}$, the spectrum has a maximum around $k_x=4$, while at large $\mathcal{D}$ the maximum is close to $k_x=10$. This indicates that for a given integral of viscous dissipation, the fraction of heat flux carried by the work done by viscous stresses becomes smaller and smaller as the dissipation number $\mathcal{D}$ is increased. 

This idea of smaller scales for velocity gradients at larger superadiabatic Rayleigh numbers can also give a hint on why AA, FC and ALA results converge at large $Ra_{sa}$, as seen in Fig.~\ref{RaexcessDiss}. An estimate of pressure contributions to entropy fluctuations in \citet{ajs05} -- obtained from the analysis of the order of magnitude of the forces in the momentum equation -- is adapted to a lengthscale of convection $l$: from Stokes' equation, an order of magnitude of pressure is $\alpha \rho g (\delta T) l$ and from $\md s = c_p/T \md T - \alpha / \rho \md P$, we evaluate the ratio of pressure over temperature to be of order $\alpha T \mathcal{D} l/H$. In \citet{ajs05}, the length-scale $l$ is taken to be of order $H$ and the condition of validity for the ALA approximation is given as $\alpha T \mathcal{D} \ll 1$. If however, a smaller lengthscale $l$ prevails, that estimate must be smaller proportionally to $l/H$ making the ALA approximation more valid.  

At large dissipation number and large superadiabatic Rayleigh number -- typically our last case of Fig.~\ref{example} -- we thus propose that the time and horizontal average of viscous dissipation is linked to the time and horizontal average of the product $\tilde{s} u_z$ (see equation (\ref{dissEntAA2}) with negligible term $\md G / \md z$). However, both quantities $\dot{\epsilon} : \tau$ and 
$\tilde{s} u_z$ are not pointwise (and timewise) correlated. This is illustrated in the close-up views in Fig.~\ref{example} where $\dot{\epsilon} : \tau$ and $\tilde{s} u_z$ are plotted in a small region of the fluid domain. The quantity $\tilde{s} u_z$ (right close-up) represents well the plumes while dissipation $\dot{\epsilon} : \tau$ (left close-up) looks more like a halo around the plumes. This is also a consequence of the different symmetries concerning each quantity: for a supposedly straight descending plume, $\tilde{s} u_z$ is maximum on the centerline of the plume, however dissipation must be zero on that centerline (where vertical gradients are smaller than horizontal gradients, {\it i.e.} not ahead of the tip of the plume).  

\section{Effect of confinement and inertia}
\label{conf}

Different results were obtained in \citet{cb2017}, where a larger dissipation than our limit (\ref{entropyfluxdiss0}) is obtained. There are actually several differences in the configuration they have studied: i) they use an EoS of an ideal gas, ii) they model conduction using the gradient of entropy, iii) they use a boundary condition of a fixed flux (bottom), iv) they consider inertia ($Pr = 1\ \mathrm{or}\ 10$), v) they have a square domain, vi) they have non-penetrative conditions on lateral walls. In this section, we test changes in our configuration regarding the last three points iv), v) and vi) that can be implemented easily in our code. It turns out that we need to make all three changes to recover the results of \citet{cb2017}. 

In the results presented here, we have kept the same equation of state as in the beginning of the paper, with $s=s(\rho)$. We have kept the same top and bottom boundary conditions. We have included inertia with a Prandtl number equal to ten, $Pr = 10$. We have changed the aspect ratio from $4 \sqrt{2}$ to $1$ (square domain). Then, we consider two cases, one with periodic lateral boundary conditions as before in this paper, one with a non-penetrative boundary condition. That last case is that considered in \citet{cb2017} and corresponds to a vanishing perpendicular velocity component $u_x = 0$  and no shear-stress $\partial u_z / \partial x = 0$. This is achieved numerically in dedalus by choosing the so-called SinCos base of functions for horizontal decomposition, instead of the complete Fourier base for periodic boundary conditions.

\begin{figure}
\begin{center}
	\includegraphics[width=6.5 cm, keepaspectratio]{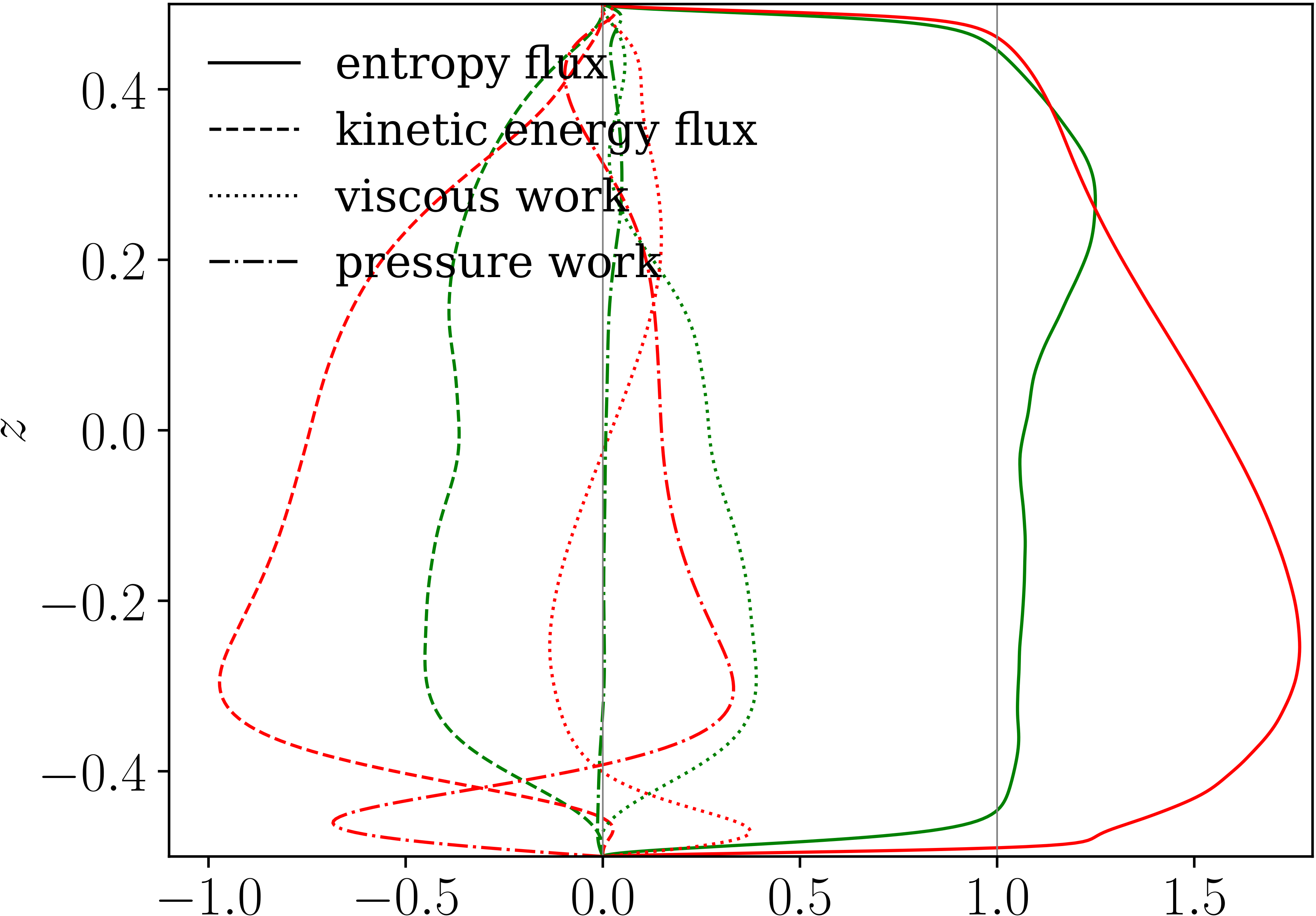} \hspace*{2 mm} \includegraphics[width=6.5 cm, keepaspectratio]{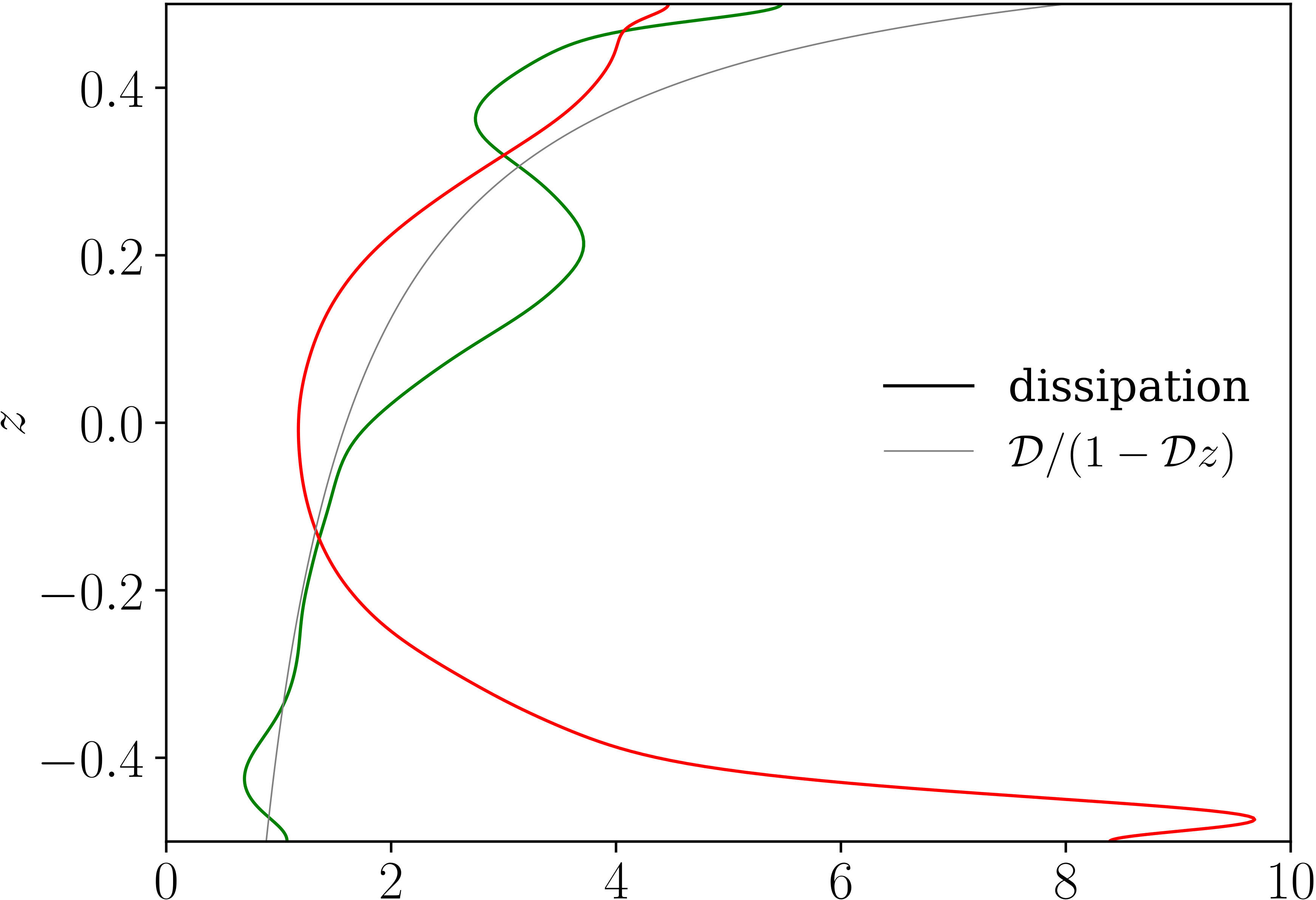}
	\caption{\N{Profiles of heat flux components from (\ref{fluxdim}) except conduction terms} (left) and dissipation profile (right), for an aspect ratio unity and a Prandtl number equal to 10, $Ra_{sa} = 10^7$, $\mathcal{D} = 1.6$. Comparison between a periodic boundary conditions (Square periodic, \N{in green}) and horizontal confinement with no shear stress (Square SinCos, \N{in red}).  }
\label{PrConfine}
\end{center}
\end{figure}

\begin{figure}
\begin{center}
        \includegraphics[width=6.7 cm, keepaspectratio]{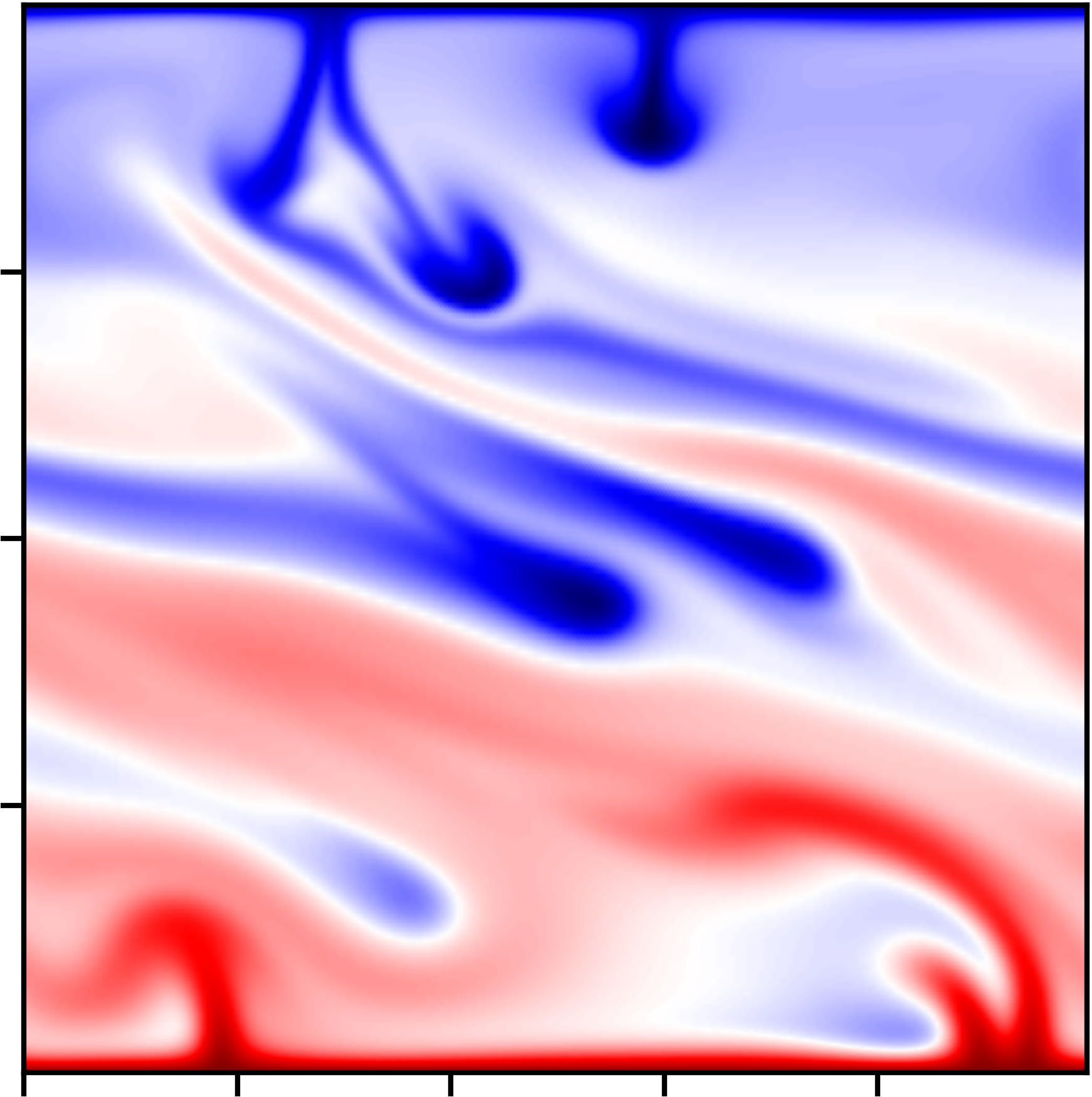} \includegraphics[width=6.7 cm, keepaspectratio]{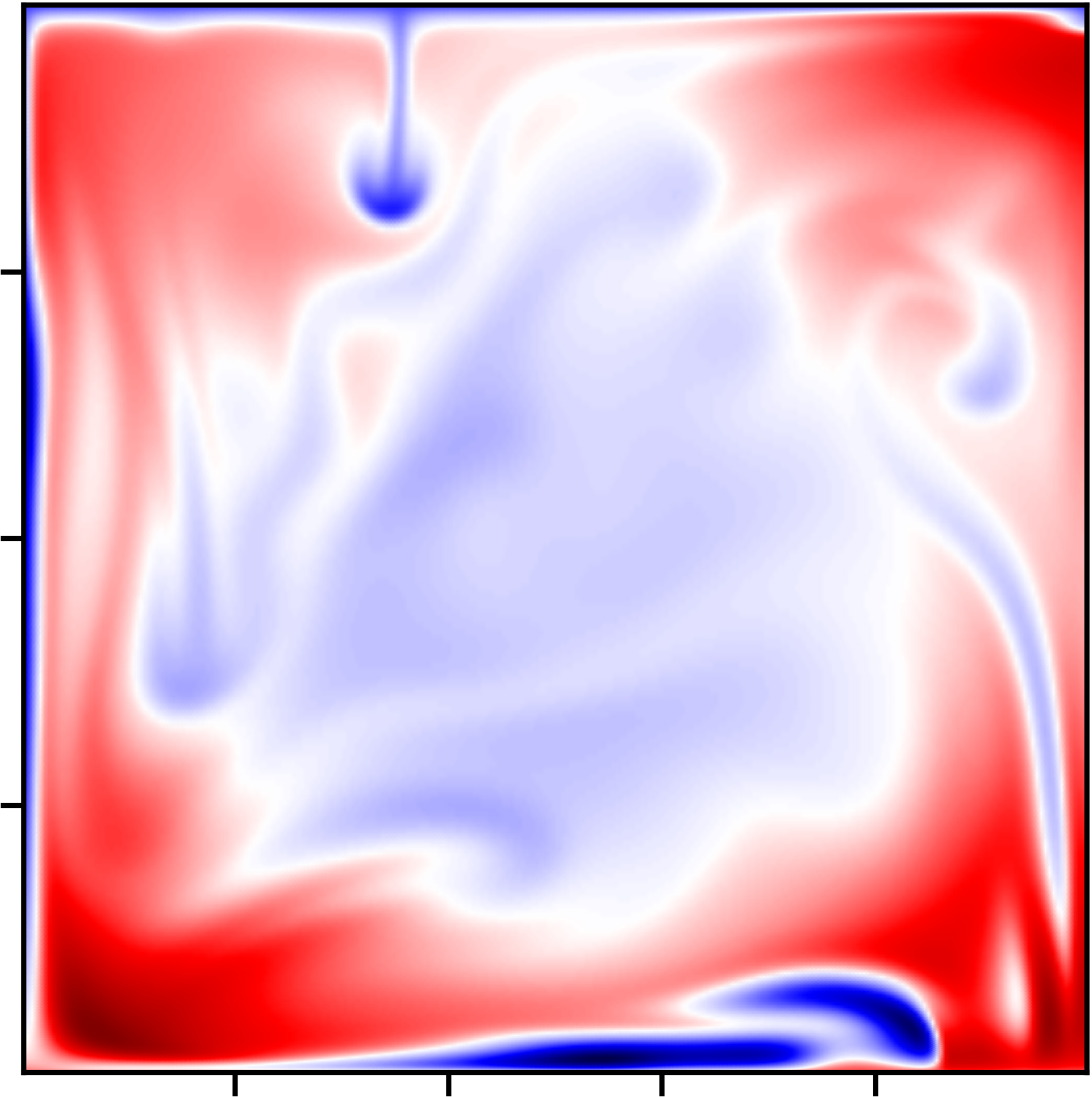}
	\caption{Snapshot of the superadiabatic temperature field for $Pr=10$, $Ra_{sa}=10^7$, $\mathcal{D}=1.6$, \N{$\alpha _0 T_0 = 1$,} in a domain of aspect ratio unity. On the left-hand side, there are periodic boundary conditions in the horizontal direction, while there are impermeable no-shear-stress walls for the calculation on the right-hand side.}
\label{PrConfineSnap}
\end{center}
\end{figure}

Figure \ref{PrConfine}, on the left-hand side, shows the averaged vertical profiles of different components of the heat flux identified in (\ref{fluxdim}): the entropy flux fraction of the heat flux $\rho _a T_a \overline{u_z \tilde{s}}$, the kinetic energy flux $1/2 \rho _a \overline{u^2 u_z} $, the viscous work $-\overline{ u_i \tau _{iz}}$ and the pressure work $\overline{\tilde{P} u_z}$. On the right-hand side of Fig.~\ref{PrConfine}, the averaged profiles of viscous dissipation are plotted for $Pr=10$, $Ra_{sa}=10^7$ and $\mathcal{D} = 1.6$. With periodic boundary conditions, we observe some departure from the results we had previously (without inertia and in a long cavity) in the top half of the cavity, but this does not change the total dissipation significantly. On the contrary, with the confined boundary condition on lateral boundaries (SinCos in dedalus), the fraction of entropy heat flux exceeds $1$ by 75\% in the bottom half, and as a consequence of (\ref{fluxEntAA2}) and (\ref{dissEntAA2}), viscous dissipation reaches much larger values there. This brings the total dissipation in the range of that obtained by \citet{cb2017}. \N{Looking into more details, the flux of kinetic energy is very significantly negative in the SinCos case, causing a significant increase of the entropy flux in the lower half of the fluid domain.} On Fig.~\ref{PrConfineSnap}, we compare a snapshot of the superadiabatic temperature field in both cases (periodic or SinCos). We can see that the vertical "walls" in the SinCos case -- and inertia -- are capable of channeling the descending plume that dissipates its kinetic energy at the bottom. A large-scale circulation is created. Note that \citet{Til11} has led numerical simulations in the fully compressible case with explicit vertical walls. In the periodic case however, descending plumes still cannot reach the bottom (even with $Pr=10$ instead of $Pr=\infty$) and finer scales develop.  Another feature of the periodic box is the large scale shear deformation, induced by Reynolds stresses (with a finite Prandtl number), which is suppressed in the wall-bounded geometry. 

So, it seems that there are two possible convection states. One that is dominated by the entropy flux, where $G(z)$ in the flux profile (\ref{fluxEntAA2}) and in the dissipation profile (\ref{dissEntAA2}) is negligible. The length scales of convection are small. In our work, this state is obtained in the limit of large superadiabatic Rayleigh numbers, although that limit might be difficult to reach for small dissipation numbers. A second state, observed by \citet{cb2017}, can be seen when inertia is introduced and when vertical walls are present. Convection is large-scale, descending plumes are stuck to a wall (the left wall in the particular snapshot in Fig.~\ref{PrConfineSnap}) and viscous dissipation is enhanced compared to the first state.  
Our investigations concerning the effect of inertia and confinement have not been systematic and complete and they only concern the anelastic approximation AA. We have tested all combinations of two aspect ratios (our usual $4 \sqrt{2}$ and $1$), two types of boundary conditions in the horizontal direction (periodic or SinCos) and two values of the Prandtl number ($Pr=\infty$ and $Pr = 10$). Yet, that second state was only seen when we had simultaneously vertical walls (SinCos boundary condition), inertia ($Pr=10$) and an aspect ratio equal to one. Incidentally, this configuration is that of most experiments where the `ultimate' regime of thermal convection is investigated, for instance in \citet{Roche_2002}.  

In their paper, \citet{cb2017} have tested the model (that we call 'entropy flux') leading to the global dissipation (\ref{entropyfluxdiss}), but they discard it on the basis of their numerical results. However, they write {\it "Often it is assumed that in the bulk of the convection zone, the
total heat flux is just equal to the convective flux..."}, where they call "convective flux" that part of flux we call here "entropy flux", $T_a \overline{\tilde{s} u_z}$. This corresponds precisely to the assumption of negligible contribution of $G(z)$ to the heat flux in (\ref{fluxEntAA2}). So the idea has been expressed already and is seemingly well accepted in astrophysics, but we could not find a precise reference for it until now. 

\N{At this point, it is important to have in mind the specific nature of the present study. The equation of state considered here is a peculiar one, a sort of limit case retaining the thermal features of compressibility (adiabatic gradient, adiabatic cooling) but minimizing the actual changes in density (nearly uniform density). 
Previous works on stellar or gas planet convection with an ideal gas equation of state have mostly shown that the flux of kinetic energy corresponds to a significant fraction of the total heat flow \citep{ChanSofia1989,Viallet2013,featherstone2016,kapyla2019}. These studies report the existence of deep convective plumes crossing the whole adiabatic layer. The impact of density stratification in an adiabatic region is highlighted by \citet{Anders2019} who point out its effect on descending plumes that can be sometimes compacted and accelerated downward. However, their authors also insist on the gap between the values of the stellar Rayleigh number and those actually accessible numerically. In their three-dimensional calculations, the Rayleigh number based on the heat flux is restricted to be less than $Ra_F=10^{7.5}$ (under a classical scaling $Nu \sim Ra^{1/3}$, this  corresponds to a Rayleigh number based on a temperature difference of $Ra=Ra_F^{3/4}=10^{5.625}$). In these papers, it is also reported that the numerical models overestimate stellar convective velocities compared to the observations: this might be related to the relative small values of the numerically accessible Rayleigh numbers. For instance, \citet{featherstone2016} report a slow tendency of the spectrum of convection to shift to larger wavenumbers as the Rayleigh number is increased. This is also something we observe (see Fig.~\ref{spectrum}) and we associate this to a slow convergence toward a regime of heat flow dominated by the entropy flux. In the context of the Boussinesq approximation, \citet{goluskin2014} show how shear flow is generated by convection in a periodic domain, leading to a reduction of the vertical heat transfer. In their two-dimensional case, they can reach Rayleigh numbers of $10^{10}$. In addition to the role of the equations of state, other features of the model can potentially affect the final structure of the flow: imposed temperatures versus imposed heat flux, sub-grid-scale models in particular under the form of a Fourier heat flux proportional to the gradient of entropy... Finally, the very important effects of rotation and magnetic field are not considered here: for instance, the dynamics of the Earth's core is dominated by the influence of the Coriolis force and one consequence is that kinetic energy is negligible despite the small value of the Prandtl number \citep{schaeffer2017}.  
}

\section{Conclusions}
\label{conclusions}

In this paper, we have taken the limit case of a class of equations of state such that entropy is a function of density. In the assumption of an infinite Prandtl number, we have written the fully compressible governing equations of convection as well as anelastic approximations of increasing simplification AA, ALA and SCA (anelastic, anelastic liquid and simple compressible). Under that choice, a nearly uniform entropy field implies that density is nearly uniform. A consequence is that, with a uniform (dynamic) viscosity, we also have a nearly uniform kinematic viscosity and a uniform thermal diffusivity. With such a class of equations of state, there is still an adiabatic temperature gradient and its effect in heat transport is still present. The idea is to keep features of compressible convection and to discard effects related to non-uniform fluid properties, sometimes called non-Oberbeck-Boussinesq (NOB) effects. The anelastic approximation AA is based on a linear expansion about a state of uniform entropy, in the anelastic liquid approximation ALA pressure variations are neglected on all thermodynamic quantities, and in our simple compressible approximation SCA even the adiabatic gradient of temperature is eliminated. In that last approximation, thermodynamics is badly treated, however adiabatic heating and cooling is retained and the mathematical structure of the equations contains the key ingredients of compressible convection. It has the advantage of simplicity with just two scalar parameters $Ra_{sa}$ and $\mathcal{D}$. There is no need to determine a profile of adiabatic temperature. Applied mathematicians might want to play with that system (see also appendix \ref{A2SCA}) and determine fundamental properties of its solutions, which might then have applications in the more physical models.  

The genuine compressible effects are governed by the dissipation parameter $\mathcal{D}$. Around $\mathcal{D} = 0.1$ (between 0.05 and 0.2), compressible effects create a top-bottom asymmetry. Ascending plumes starting from the bottom thermal boundary layer do no longer reach the top. This leads to a change in the temperature profile with the loss of the overshoot near the top of the superadiabatic temperature profile. 
At larger values of the dissipation number, the change in the global heat flux under a constant superadiabatic Rayleigh number is compatible with the model of critical boundary layer of \citet{malkus1954}. As we have shown in section \ref{HeatFlux}, increasing the dissipation number leads to an increase of heat transfer owing to the asymmetry of heat transfer resistance of the top and bottom thermal boundary layers. 

With significant compressible effect ($\mathcal{D}$ large enough) and in the limit of very large superadiabatic Rayleigh numbers, we have shown that a state of "local equilibrium" is reached, where the heat flow due to the flux of entropy fluctuations is accompanied with the corresponding viscous dissipation at the same height. This is a small-scale process which is consistent with the observation that heat flux components such as shear-stress or pressure fluxes are negligible. In that limit, we can predict the vertical profile of viscous dissipation, as soon as the profile of $\alpha$, $g$ and $c_p$ are known from equation (\ref{dissprofiledim}). A similar process takes place in the simplest SCA model: however, due to its extreme simplicity, it ignores the depth dependence of the expansion coefficient and a constant heat flux with depth leads to a constant profile of viscous dissipation (see equations (\ref{fluxEntSCA2}) and (\ref{dissEntSCA2})). Although the result is different -- and certainly less relevant to geophysics and astrophysics -- it is still mathematically interesting to investigate the consequences of the SCA model on dissipation distribution. \N{It still contains viscous heating and adiabatic cooling in the thermal equation. These terms -- particularly at high superadiabatic Rayleigh numbers -- are capable to drive the system in a state of mesoscale equilibrium between themselves and lead to a dominant mode of heat transfer due to the flux of entropy, while other modes (kinetic energy flux, pressure and stress terms) become small. If this process is understood in the simpler SCA model, then it could certainly help understand the behaviour of the other models.} 

Other works, in particular in \citet{cb2017}, find that a different type of flows can exist, with large-scale circulation and more dissipation for the same heat flux. In that case, the vertical profile of dissipation has larger values in the lower half of the domain where temperature is large (and less in the upper half): this explains how the entropy budget is balanced despite an increased global dissipation. We have observed that this type of convection can be reached only when the Prandtl number is finite, vertical walls are present and the aspect ratio of the domain is not large. In geophysical or astrophysical contexts, vertical walls certainly do not exist and that second type of flow appears unlikely to develop. 

\N{Regarding the different models, we find a good agreement between FC and AA, as expected, except at large dissipation numbers and small superadiabatic Rayleigh numbers (above threshold). We also find the ALA approximation to be good, especially for small values of $\alpha _0 T_0$ (as expected again), and at large superadiabatic Rayleigh numbers: that last feature may be due to the fact that we get solutions with small convective scales (mesoscale equilibrium) for which pressure variations are small. The differences would be larger in the case of large-scale circulations. The SCA model gives different results, but there are good reasons for that. As explained above, this model is not meant to provide realistic results. It should be seen as a simple set of equations where compressible effects can still be studied. }

In future works, it seems important to investigate precisely the conditions of existence of both types of flow. The following features should be studied: \N{realistic equations of state}, three-dimensional flows, electromagnetic forces, \N{effect of rotation}. If the flux of kinetic energy or the Poynting flux cannot compete with the entropy flux, then it is likely that the model of ``local equilibrium'' applies, embodied by the vertical distribution of dissipation (\ref{dissprofiledim}).

\appendix

\section{\N{Equations for FC, AA, ALA and SCA models}}
\label{A2}

\N{
The velocity boundary conditions are common to all models: stress-free, non-penetrative conditions, with an additional constraint on the average horizontal velocity since Galilean invariance does not constrain it. 
\begin{align}
	& \frac{\partial u_x}{\partial z} = 0, \hspace{1 cm} \mathrm{when} \hspace{1 cm} z=\pm \frac{1}{2}, \label{uxbcA2}\\
 	& u_z = 0, \hspace{1 cm} \mathrm{when} \hspace{1 cm} z=\pm \frac{1}{2}. \label{uzbcA2} \\
	& \int _{-L/(2H)}^{L/(2H)} u_x \left( x,  z=\frac{1}{2} \right) \md x =0, \label{conduxA2}
\end{align}
The non-dimensional equations of the different models studied -- fully compressible (FC), anelastic  approximation (AA), anelastic liquid approximation (ALA) and simple compressible approximation model (SCA) -- and other boundary conditions are as follows:
}

\N{\subsection{Fully compressible model FC}}
\label{A2FC}

\N{The governing equations are
\begin{align}
	\frac{\mathrm{D} \rho}{\mathrm{D} t} &= - \rho {\bnabla} \cdot {\bf u}, \label{continuityaA2}\\
	{\mathbf 0} &= - \frac{Ra_{sa} (n+1)}{\varepsilon } \left( \frac{ \rho ^n }{ \mathcal{D}}  \left[ (n+1) T {\mathbf \bnabla} \rho + \rho {\mathbf \bnabla} T  \right] + \rho {\mathbf e}_z \right) + {\mathbf \bnabla}^2 {\bf u} + \frac{1}{3} {\mathbf \bnabla} \left(\bnabla \cdot {\bf u} \right) , \label{momentumaA2} \\
	0 &= - (n+1) \rho ^{n+1} T \bnabla \cdot {\bf u} + \frac{\varepsilon \mathcal{D}}{Ra_{sa}} \dot{\epsilon} : \tau + \nabla ^2 T . \label{entropyaA2}
\end{align}
Temperatures are imposed at the top and bottom 
\begin{align}
	T &= 1 + \frac{\mathcal{D} + \varepsilon }{2} , \hspace{1 cm} \mathrm{when} \hspace{1 cm} z= - \frac{1}{2}, \label{temphA2} \\ 
	T &= 1 - \frac{\mathcal{D} + \varepsilon }{2} , \hspace{1 cm} \mathrm{when} \hspace{1 cm} z= \frac{1}{2}. \label{tempcA2}
\end{align}
}

\N{\subsection{Anelastic Approximation AA}}
\label{A2AA}

\N{The governing equations are
\begin{align}
        {\mathbf \bnabla} \cdot {\bf u} &= 0 , \label{AAcontA2} \\
        {\bf 0} &= -\frac{Ra_{sa}}{\mathcal{D}} {\mathbf \bnabla} \tilde{P} + Ra_{sa} \left( \frac{\tilde{T}}{T_a} - \frac{\tilde{P}}{(n+1) T_a}  \right) {\hat{\mathbf{e}}}_z + {\mathbf \bnabla}^2 {\bf u} , \label{AAmomentumA2} \\
        \frac{ \mathrm{D} }{\mathrm{D} t} \left( \tilde{T} - \frac{\tilde{P}}{n+1} \right) &= - \mathcal{D} u_z  \left( \frac{\tilde{T}}{T_a} - \frac{\tilde{P}}{(n+1) T_a}  \right) + \frac{\mathcal{D}}{Ra_{sa}} \dot{\epsilon} : \tau + \nabla ^2  \tilde{T}, \label{AAentropyA2}
\end{align}
and boundary conditions
\begin{align}
	&\tilde{T} \left( z = \pm \frac{1}{2} \right) = \mp \frac{1}{2} . \label{AAbcA2} \\
	&\int _0^{L/H} \tilde{P} \left( x, z=\frac{1}{2} \right) - \tilde{P} \left( x, z=- \frac{1}{2} \right) \md x = 0, \label{AApressureA2}
\end{align}
}

\N{\subsection{Anelastic Liquid Approximation ALA}}
\label{A2ALA}

\N{The governing equations are
\begin{align}
        {\mathbf \bnabla} \cdot {\bf u} &= 0 , \label{ALAcontA2} \\
        {\bf 0} &= -\frac{Ra_{sa}}{\mathcal{D}} {\mathbf \bnabla} \tilde{P} + Ra_{sa} \frac{\tilde{T}}{T_a} {\hat{\mathbf{e}}}_z + {\mathbf \bnabla}^2 {\bf u} , \label{ALAmomentumA2} \\
        \frac{ \mathrm{D} \tilde{T} }{\mathrm{D} t} &= - \mathcal{D} u_z  \frac{\tilde{T}}{T_a} + \frac{\mathcal{D}}{Ra_{sa}} \dot{\epsilon} : \tau + \nabla ^2  \tilde{T}, \label{ALAentropyA2}
\end{align}
and boundary conditions
\begin{align}
        &\frac{\partial  \tilde{T}}{\partial x} = 0 \hspace*{1 cm} z = \pm \frac{1}{2}, \label{ALAtempA2} \\
        &\tilde{T} \left( x,z= - \frac{1}{2} \right) - \tilde{T} \left( x,z= \frac{1}{2} \right) = 1 , \label{ALAtempdeltaA2} \\
	&\int _0^{L/H} \tilde{P} \left( x, z= \pm \frac{1}{2} \right)    \md x = 0 . \label{ALAPA2}
\end{align}
}

\N{\subsection{Simple Compressible Approximation SCA}
\label{A2SCA}}

\N{The governing equations are
\begin{align}
        {\mathbf \bnabla} \cdot {\bf u} &= 0 , \label{SCAcont} \\
        {\bf 0} &= -\frac{Ra_{sa}}{\mathcal{D}} {\mathbf \bnabla} \tilde{P} + Ra_{sa} \tilde{T} {\hat{\mathbf{e}}}_z + {\mathbf \bnabla}^2 {\bf u} , \label{SCAmomentum} \\
        \frac{ \mathrm{D} \tilde{T} }{\mathrm{D} t} &= - \mathcal{D} u_z  \tilde{T} + \frac{\mathcal{D}}{Ra_{sa}} \dot{\epsilon} : \tau + \nabla ^2  \tilde{T}, \label{SCAentropy}
\end{align}
and boundary conditions
\begin{align}
        &\frac{\partial  \tilde{T}}{\partial x} = 0 \hspace*{1 cm} z = \pm \frac{1}{2}, \label{SCAtempA2} \\
        &\tilde{T} \left( x,z= - \frac{1}{2} \right) - \tilde{T} \left( x,z= \frac{1}{2} \right) = 1 , \label{SCAtempdeltaA2} \\
        &\int _0^{L/H} \tilde{P} \left( x, z= \pm \frac{1}{2} \right)    \md x = 0 . \label{SCAPA2}
\end{align}
}

\section{\N{Tables of simulation parameters}}
\label{A3}

\N{
The parameters of infinite Prandtl number, $4 \sqrt{2}$ aspect ratio, simulations are the following }

{\color{black}\begin{tabular}{lllllllll}
	& $Ra_{sa}$ & $\mathcal{D}$ & $\alpha _0 T_0 $ & $\epsilon$ &  $n_x$ & $n_z$ & model & duration \\ \hline
	12 runs & $10^4$    &  $1.5$          & $1.0$/$0.5$ & 0.1 & 128 &32 & FC/AA & 0.3 \\ 
	& & & 0.1 & & & & ALA/SCA & \\ \hline
	5 runs & $10^6$ & 0/0.05/0.1 & 1.0 & & 256 & 64 & AA & 0.5 \\
	& & 0.2/0.4 & & & & & & \\ \hline
	5 runs & $10^8$ & 0/0.05/0.1 & 1.0 & & 512 & 128 & AA & $10^{-2}$ \\ 
	& & 0.2/0.4 & & & & & &  \\ \hline
	2 runs & $10^7$ & 0.05/0.2 & 1.0 & & 256 & 64 & AA & $0.1$ \\ \hline
	4 runs & $3.0 \, \times \, 10^5$ & 1.2 & 1.0 & 0.1 & 256 & 64 & FC/AA & $0.5$ \\ 
	& & &  & & & & ALA/SCA & \\ \hline
	192 runs & $10^{3/3.5/4/4.5}$ & 0.25/0.5/0.75 & 1.0 & 0.1 & 256 & 64 & FC/AA & $1.0$ \\ 
	& $10^{5/5.5/6/6.5}$ & 1.0/1.25/1.5 & & & & & ALA/SCA & \\ \hline
	17 runs & $10^7$ & 0.001/0.01/0.02 & 1.0 & & 512 & 128 & AA & 0.1 \\
	& & 0.03/0.04/0.05 & & & & & & \\
	& & 0.07/0.1/0.2/0.3 & & & & & & \\
	& & 0.5/0.7/1.0/1.2 & & & & & & \\
	& & 1.4/1.6/1.8 & & & & & & \\ \hline
	8 runs  & $10^8$ & 0.05/0.1/0.2/0.4 & 1.0 & &  1024 & 256 & AA & 0.03 \\
	& & 0.8/1.2/1.6/1.8 & & & & & & \\ \hline
	8 runs  & $10^9$ & 0.05/0.1/0.2/0.4 & 1.0 & &  2048 & 512 & AA & 0.01 \\
	& & 0.8/1.2/1.6/1.8 & & & & & & 
\end{tabular}}

\N{Two additional runs have been performed with a Prandlt number equal to $10$ in a cavity of aspect ratio $1$, with the following parameters}

{\color{black}\begin{tabular}{llllllllll}
		& $Ra_{sa}$ & $\mathcal{D}$ & $\alpha _0 T_0 $ &  $n_x$ & $n_z$ & model & duration & lateral \\[-2mm] 
	& & & & & &  & & condition \\ \hline
	1 run & $10^7$ & 1.6 & 1.0  & 512 & 128 & AA & 0.1 & x-periodic \\[-2mm] 
	 & & & & & & & & (Fourier) \\ \hline
	1 run & $10^7$ & 1.6 & 1.0  & 512 & 128 & AA & 0.1 & wall-bounded \\[-2mm]
	 & & & & & & & & (SinCos) 
\end{tabular}}

\section{Dimensional anelastic heat flux and dissipation profiles}
\label{A1}

From the dimensional anelastic equations for a general equation of state, we derive expressions for the horizontal and time average of the vertical heat flux and dissipation profile. The anelastic equations
are
\begin{align}
	{\bf \bnabla } \cdot \left( \rho _a {\bf u} \right) &= 0 , \label{dimcont} \\
	\rho _a \frac{\mathrm{D} {\bf u} }{\mathrm{D} t} &= - \rho _a {\bf \bnabla } \left( \frac{\tilde{P}}{\rho _a } \right) + \frac{\rho _a \alpha _a T_a g}{c_{pa}} \tilde{s} {\bf \hat{e}}_z  + {\bf \bnabla} \cdot \tau . \label{dimNS} \\
	\rho _a \frac{\mathrm{D} \left( T_a \tilde{s} \right) }{\mathrm{D} t}  &= - \frac{\rho _a \alpha _a T_a g}{c_{pa}} u_z \tilde{s} + \dot{\epsilon}:\tau - {\bf \bnabla} \cdot \left( \phi _a + \tilde{\phi} \right), \label{dimS}
\end{align}
where $\phi _a$ and $\tilde{\phi}$ are the conduction heat flux along the adiabat and the superadiabatic temperature, respectively. 
The scalar product of the Navier-Stokes equation is averaged horizontally and temporally (denoted by an overline) in the assumption of a statistically stationary flow, to obtain the dissipation profile (after integrating by parts the last term)
\begin{equation}
	\overline{ \dot{\epsilon}:\tau } (z) = \frac{\rho _a \alpha _a T_a g}{c_{pa}} \overline{ u_z \tilde{s} } - \frac{\mathrm{d}}{\mathrm{d} z} \left[ \rho _a \overline{\frac{u^2}{2} u_z} + \overline{ \tilde{P} u_z } - \overline{u_i \tau _{iz}} \right]. \label{dissdim}
\end{equation}
Taking the horizontal and temporal average of the energy equation (\ref{dimS}), eliminating $\overline{\dot{\epsilon}:\tau } $ using (\ref{dissdim}), shows that the following function is independent of $z$ while it is obviously equal to the heat flux at the top and bottom 
\begin{equation}
	Q_{AA} (z) = \rho _a T_a \overline{ u_z \tilde{s}} + \left[ \rho _a \overline{\frac{u^2}{2} u_z} + \overline{ \tilde{P} u_z } - \overline{u_i \tau _{iz}} \right] + \phi _a + \overline{\phi } . \label{fluxdim}
\end{equation}
If the heat flux components in brackets converge toward zero, for instance when the Rayleigh number increases to large values, then the main part of the flux is carried by the entropy flux $\rho _a T_a \overline{ u_z \tilde{s}}$, except in small layers at the top and bottom where thermal conduction can compete. In the statically stationary case, the heat flux is uniform along $z$. From (\ref{dissdim}), it can be seen that the dissipation profile converges toward ${\rho _a \alpha _a T_a g}/{c_{pa}} \overline{ u_z \tilde{s} }$, so that the dissipation profile becomes a well-defined function of height
\begin{equation}
	\overline{ \dot{\epsilon}:\tau } (z) \simeq  \frac{\alpha _a g }{c_{pa}} Q_{AA} , \label{dissprofiledim}
\end{equation}
depending on the vertical profiles of $\alpha _a$, $c_{pa}$ and $g$. \\

\begin{acknowledgements}
{\bf Acknowledgements}: thanks are due to Daniel Lecoanet who provided help on how to use the Dedalus software. The authors are grateful to the LABEX Lyon Institute of Origins (ANR-10-LABX-0066) of the Universit\'e de Lyon for its financial support within the program ``Investissements d'Avenir'' (ANR-11-IDEX-0007) of the French government operated by the National Research Agency (ANR). Support was provided by the ICMAT Severo Ochoa Project No. SEV-2015‐0554. J. Curbelo also acknowledges the support of the RyC project RYC2018-025169, the Spanish grant [PID2020-114043GB-I00] and the Catalan Grant [No. 2017SGR1049].
\end{acknowledgements}

{\bf Declaration of Interests}. The authors report no conflict of interest.

\bibliography{../local}

\begin{thebibliography}{56}
\providecommand{\natexlab}[1]{#1}
\providecommand{\url}[1]{\texttt{#1}}
\expandafter\ifx\csname urlstyle\endcsname\relax
  \providecommand{\doi}[1]{doi: #1}\else
  \providecommand{\doi}{doi: \begingroup \urlstyle{rm}\Url}\fi

\bibitem[Alboussi\`ere and Ricard(2013)]{AlRi13}
T.~Alboussi\`ere and Y.~Ricard.
\newblock Reflections on dissipation associated with thermal convection.
\newblock \emph{Journal of Fluid Mechanics}, 725:\penalty0 1469--7645, 2013.

\bibitem[Alboussi\`ere and Ricard(2017)]{AR2017}
T.~Alboussi\`ere and Y.~Ricard.
\newblock Rayleigh-{B}\'enard stability and the validity of quasi-{B}oussinesq
  or quasi-anelastic liquid approximations.
\newblock \emph{Journal of Fluid Mechanics}, 817:\penalty0 264--305, 2017.

\bibitem[Anders et~al.(2019)Anders, Lecoanet, and Brown]{Anders2019}
E.H. Anders, D.~Lecoanet, and B.~Brown.
\newblock Entropy rain: dilution and compression of thermals in stratified
  domains.
\newblock \emph{The Astrophysical Journal}, 884\penalty0 (65), 2019.

\bibitem[Anufriev et~al.(2005)Anufriev, Jones, and Soward]{ajs05}
A.P. Anufriev, C.A. Jones, and A.M. Soward.
\newblock {The {B}oussinesq and anelastic liquid approximations for convection
  in the {E}arth’s core}.
\newblock \emph{Phys. {E}arth and {P}lanet. {I}nt.}, 152:\penalty0 163--190,
  2005.

\bibitem[Backus(1975)]{ba1975}
G.E. Backus.
\newblock {Gross thermodynamics of heat engines in deep interior of Earth}.
\newblock \emph{Proceedings of the National Academy of Sciences}, 72\penalty0
  (4):\penalty0 1555--1558, 1975.

\bibitem[Bazarov(1989)]{bazarov}
I.~P. Bazarov.
\newblock \emph{Thermodynamique}.
\newblock Ed. MIR, 1989.

\bibitem[B\'enard(1901)]{benard}
H.~B\'enard.
\newblock \emph{Les Tourbillons cellulaires dans une nappe liquide propageant
  de la chaleur par convection, en r\'egime permanent}.
\newblock Th\`ese de la Facult\'e des Sciences de Paris, 1901.

\bibitem[Bormann(2001)]{bormann}
A.S. Bormann.
\newblock The onset of convection in the {R}ayleigh-{B}\'enard problem for
  compressible fluids.
\newblock \emph{Continuum Mech. Thermodyn.}, 13:\penalty0 9--23, 2001.

\bibitem[Boussinesq(1903)]{boussinesq}
J.~Boussinesq.
\newblock \emph{Th\'eorie analytique de la chaleur, tome 2}, pages 157--161.
\newblock Gauthier-Villars, Paris, 1903.

\bibitem[Braginsky and Roberts(1995)]{br1995}
S.I. Braginsky and P.H. Roberts.
\newblock Equations governing convection in earth's core and the geodynamo.
\newblock \emph{Geophys. Astrophys. Fluid Dynam.}, 79:\penalty0 1--97, 1995.

\bibitem[Burns et~al.(2020)Burns, Vasil, Oishi, Lecoanet, and Brown]{dedalus}
K.J. Burns, G.M. Vasil, J.S. Oishi, D.~Lecoanet, and B.P. Brown.
\newblock {D}edalus: A flexible framework for numerical simulations with
  spectral methods.
\newblock \emph{{P}hys. {R}ev. {R}esearch}, 2:\penalty0 023068, 2020.

\bibitem[Busse(1967)]{busse1967}
F.~H. Busse.
\newblock The stability of finite amplitude cellular convection and its
  relation to an extremum principle.
\newblock \emph{Journal of Fluid Mechanics}, 30\penalty0 (4):\penalty0
  625--649, 1967.

\bibitem[Carnot(1824)]{carnot1824}
S.~Carnot.
\newblock \emph{R\'eflexions sur la puissance motrice du feu}.
\newblock Bachelier Libraire, 1824.

\bibitem[Chan and Sofia(1989)]{ChanSofia1989}
K.L. Chan and S.~Sofia.
\newblock Turbulent compressible convection in a deep atmosphere. {IV.}
  {R}esults of three-dimensional computations.
\newblock \emph{The Astrophysical Journal}, 336:\penalty0 1022--1040, 1989.

\bibitem[Curbelo et~al.(2019)Curbelo, Duarte, Alboussi\`ere, Dubuffet,
  Labrosse, and Ricard]{cdadlr2019}
J.~Curbelo, L.~Duarte, T.~Alboussi\`ere, F.~Dubuffet, S.~Labrosse, and
  Y.~Ricard.
\newblock {N}umerical solutions of compressible convection with an infinite
  {P}randtl number: comparison of the anelastic and anelastic liquid models
  with the exact equations.
\newblock \emph{Journal of Fluid Mechanics}, 873:\penalty0 646--687, 2019.

\bibitem[Currie and Browning(2017)]{cb2017}
L.K. Currie and M.K. Browning.
\newblock The magnitude of viscous dissipation in strongly stratified
  two-dimensional convection.
\newblock \emph{{T}he {A}strophysical {J}ournal {L}etters}, 845\penalty0 (L17),
  2017.

\bibitem[Durran(1989)]{durran1989}
D.R. Durran.
\newblock Improving the anelastic approximation.
\newblock \emph{J. of the Atmosph. Sci.}, 46\penalty0 (11):\penalty0
  1453--1461, 1989.

\bibitem[Featherstone and Hindman(2016)]{featherstone2016}
N.A. Featherstone and B.W. Hindman.
\newblock The spectral amplitude of stellar convection and its scaling in the
  high-{R}ayleigh-number regime.
\newblock \emph{The Astrophysical Journal}, 818\penalty0 (1), 2016.

\bibitem[Fr\"ohlich et~al.(1992)Fr\"ohlich, Laure, and Peyret]{flp1992}
J.~Fr\"ohlich, P.~Laure, and R.~Peyret.
\newblock Large departure from {B}oussinesq approximation in the
  {R}ayleigh-{B}\'enard problem.
\newblock \emph{Phys. Fluids}, 4\penalty0 (7):\penalty0 1355--1372, 1992.

\bibitem[Fuentes and Cumming(2020)]{fuentes2020}
J.R. Fuentes and A.~Cumming.
\newblock Penetration of a cooling convective layer into a stably-stratified
  composition gradient: Entrainment at low prandtl number.
\newblock \emph{Physical Review Fluids}, 5\penalty0 (124501), 2020.

\bibitem[Garaud(2020)]{garaud2020}
P.~Garaud.
\newblock Horizontal shear instabilities at low prandtl number.
\newblock \emph{The Astrophysical Journal}, 901\penalty0 (2), 2020.

\bibitem[Gebhart(1962)]{gebhart1962}
B.~Gebhart.
\newblock Effects of viscous dissipation in natural convection.
\newblock \emph{Journal of Fluid Mechanics}, 14\penalty0 (2):\penalty0
  225--232, 1962.

\bibitem[Gilbert(1991)]{sylvester}
G.~T. Gilbert.
\newblock Positive definite matrices and sylvester's criterion.
\newblock \emph{{T}he {A}merican {M}athematical {M}onthly}, 98\penalty0 (1),
  1991.

\bibitem[Giterman and Shteinberg(1970)]{gs1970}
M.Sh. Giterman and V.A. Shteinberg.
\newblock Criteria of occurrence of free convection in a compressible viscous
  heat-conducting fluid.
\newblock \emph{J. Appl. Math. Mech.}, 34\penalty0 (2):\penalty0 305--311,
  1970.

\bibitem[Goluskin et~al.(2014)Goluskin, Johnston, Flierl, and
  Spiegel]{goluskin2014}
D.~Goluskin, H.~Johnston, G.R. Flierl, and E.A. Spiegel.
\newblock Convectively driven shear and decreased heat flux.
\newblock \emph{Journal of Fluid Mechanics}, 759:\penalty0 360--385, 2014.

\bibitem[Grandi and Passerini(2020)]{gp2020}
D.~Grandi and A.~Passerini.
\newblock Approximation \`a la {O}berbeck-{B}oussinesq for fluids with
  pressure-induced stratified density.
\newblock \emph{{G}eophysical \& {A}strophysical {F}luid {D}ynamics}, pages
  1--24, 2020.

\bibitem[Grossmann and Lohse(2000)]{GrLo00}
S.~Grossmann and D.~Lohse.
\newblock Scaling in thermal convection: a unifying theory.
\newblock \emph{J.\,Fluid Mech.}, 407:\penalty0 27--56, 2000.

\bibitem[Hewitt et~al.(1975)Hewitt, McKenzie, and Weiss]{HMKW75}
J.~M. Hewitt, D.~P. McKenzie, and N.~O. Weiss.
\newblock Dissipative heating in convective flows.
\newblock \emph{J.\,Fluid Mech.}, 68\penalty0 (4):\penalty0 721--738, 1975.

\bibitem[Horn et~al.(2013)Horn, Shishkina, and Wagner]{HSW13}
S.~Horn, O.~Shishkina, and C.~Wagner.
\newblock On non-{O}berbeck-{B}oussinesq effects in three-dimensional
  {R}ayleigh-{B}\'enard convection in glycerol.
\newblock \emph{J.\,Fluid Mech.}, 724:\penalty0 175--202, 2013.

\bibitem[Howard(1963)]{howard63}
L.N. Howard.
\newblock Heat transport by turbulent convection.
\newblock \emph{Journal of Fluid Mechanics}, 17\penalty0 (3):\penalty0
  405--432, 1963.

\bibitem[Howard(1964)]{howard64}
L.N. Howard.
\newblock {C}onvection at high {R}ayleigh number.
\newblock \emph{G\"ortler H., {A}pplied {M}echanics}, 1964.

\bibitem[Jeffreys(1930)]{jeffreys}
H.~Jeffreys.
\newblock The instability of a compressible fluid heated below.
\newblock \emph{Proc. of the Cambridge Phil. Soc.}, 26\penalty0 (2):\penalty0
  170--172, 1930.

\bibitem[Jones et~al.(2022)Jones, Mizerski, and Kessar]{jmk2022}
C.A. Jones, K.~Mizerski, and M.~Kessar.
\newblock Fully developed anelastic convection with no-slip boundaries.
\newblock \emph{Journal of Fluid Mechanics}, 930\penalty0 (A13), 2022.

\bibitem[K\"apyl\"a et~al.(2019)K\"apyl\"a, Viviani, K\"apyl\"a, Brandenburg,
  and Spada]{kapyla2019}
P.J. K\"apyl\"a, M.~Viviani, M.J. K\"apyl\"a, A.~Brandenburg, and F.~Spada.
\newblock Effects of a subadiabatic layer on convection and dynamos in
  spherical wedge simulations.
\newblock \emph{Geophysical and Astrophysical Fluid Dynamics}, 113\penalty0
  (1--2), 2019.

\bibitem[King et~al.(2010)King, Lee, van Keken, Leng, Zhong, Tan, Tosi, and
  Kameyama]{king}
S.~D. King, C.~Lee, P.~E. van Keken, W.~Leng, S.~Zhong, E.~Tan, N.~Tosi, and
  M.~C. Kameyama.
\newblock {A} community benchmark for 2-{D} {C}artesian compressible convection
  in the {E}arth's mantle.
\newblock \emph{Geophysical Journal International}, 180:\penalty0 73--87, 2010.

\bibitem[Lantz and Fan(1999)]{lf1999}
S.R. Lantz and Y.~Fan.
\newblock Anelastic magnetohydrodynamic equations for modeling solar and
  stellar convection zones.
\newblock \emph{Astrophys. Journal}, 121:\penalty0 247--264, 1999.

\bibitem[Lecoanet et~al.(2014)Lecoanet, Brown, Zweibel, Burns, Oishi, and
  Vasil]{lecoanet2014}
D.~Lecoanet, B.P. Brown, E.G. Zweibel, K.J. Burns, J.S. Oishi, and G.M. Vasil.
\newblock Conduction in low {M}ach number flows. {I}. {L}inear and weakly
  nonlinear regimes.
\newblock \emph{The Astrophysical Journal}, 797\penalty0 (2), 2014.

\bibitem[Lipps(1990)]{lipps1990}
F.B. Lipps.
\newblock On the anelastic approximation for deep convection.
\newblock \emph{J. of the Atmosph. Sci.}, 47\penalty0 (14):\penalty0
  1794--1798, 1990.

\bibitem[Malkus(1954)]{malkus1954}
W.~V.~R. Malkus.
\newblock The heat transport and spectrum of thermal turbulence.
\newblock \emph{Proceedings of the Royal Society A}, 225:\penalty0 196--212,
  1954.

\bibitem[Oberbeck(1879)]{oberbeck}
A.~Oberbeck.
\newblock {\"U}ber die {W}\"armeleitung des {F}l\"ussigkeiten bei
  {B}er\"ucksichtigung des {S}tr\"omungen infolge von {T}emperaturdifferenzen.
\newblock \emph{Ann. Phys. Chem.}, 7:\penalty0 271--292, 1879.

\bibitem[Ogura and Phillips(1961)]{op1961}
Y.~Ogura and N.A. Phillips.
\newblock Scale analysis of deep and shallow convection in the atmosphere.
\newblock \emph{J. Atm. Sci.}, 19:\penalty0 173--179, 1961.

\bibitem[Paolucci and Chenoweth(1987)]{pc1987}
S.~Paolucci and D.~R. Chenoweth.
\newblock Departures from the {B}oussinesq approximation in laminar {B}\'enard
  convection.
\newblock \emph{Physics of Fluids}, 30\penalty0 (5):\penalty0 1561--1564, 1987.

\bibitem[Rayleigh(1916)]{rayleigh}
J.W.S. Rayleigh.
\newblock On convection currents in a horizontal layer of fluid, when the
  higher temperature is on the under side.
\newblock \emph{Phil. Mag. S.}, \begin{bf} 32\end{bf}\penalty0 (192):\penalty0
  529--546, 1916.

\bibitem[Ricard(2015)]{RicardTOG}
Y.~Ricard.
\newblock Vol. 7, physics of mantle convection.
\newblock In Bercovici D. and Schubert G., editors, \emph{Treatise on
  Geophysics}. Cambridge University Press, 2015.

\bibitem[Ricard et~al.(2022)Ricard, Alboussi\`ere, Labrosse, Curbelo, and
  Dubuffet]{ralcd22}
Y.~Ricard, T.~Alboussi\`ere, S.~Labrosse, J.~Curbelo, and F.~Dubuffet.
\newblock Fully compressible convection for planetary mantles.
\newblock \emph{GJI}, 2022.

\bibitem[Roche et~al.(2002)Roche, Castaing, Chabaud, and
  H{\'{e}}bral]{Roche_2002}
P.-E Roche, B~Castaing, B~Chabaud, and B~H{\'{e}}bral.
\newblock Prandtl and rayleigh numbers dependences in rayleigh-b{\'{e}}nard
  convection.
\newblock \emph{Europhysics Letters ({EPL})}, 58\penalty0 (5):\penalty0
  693--698, jun 2002.

\bibitem[Schaeffer et~al.(2017)Schaeffer, Jault, Nataf, and
  Fournier]{schaeffer2017}
N.~Schaeffer, D.~Jault, H.-C. Nataf, and A.~Fournier.
\newblock Turbulent geodynamo simulations: a leap towards earth's core.
\newblock \emph{Geophysical Journal International}, 211:\penalty0 1--29, 2017.

\bibitem[Schwarzschild(1906)]{schwarzschild}
K.~Schwarzschild.
\newblock {\"U}ber das {G}leichgewicht des {S}onnenatmosph\"are.
\newblock \emph{Nachr. Kgl. Ges. d. Wiss. zu G\"ott. Math. Phys. Klasse},
  1:\penalty0 41--53, 1906.

\bibitem[Scott(2001)]{scott2001}
N.H. Scott.
\newblock Thermoelasticity with thermomechanical constraints.
\newblock \emph{{I}nternational {J}ournal of {N}on-{L}inear {M}echanics},
  36:\penalty0 549--564, 2001.

\bibitem[Sotin and Labrosse(1999)]{sl99}
L.~Sotin and S.~Labrosse.
\newblock {T}hree-dimensional thermal convection in an iso-viscous, infinite
  {P}randtl number fluid heated from within and from below: applications to the
  transfer of heat through planetary mantles.
\newblock \emph{Phys. {E}arth and {P}lanet. {I}nt.}, 112:\penalty0 171--190,
  1999.

\bibitem[Spiegel and Veronis(1971)]{sv1960}
E.~A. Spiegel and G.~Veronis.
\newblock On the boussinesq approximation for a compressible fluid.
\newblock \emph{Astrophys.\ J.}, 131:\penalty0 442--447, 1971.

\bibitem[Spiegel(1965)]{spiegel}
E.A. Spiegel.
\newblock Convective instability in a compressible atmosphere. {I}.
\newblock \emph{Astrophys. Journal}, 141\penalty0 (3):\penalty0 1068--1090,
  1965.

\bibitem[Tilgner(2011)]{Til11}
A~Tilgner.
\newblock Convection in an ideal gas at high {R}ayleigh numbers.
\newblock \emph{Physical Review E}, 84\penalty0 (2):\penalty0 026323, 2011.

\bibitem[Vasil et~al.(2013)Vasil, Lecoanet, Brown, Wood, and
  Zweibel]{vlbwz2013}
G.M. Vasil, D.~Lecoanet, B.P. Brown, T.~Wood, and E.G. Zweibel.
\newblock {E}nergy conservation and gravity waves in sound-proof treatments of
  stellar interiors. {II}. {L}agrangian constrained analysis.
\newblock \emph{{T}he {A}strophysical {J}ournal}, 773\penalty0 (169), 2013.

\bibitem[Viallet et~al.(2013)Viallet, Meakin, Arnett, and Moc\'ak]{Viallet2013}
M.~Viallet, C.~Meakin, D.~Arnett, and M.~Moc\'ak.
\newblock Turbulent convection in stellar interiors. {III}. {M}ean-field
  analysis and stratification effects.
\newblock \emph{The Astrophysical Journal}, 769\penalty0 (1), 2013.

\bibitem[Wang(2004)]{wang2004}
X.~Wang.
\newblock Infinite {P}randtl number limit of {R}ayleigh-{B}\'enard convection.
\newblock \emph{Communications on pure and applied mathematics}, 57\penalty0
  (10):\penalty0 1265--1282, 2004.

\end{thebibliography}

\end{document}